\documentclass[%
twocolumns,
superscriptaddress,
showpacs,preprintnumbers,
 amsmath,amssymb,
 aps,
prc,
longbibliography,
floatfix,
lengthcheck,%
]{revtex4}

\usepackage{wrapft}
\usepackage{subfigure}
\usepackage{ulem}
\usepackage[dvipdfm]{graphicx,psfrag,color}
\usepackage{comment}
\usepackage{amsmath,amssymb}
\usepackage{type1cm}
\usepackage{bm}

\usepackage{graphicx}
\usepackage{dcolumn}
\usepackage{bm}
\usepackage{hyperref}

\newcommand{\bra}[1]{\left\langle #1 \right|}

\newcommand{\ket}[1]{\left| #1 \right\rangle}

\newcommand{\Dhatp}{\hat{D}^{(+)}}
\newcommand{\Dp}{D^{(+)}}
\newcommand{\Ddotp}{\dot{D}^{(+)}}

\newcommand{\Hhat}{\hat{H}}

\newcommand{\Ihat}{\hat{I}}

\newcommand{\Jc}{{\cal J}}

\renewcommand{\b}{\beta}
\newcommand{\g}{\gamma}

\newcommand{\bg}{\beta,\gamma}

\newcommand{\Nhat}{\hat{N}}

\newcommand{\Dc}{{\cal D}}

\newcommand{\Dhat}{\hat{D}}

\newcommand{\Qhat}{\hat{Q}}

\newcommand{\Phat}{\hat{P}}

\newcommand{\del}{\partial}

\newcommand{\beq}{\begin{equation}}
\newcommand{\beqa}{\begin{eqnarray}}
\newcommand{\eeq}{\end{equation}}
\newcommand{\eeqa}{\end{eqnarray}}

\renewcommand{\thanks}{\footnote}

\begin{document}

\preprint{APS/123-QED}

\title{Shape transition and fluctuation in neutron-rich Cr isotopes around $N=40$}

\author{Koichi Sato}
 \affiliation{RIKEN Nishina Center, Wako 351-0198, Japan}
\author{Nobuo Hinohara}
\affiliation{Department of Physics and Astronomy, University of North Carolina, Chapel Hill, North Carolina 27599-3255, USA}
\affiliation{RIKEN Nishina Center, Wako 351-0198, Japan}
\author{Kenichi Yoshida}
\affiliation{Graduate School of Science and Technology, Niigata University, Niigata 950-2181, Japan}
\affiliation{Department of Physics, Faculty of Science, Niigata University, Niigata 950-2181, Japan}
\author{Takashi~Nakatsukasa}
\affiliation{RIKEN Nishina Center, Wako 351-0198, Japan}
\author{Masayuki Matsuo}
\affiliation{Graduate School of Science and Technology, Niigata University, Niigata 950-2181, Japan}
\affiliation{Department of Physics, Faculty of Science, Niigata University, Niigata 950-2181, Japan}
\author{Kenichi Matsuyanagi}
\affiliation{RIKEN Nishina Center, Wako 351-0198, Japan}
\affiliation{%
Yukawa Institute for Theoretical Physics, Kyoto University, 
Kyoto 606-8502, Japan
}%
\date{\today}

\begin{abstract}
The spherical-to-prolate shape transition in neutron-rich Cr isotopes 
from $N=34$ to 42 is studied 
by solving the collective Schr\"odinger equation for the five-dimensional
quadrupole collective Hamiltonian. The collective potential and inertial functions are 
microscopically derived with use of the constrained Hartree-Fock-Bogoliubov 
plus local quasiparticle random-phase approximation method. 
Nature of the quadrupole collectivity of low-lying states is discussed 
by evaluating excitation spectra and electric quadrupole moments and transition strengths. 
The result of calculation indicates that Cr isotopes around $^{64}$Cr are  
prolately deformed but still possess transitional character;  
large-amplitude shape fluctuations dominate in their low-lying states.   
\end{abstract}

\maketitle

\section{Introduction}

Recent experiments on neutron-rich Cr isotopes show that 
quadrupole collectivity appreciably develops toward $^{64}$Cr with $N=40$  
\cite{Sorlin2003, Burger2005, Zhu2006, aoi08, Aoi2009, Gade2010}. 
Going from $^{58}$Cr to $^{64}$Cr, 
the excitation energy of the first excited $2_1^+$ state decreases and  
$R_{4/2}$, the ratio of the excitation energy of 
the $4_1^+$ state to that of the $2_1^+$ state, increases. 
These data seem to indicate that 
a quantum phase transition from the spherical to deformed shapes 
takes place near $N=40$. 
The microscopic origin of the enhanced quadrupole collectivity toward $N=40$ 
has been actively discussed from various theoretical approaches: 
the Hartree-Fock-Bogoliubov (HFB) mean-field calculations 
using the Skyrme force \cite{oba08} or the Gogny force \cite{gau09}, 
the spherical shell-model \cite{Lenzi2010, kan08}, 
and the projected deformed shell model \cite{yan10}. 
These calculations have clarified the important role of 
the neutron $g_{9/2}$ and $d_{5/2}$ single-particle levels 
in the emergence of the quadrupole collectivity near $N=40$. 
Although the spherical shell model calculations reproduce the 
experimental data rather well, the character of the quadrupole deformation, 
especially, the distinction between the equilibrium shape and 
shape fluctuations around it is not sufficiently clear.  
  
In this paper, we investigate the nature of the quadrupole collectivity 
in low-lying states of the neutron-rich Cr isotopes $^{58-64}$Cr 
using an approach that treats the quadrupole deformations
as dynamical variables.
Thus, the distinction of
the equilibrium shape and shape fluctuations is transparent. 
The deformation energy curve with respect to the axial quadrupole deformation
was obtained in the Skyrme HFB mean-field 
calculation \cite{oba08}, which shows that the quadrupole instability
occurs around $N=38-42$.
However, the deformed minima are extremely shallow in these nuclei,
suggesting a transitional character.
In such transitional situations, one naturally expects that 
large-amplitude shape fluctuations play an important role 
in  determining the properties of low-lying excited states. 
Therefore, we take the five-dimensional (5D) quadrupole collective 
Hamiltonian approach \cite{Baranger1968}, 
which is capable of describing the large-amplitude quadrupole shape fluctuations
associated with the quantum shape transition.
It enables us to treat a variety of quadrupole deformation phenomena  
(vibrational, spherical-prolate transitional, rotational, $\gamma$-unstable,  
triaxial, oblate-prolate shape-coexistent situations, etc.) on an equal footing.
Dynamical variables of the 5D quadrupole collective Hamiltonian approach are  
the magnitude and triaxiality of quadrupole deformation $(\beta,\gamma)$ 
and the three Euler angles. 
To explicitly indicate the two vibrational and 
three rotational degrees of freedom,
we call it (2+3)D, hereafter.
The 5D collective Hamiltonian is characterized by seven functions: the collective potential,
three vibrational inertial functions (also called vibrational masses), 
and three rotational inertial functions.
To evaluate the inertial functions, the Inglis-Belyaev (IB) cranking formula
has been conventionally used.
However, it is well known that the contribution of the time-odd components of 
the moving mean field is ignored in the IB cranking formula,
which leads to the overestimation of excitation energies \cite{hin08, hin09}.

The constrained Hartree-Fock-Bogoliubov plus local 
quasiparticle random-phase approximation (CHFB+LQRPA) method \cite{Hinohara2010} 
is a method which can overcome the shortcoming of the IB cranking formula.
It is derived on the basis of the adiabatic self-consistent collective coordinate 
(ASCC) method \cite{hin09, mat00, hin07}  
by assuming that there is a one-to-one mapping from a point  
on the collective submanifold embedded
in the large-dimensional time-dependent HFB phase space 
to a point in the $(\beta, \gamma)$ deformation space.
In the CHFB+LQRPA method, the inertial functions are derived
by transforming the local canonical coordinates determined by 
the LQRPA normal modes to the $(\beta, \gamma)$ degrees of freedom.
This method has been successfully applied to several phenomena:
shape coexistence/fluctuation in Se and Kr isotopes \cite{hin08, Hinohara2010, Sato2011}, 
development of triaxial deformation in $^{110}$Mo \cite{Watanabe2011},
and shape fluctuations in neutron-rich Mg isotopes \cite{Hinohara2011}.  
Use of the Skyrme energy density functional in solving the CHFB+LQRPA equations   
has also been initiated for the axially symmetric quadrupole Hamiltonian \cite{Yoshida2011}.
In this paper,  we solve the LQRPA equations  
with use of the pairing-plus-quadrupole (P+Q) model \cite{Baranger1968}
including the quadrupole pairing interaction. 
For the collective Hamiltonian quantized according to the Pauli prescription,
we solve the collective Schr\"odinger equation to obtain the excitation
energies, vibrational wave functions, $E2$-transition strengths and moments.

This paper is organized as follows. 
We recapitulate the theoretical framework in Sec. \ref{Theoretical Framework}.
In Sec. \ref{Results and Discussions}, 
we present results of calculation for $^{58-66}$Cr and discuss 
the nature of quadrupole collectivity in their low-lying states.  
To examine the role of the shape fluctuations toward the triaxial shape  
($\gamma$ degree of freedom), we also carry out  
calculations imposing the axial symmetry on the nuclear shape. 
We call it (1+2)D calculation indicating that the one vibrational 
degree of freedom in the $\beta$ direction and the two Euler angles 
describing rotations perpendicular to the symmetry axis are taken into account. 
Comparing results of such a restricted calculation with those of 
the (2+3)D calculation, we shall see the importance of the triaxial shape fluctuations.   
We carry out a similar calculation also for $^{66}$Fe to compare with $^{64}$Cr. 
We then discuss similarities and differences of the quadrupole shape transition 
near $^{64}$Cr with $N=40$ and that near $^{32}$Mg with $N=20$. 
Conclusions are given in Sec. \ref{Conclusions}.

\section{Theoretical Framework}
\label{Theoretical Framework}
In this section, we briefly summarize the framework of our collective Hamiltonian
approach. See Ref.~\cite{Hinohara2010} for details.

\subsection{5D quadrupole collective Hamiltonian}
The 5D quadrupole collective Hamiltonian is given by 
\begin{align}
 \mathcal{H}_{\rm coll}&= T_{\rm vib} + T_{\rm rot} + V(\bg), \label{eq:Hc} \\
 T_{\rm vib}   &= \frac{1}{2} D_{\beta\beta}(\bg)\dot{\beta}^2 +
 D_{\beta\gamma}(\bg)\dot{\beta}\dot{\gamma}
+ \frac{1}{2}D_{\gamma\gamma}(\bg)\dot{\gamma}^2, \label{eq:Tvib}\\
 T_{\rm rot} &= \frac{1}{2}\sum_{k=1}^3 \Jc_k(\bg) \omega^2_k, 
\label{eq:collH_BM} 
\end{align}
where  $T_{\rm vib}$ and $T_{\rm rot}$ represent  
the vibrational and rotational kinetic energies, 
while $V$ the collective potential energy.
The moments of inertia are parametrized as
$\mathcal{J}_{k}(\beta, \gamma)=4\beta^2 D_{k}(\beta,\gamma) \sin^2\gamma_k$ with $\gamma_k=\gamma -2\pi k/3$.

The collective potential and inertial functions are determined 
with the CHFB+LQRPA method as explained in the next subsection.
Once they are determined as functions of $(\beta,\gamma)$, 
we quantize the collective Hamiltonian according to the Pauli prescription.
The collective Schr\"odinger equation for the quantized collective Hamiltonian 
is given by
\begin{align}
 \{\hat{T}_{\rm vib} + \hat{T}_{\rm rot} + V \}
 \Psi_{\alpha IM}(\bg,\Omega) = E_{\alpha I} \Psi_{\alpha IM}(\bg,\Omega), 
\label{eq:Schroedinger}
\end{align}
where  
\begin{align}
& \hat{T}_{\rm vib} = \frac{-1}{2\sqrt{WR}} 
 \left\{ 
   \frac{1}{\beta^4} 
   \left[
     \left(
       \del_\beta 
       \beta^2 \sqrt{\frac{R}{W}} D_{\gamma\gamma} \del_\beta 
     \right)
   \right.  
 \right. \notag \\
&   \left.
 \left.
     - \del_\beta 
     \left(
       \beta^2 \sqrt{\frac{R}{W}} D_{\beta\gamma} \del_\gamma
     \right)
   \right] 
 \right.  \notag \\
& 
  +  \frac{1}{\beta^2 \sin 3\gamma} 
 \left[
   -\del_\gamma 
   \left( 
     \sqrt{\frac{R}{W}} \sin 3\gamma D_{\beta\gamma} \del_\beta 
   \right) 
 \right. \notag \\
&
\left.
 \left.
   + \del_\gamma 
   \left(
     \sqrt{\frac{R}{W}} \sin 3\gamma
     D_{\beta\beta} \del_\gamma 
   \right)
 \right] 
\right\}
\end{align}
and
\begin{align}
\hat{T}_{\rm rot} = \sum_k \frac{ \Ihat_k^2}{2\Jc_k}. 
\end{align}
Here, $R(\b,\g)$ and $W(\b,\g)$ are defined as
\begin{align}
 R(\bg) =& D_1(\bg) D_2(\bg) D_3(\bg),\\
 W(\bg) =& \left\{
 D_{\beta\beta}(\bg) D_{\gamma\gamma}(\bg) - [ D_{\beta\gamma}(\bg)]^2
\right\} \beta^{-2}. 
\end{align}
The collective wave function $\Psi_{\alpha I M}(\b,\g,\Omega)$ is specified by the total
angular momentum $I$, its projection onto the $z$-axis of the laboratory frame $M$, and
$\alpha$ distinguishing the states with the same $I$ and $M$.
It can be written as a sum of products of the vibrational and rotational wave functions: 
\begin{align}
 \Psi_{\alpha IM}(\bg,\Omega) = 
 \sum_{K={\rm even}}\Phi_{\alpha IK}(\bg)\langle\Omega|IMK\rangle, \label{eq:Psi}
\end{align}
where
\begin{align}
\langle \Omega | IMK \rangle = 
\sqrt{\frac{2I+1}{16\pi^2 (1+\delta_{K0})}} [\Dc^{I}_{MK}(\Omega) + (-)^I \Dc^I_{M-K}(\Omega)].
\end{align}
$\Dc^I_{MK}$ is the Wigner rotation matrix 
and $K$ is the projection of the angular momentum
onto the $z$-axis in the body-fixed frame.
The summation over $K$ is taken from 0 to $I$ for even $I$ and from 2 to $I-1$ for odd $I$. 

The vibrational wave functions in the body-fixed frame, $\Phi_{\alpha IK}(\bg)$,  
are normalized as
\begin{align} \label{eq:normalization}
 \int d\beta d\gamma  |\Phi_{\alpha I}(\bg)|^2 |G(\bg)|^{\frac{1}{2}} = 1, 
\end{align}
where  
\begin{align}
 |\Phi_{\alpha I}(\bg)|^2 \equiv \sum_{K={\rm even}} |\Phi_{\alpha IK}(\bg)|^2,
\end{align}
and the volume element $|G(\bg)|$ is given by
\begin{align} \label{eq:metric}
 |G(\bg)| = 4\beta^8 W(\bg)R(\bg) \sin^2 3\gamma.
\end{align}
The symmetries and boundary conditions of the collective Hamiltonian and 
wave functions are discussed in Ref. \cite{Kumar1967}.

\subsection{The CHFB+LQRPA method}

We determine the collective potential and inertial functions
with the CHFB+LQRPA method. 
We first solve the CHFB equation
\begin{align}
 \delta \bra{\phi(\bg)} \Hhat_{\rm CHFB}(\bg) \ket{\phi(\bg)} = 0,  \label{eq:CHFB}  \\
 \Hhat_{\rm CHFB}=\Hhat-\sum_\tau\lambda^{(\tau)}\hat N^{(\tau)}
 -\sum_m\mu^{(m)}\hat D_{2m}^{(+)}
\end{align}
with the constraints on the particle numbers and quadrupole deformation parameters:
\begin{align}
 \bra{\phi(\bg)} \Nhat^{(\tau)} \ket{\phi(\bg)} = N^{(\tau)}_0,  
\quad (\tau = n, p) \\
 \bra{\phi(\bg)} \Dhatp_{2m} \ket{\phi(\bg)} = \Dp_{2m}, \quad (m = 0, 2). \label{eq:CHFB4}
\end{align}
Here, $\Dhatp_{2m}$ denotes Hermitian quadrupole operators, 
$\Dhat_{20}$ and $(\Dhat_{22} + \Dhat_{2-2})/2$ 
for  $m = 0$ and  2, respectively. 
Then, we solve the LQRPA equations for vibration on top of the CHFB states obtained above,
\begin{align}
 \delta  \bra{\phi(\bg)} [ \Hhat_{\rm CHFB}(\bg), & \Qhat^i(\bg) ] \notag \\
  - \frac{1}{i}& \Phat_i(\bg) \ket{\phi(\bg)}  = 0, \label{eq:LQRPA1} \\
 \delta  \bra{\phi(\bg)} [ \Hhat_{\rm CHFB}(\bg), \frac{1}{i} & \Phat_i(\bg)] \notag \\
  - C_i(\bg) \Qhat^i(\bg) \ket{\phi(\bg)}  & = 0,  \quad\quad (i=1, 2). 
\label{eq:LQRPA2}
\end{align}
The infinitesimal generators, 
$\Qhat^i(\bg)$ and $\Phat_i(\bg)$, 
are locally defined at every point of the $(\beta, \gamma)$ deformation space. 
The quantity $C_i(\bg)$ is related to 
the eigenfrequency $\omega_i(\bg)$ of the local normal mode  
through $\omega_i^2(\bg)=C_i(\bg)$. 
It is worth noting that these equations are valid also for regions with negative curvature 
($C_i(\bg)<0$) where $\omega_i(\bg)$ takes an imaginary value. 

We assume that a 2D collective submanifold is embedded 
in a large-dimensional TDHFB configuration space 
and define local canonical coordinates $(q_1,q_2)$ on it. 
We also assume that there is a one-to-one mapping between  
$(q_1,q_2)$ and $(\beta, \gamma)$. 
By performing a similarity transformation of $(q_1,q_2)$,
the vibrational kinetic energy can be written, without loss of generality, as  
\beq
T_{\rm vib}=\frac{1}{2}\sum_{i=1, 2}\dot q_i^2, 
\eeq
which can be transformed to 
\begin{align}
 T_{\rm vib} = \frac{1}{2} M_{00} [\Ddotp_{20}]^2
  + M_{02} \Ddotp_{20} \Ddotp_{22}+ \frac{1}{2} M_{22} [\Ddotp_{22}]^2,
\label{eq:TvibD}
\end{align}
in terms of the time derivatives of $D_{2m}^{(+)}$.
The vibrational masses with respect to $D_{2m}^{(+)}$ are given by 
\begin{align}
 M_{mm'}(\bg) = \sum_{i=1,2} \frac{\del q^i}{\del
 \Dp_{2m}} \frac{\del q^i}{\del \Dp_{2m'}}. \label{eq:M_mm}
\end{align} 
The partial derivatives can be easily evaluated as 
\begin{align}
 \frac{\del \Dp_{20}}{\del q^i} =& \frac{\del}{\del q^i}\bra{\phi(\bg)}
 \Dhatp_{20} \ket{\phi(\bg)} \nonumber \\
  =& \bra{\phi(\bg)} [\Dhatp_{20}, \frac{1}{i}\Phat_i 
 (\bg)] \ket{\phi(\bg)}, \\
 \frac{\del \Dp_{22}}{\del q^i} =& \frac{\del}{\del q^i}\bra{\phi(\bg)}
 \Dhatp_{22} \ket{\phi(\bg)} \nonumber \\
  =& \bra{\phi(\bg)} [\Dhatp_{22}, \frac{1}{i}\Phat_i
 (\bg)] \ket{\phi(\bg)}, 
\end{align}
without need of numerical derivatives. 
Through the definition 
of ($\beta, \gamma$) in terms of 
($\Dp_{20}, \Dp_{22}$), the vibrational inertial functions 
($D_{\beta\beta},  D_{\gamma\gamma}, D_{\beta\gamma}$) 
with respect to ($\beta, \gamma$) are easily obtained from 
($M_{00}, M_{02}, M_{22}$) \cite{Hinohara2010}. 
To select two collective normal modes among the LQRPA modes 
obtained by solving Eqs.~(\ref{eq:LQRPA1}) and (\ref{eq:LQRPA2}) ,
we employ the minimal metric criterion \cite{Hinohara2010}. 

The rotational moments of inertia are calculated 
by solving the LQRPA equations for rotation on each CHFB state:
\begin{align}
 \delta \bra{\phi(\bg)} [\Hhat_{\rm CHFB}, \hat{\Psi}_k(\bg)]
- \frac{1}{i} (\Jc_k)^{-1} \Ihat_k \ket{\phi(\bg)} &= 0, \\ 
\bra{\phi(\bg)} [\hat{\Psi}_k(\bg), \Ihat_{k'}]\ket{\phi(\bg)} &= i \delta_{kk'},
\end{align}
where $\Ihat_k$ and $\hat{\Psi}_k(\bg)$ represent the angular momentum 
and the rotational angle operators with respect to the principal axes
associated with the CHFB state $\ket{\phi(\bg)}$.
This is an extension of the Thouless-Valatin equation \cite{Thouless1962} 
for the HFB equilibrium state to non-equilibrium CHFB states.  
We call  $\Jc_k(\bg)$ determined by the above equations 
`LQRPA moments of inertia.' 

We solve the collective Schr\"odinger equation (\ref{eq:Schroedinger}) 
to obtain excitation energies and vibrational wave functions. 
Then, electric transition strengths and moments are readily calculated  
(see Ref.~\cite{Sato2011} for details).

\subsection{Details of the numerical calculation}

In this study, we adopt a version of the pairing-plus-quadrupole (P+Q) model 
\cite{Baranger1968} 
including the quadrupole pairing interaction as well as the monopole pairing interaction.
We take two harmonic-oscillator shells 
with $N_{sh}$=3, 4 and $N_{sh}$=2, 3 for neutrons and protons, respectively.
The single-particle energies are determined from those obtained with the
constrained Skyrme-HFB calculations at the spherical shape 
using the HFBTHO code \cite{Stoitsov2005}.
In these Skyrme-HFB calculations, we employ the SkM* functional and 
the volume-type pairing with the pairing strength $V_0=-180$ MeV fm$^{-3}$.
The pairing strength has been adjusted such that 
the calculated neutron pairing gaps at the HFB minima reproduce
the experimental gaps in $^{58-64}$Cr. 
The single-particle energies 
are scaled with the effective mass of the SkM* functional $m^*/m=0.79$, 
because the P+Q model is designed to be
used for single particles whose mass is the bare nucleon mass. 

The parameters of the P+Q model are determined as follows.
For $^{62}$Cr, the monopole pairing strengths $G_0^{\tau} (\tau=n,p)$ and 
quadrupole interaction strength $\chi$ are determined 
to approximately reproduce the HFB equilibrium deformation and 
the paring gaps at the spherical and HFB equilibrium shapes.
For the other nuclei $^{58,60,64,66}$Cr and $^{66}$Fe, we 
assume the simple mass number dependence according to
Baranger and Kumar \cite{Baranger1968}: 
$G_0^{\tau} \sim A^{-1}$ and $\chi^\prime \equiv \chi b^4 \sim A^{-5/3}$ ($b$ denotes the oscillator-length parameter).
We follow the Sakamoto-Kishimoto prescription to 
determine the strengths of the quadrupole pairing \cite{Sakamoto1990}. 
We omit the Fock term as in the conventional treatment of the P+Q model.

The CHFB+LQRPA equations are solved at $60 \times 60$ mesh points 
in the $(\beta, \gamma)$ plane defined by
\begin{align}
 \beta_i  = (i - 0.5) \times 0.01, \quad (i = 1, \cdots 60), \\
 \gamma_j = (j - 0.5) \times 1^\circ, \quad (j = 1, \cdots 60). 
\end{align}
For the calculation of the $E2$ transitions and moments, 
we use the effective charges $(e_{\rm eff}^{(n)}, e_{\rm eff}^{(p)} )=(0.5, 1.5)$.

\section{Results and Discussion}
\label{Results and Discussions}

In this section, we present the numerical results for $^{58-66}$Cr 
and discuss the nature of quadrupole collectivity in their low-lying states.
{ The results for $^{64}$Cr are compared with those for the neighboring nucleus $^{66}$Fe with $N=40$.} 
We furthermore discuss the similarities and differences with 
Mg isotopes around $N=20$. 

\subsection{Collective potentials and inertial functions}

\begin{figure}[h]
\begin{center}
\includegraphics[height=0.25\textwidth,keepaspectratio,clip,trim=60 50 60 50]{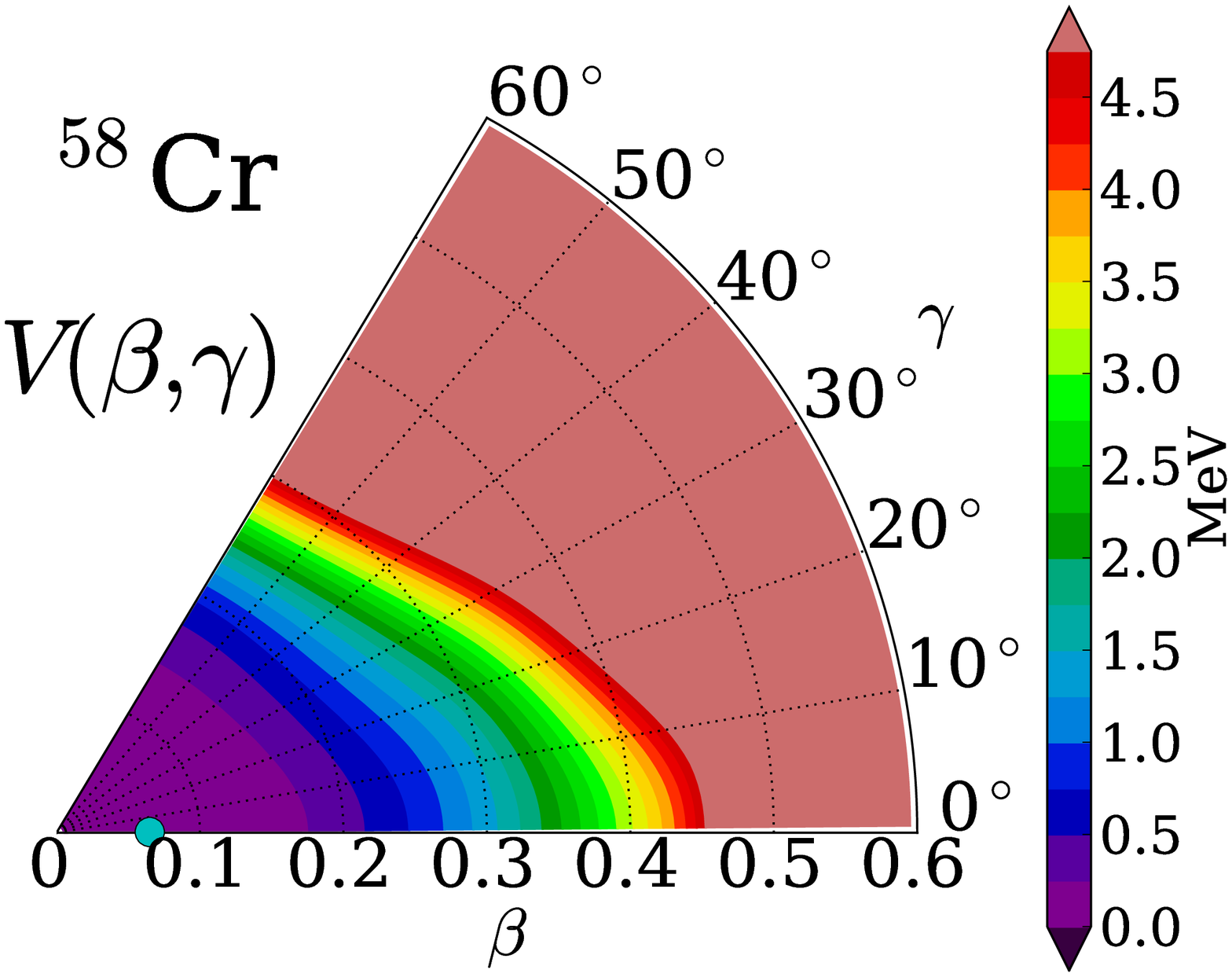}
\includegraphics[height=0.25\textwidth,keepaspectratio,clip,trim=60 50 60 50]{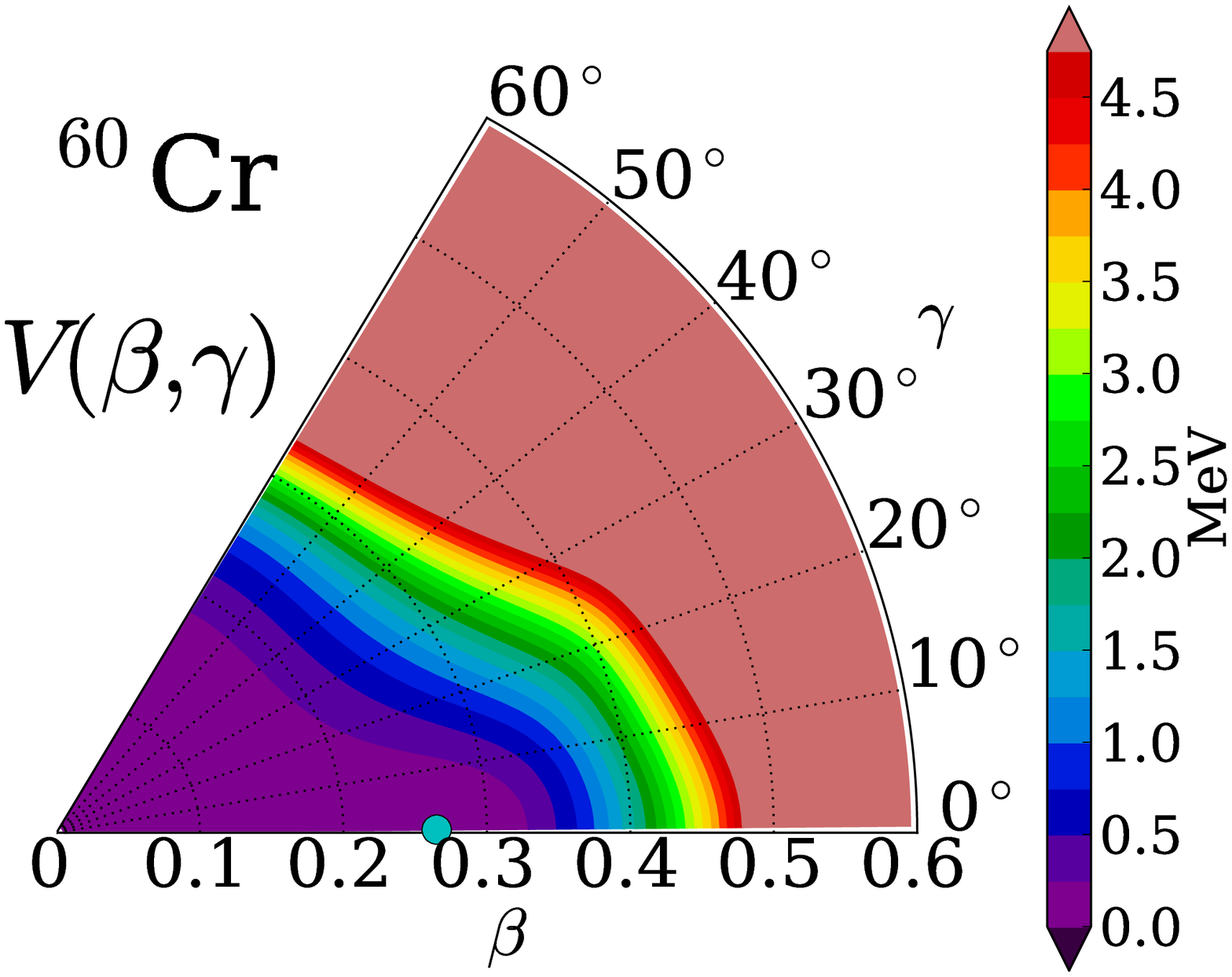} 
\includegraphics[height=0.25\textwidth,keepaspectratio,clip,trim=60 50 60 50]{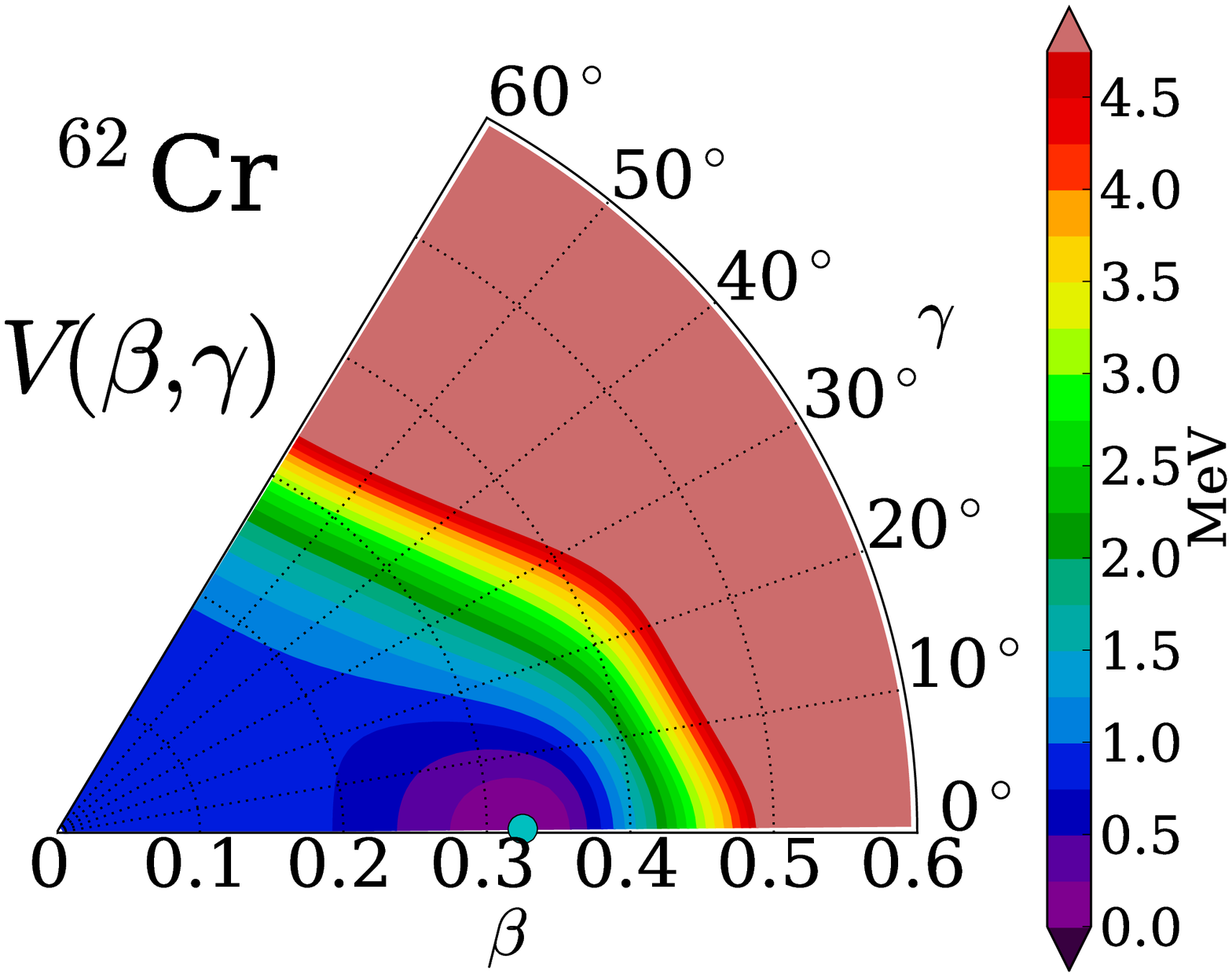} 
\includegraphics[height=0.25\textwidth,keepaspectratio,clip,trim=60 50 60 50]{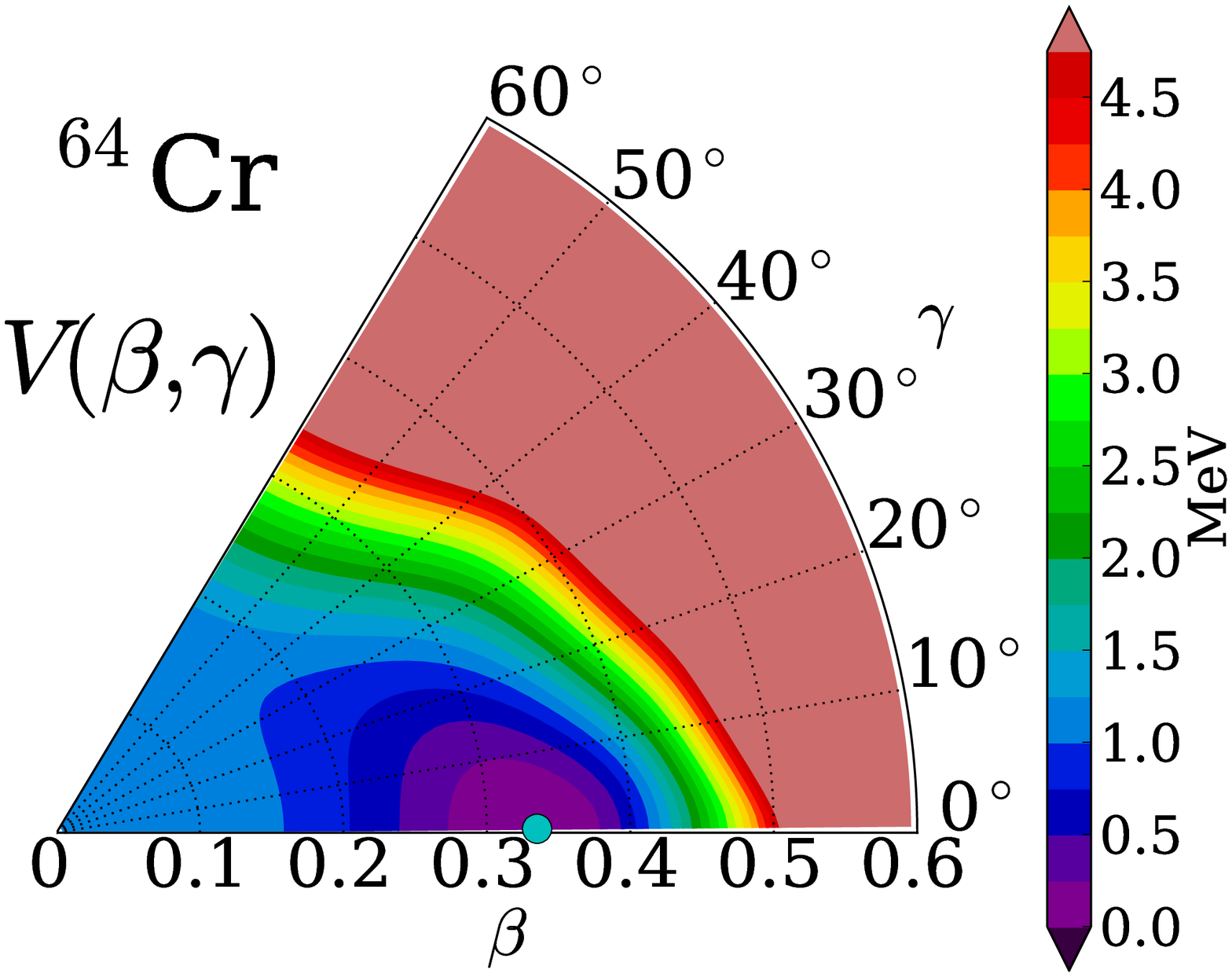} 
\includegraphics[height=0.25\textwidth,keepaspectratio,clip,trim=60 50 60 50]{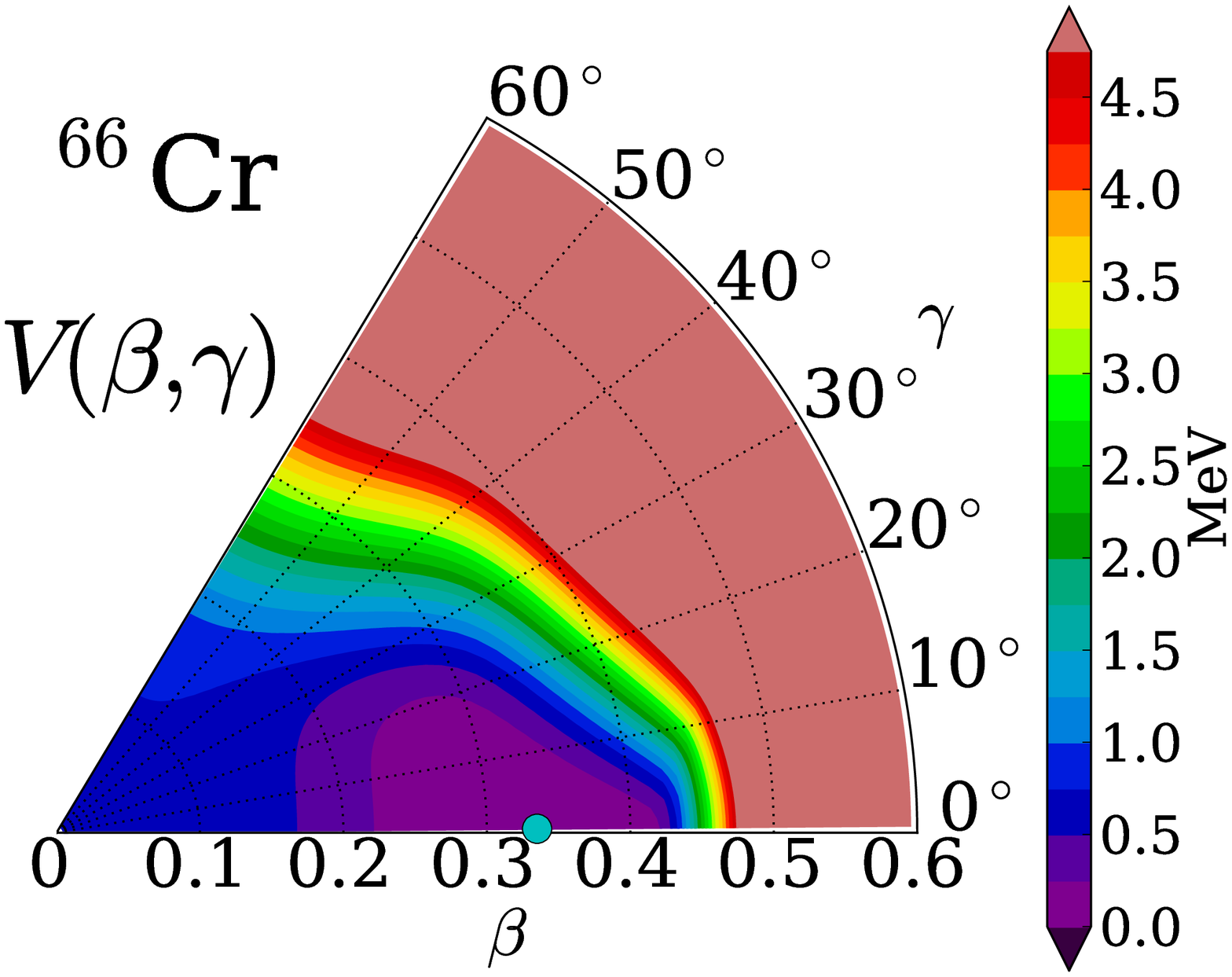} 
\end{center}
\caption{(Color online) Collective potential energy surfaces $V(\beta,\gamma)$ for $^{58-66}$Cr. 
The regions higher than 5 MeV (measured from the HFB minima) 
are colored rosy-brown. }
\label{fig:potential}
\end{figure}

\begin{figure}[h]
\includegraphics[width=0.4\textwidth,keepaspectratio,clip,trim=0 0 0 0]{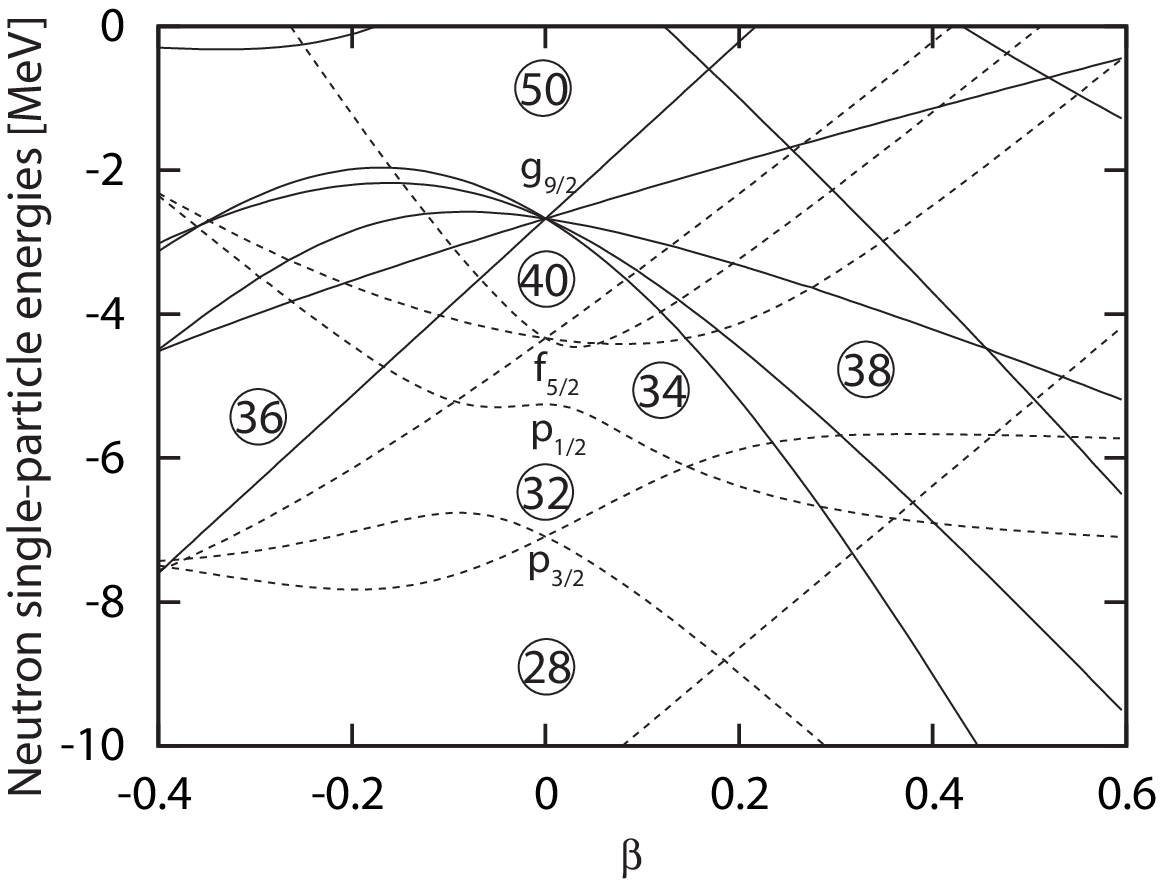} 
\includegraphics[width=0.4\textwidth,keepaspectratio,clip,trim=0 0 0 0]{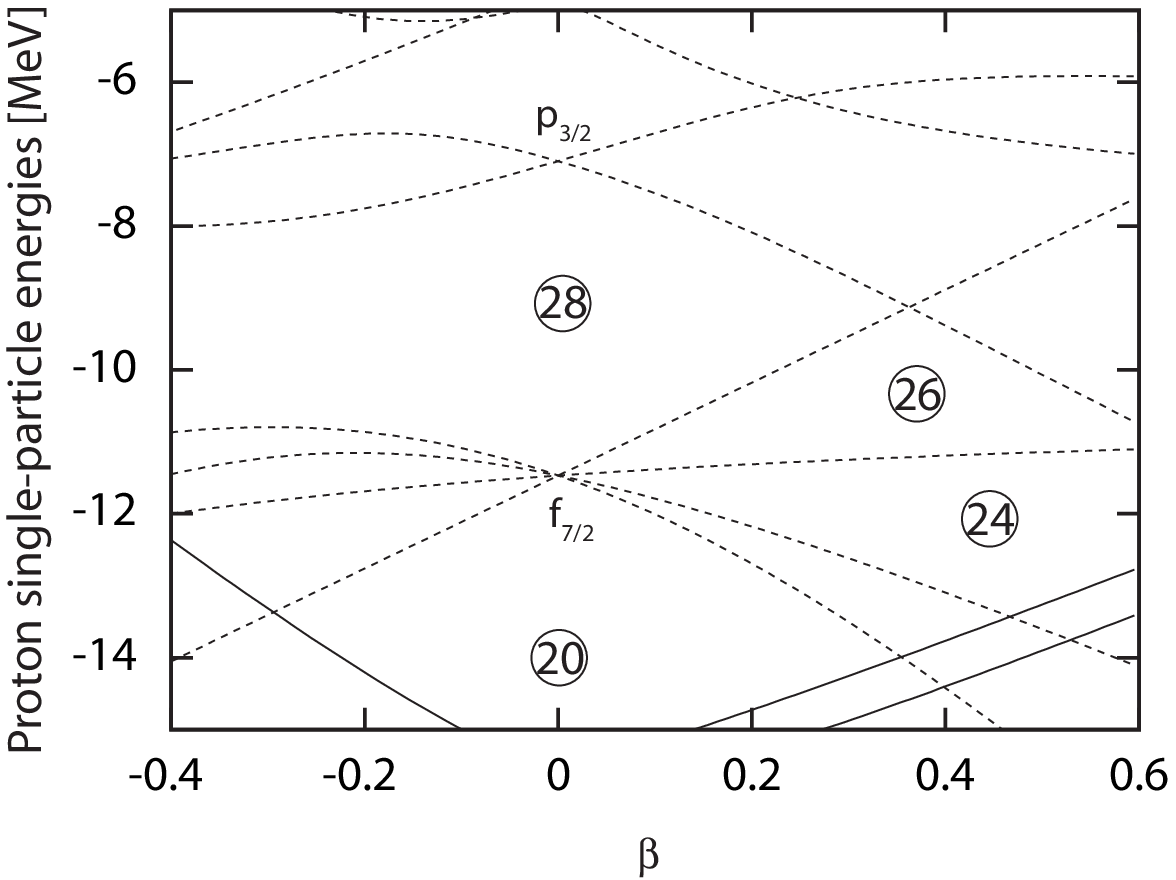} 
\caption{Nilsson diagrams for neutrons (upper) and protons (lower) in $^{62}$Cr as functions of $\beta$, 
calculated as in Ref. \cite{Kumar1968}. The levels with the positive (negative) parity are plotted with solid (dotted) lines.}
\label{fig:single-particle energies}
\end{figure}

We plot the collective potential $V(\beta,\gamma)$ calculated for $^{58-66}$Cr 
in Fig. \ref{fig:potential}.
The location of the absolute minimum is indicated by the (blue) circle.
In $^{58}$Cr, the absolute minimum is located at a nearly spherical shape.
Although the minimum shifts to larger deformation in $^{60}$Cr, 
the collective potential is extremely soft in the $\beta$ direction.
A more pronounced local minimum appears at larger deformation in $^{62}$Cr, 
and the minimum becomes even deeper in $^{64}$Cr.
In $^{66}$Cr, the collective potential becomes slightly softer than in $^{64}$Cr.
These potential energy surfaces indicate that a quantum shape transition 
from a spherical to a prolately deformed shape takes place 
along the isotopic chain toward $N=40$.
In Fig.~\ref{fig:single-particle energies}, we plot the Nilsson diagrams of neutrons and protons as functions of $\beta$ calculated for $^{62}$Cr as in 
Ref. \cite{Kumar1968}.
This is similar to Figs. 5(a) and 5(b) in Ref.~\cite{oba08}.

\begin{figure}[h]
\begin{center}
\includegraphics[width=0.3\textwidth,keepaspectratio,clip,trim=20 20 20 25]{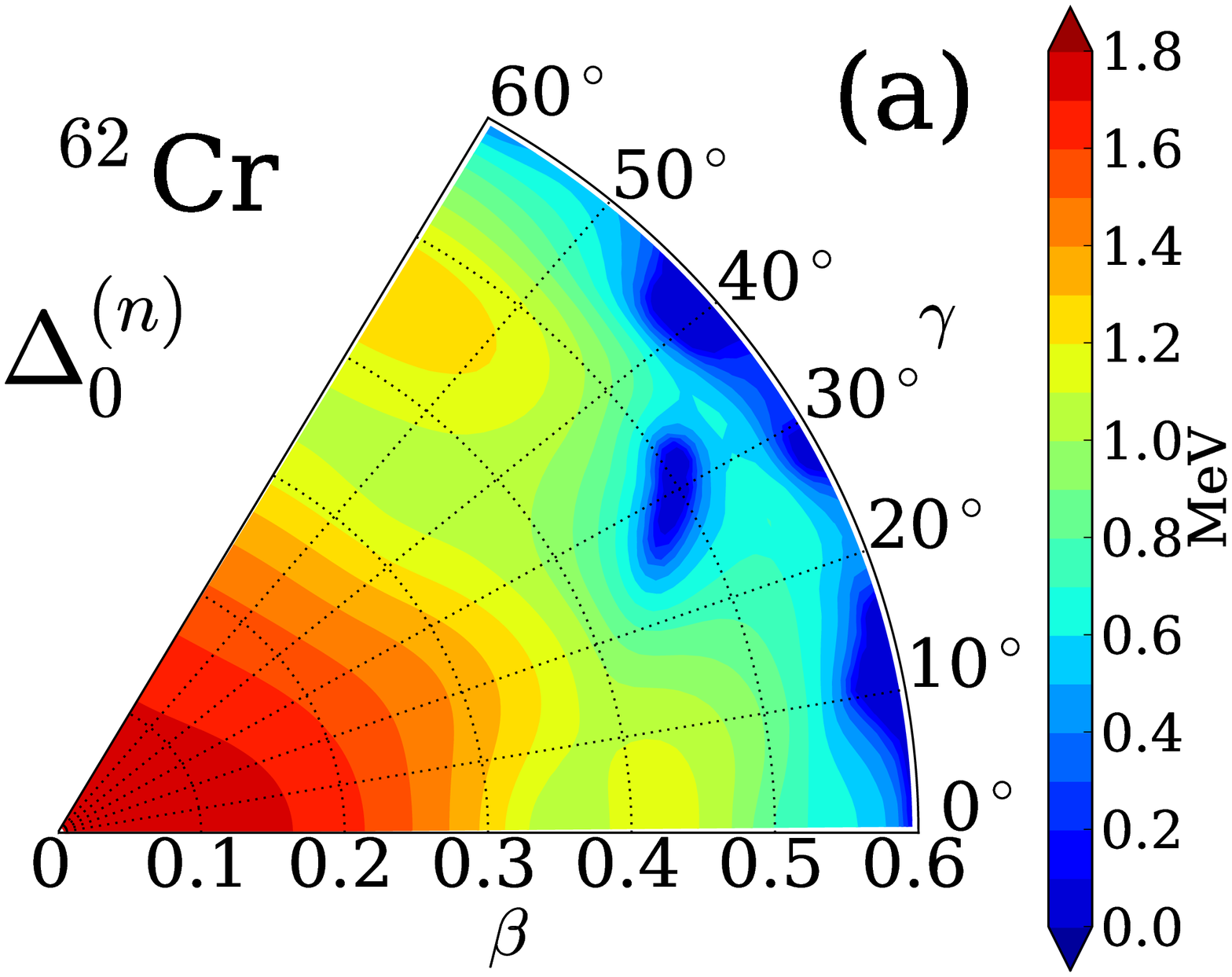}  
\includegraphics[width=0.3\textwidth,keepaspectratio,clip,trim=20 20 20 25]{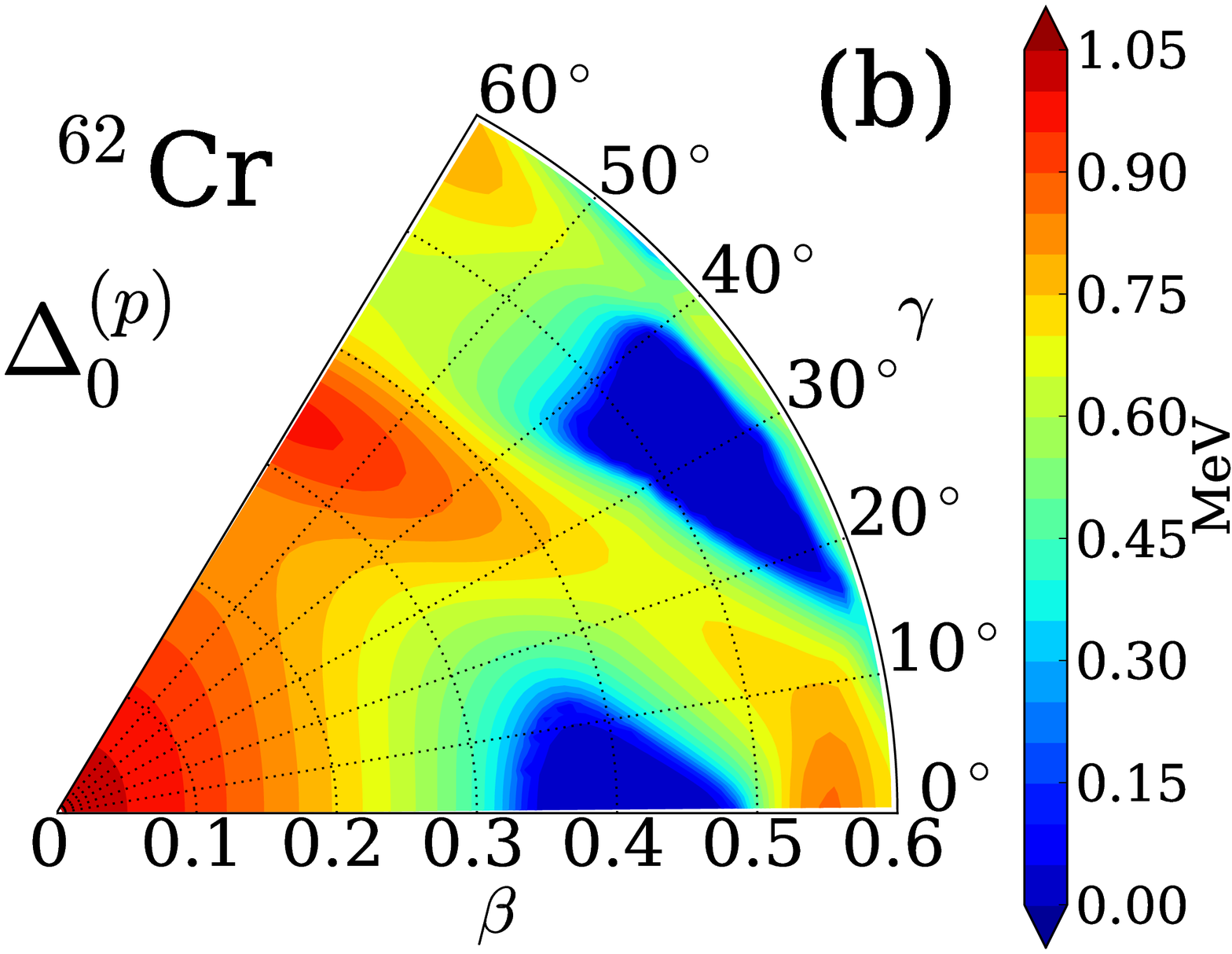}  
\includegraphics[width=0.3\textwidth,keepaspectratio,clip,trim=20 20 20 25]{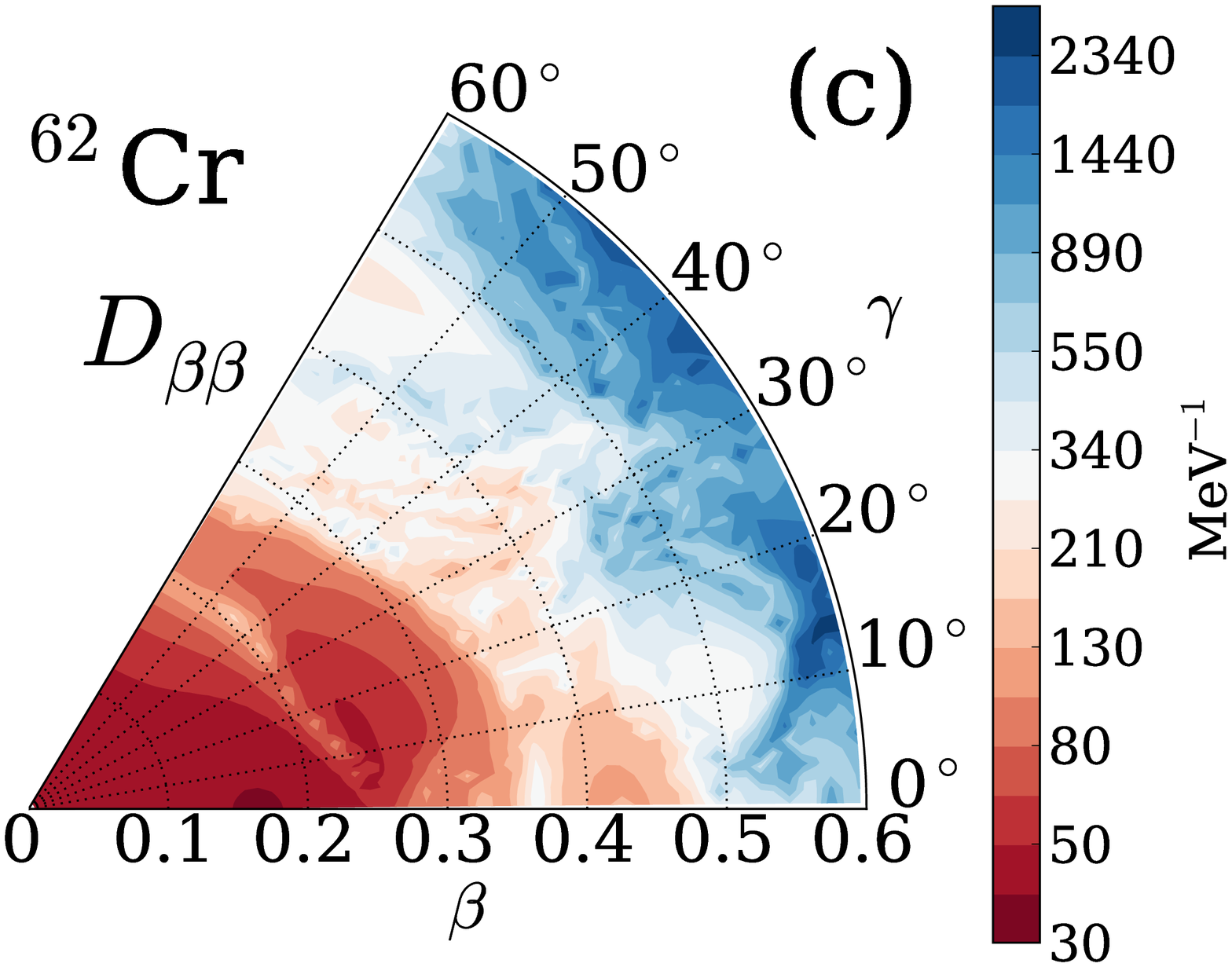} 
\includegraphics[width=0.3\textwidth,keepaspectratio,clip,trim=20 20 20 25]{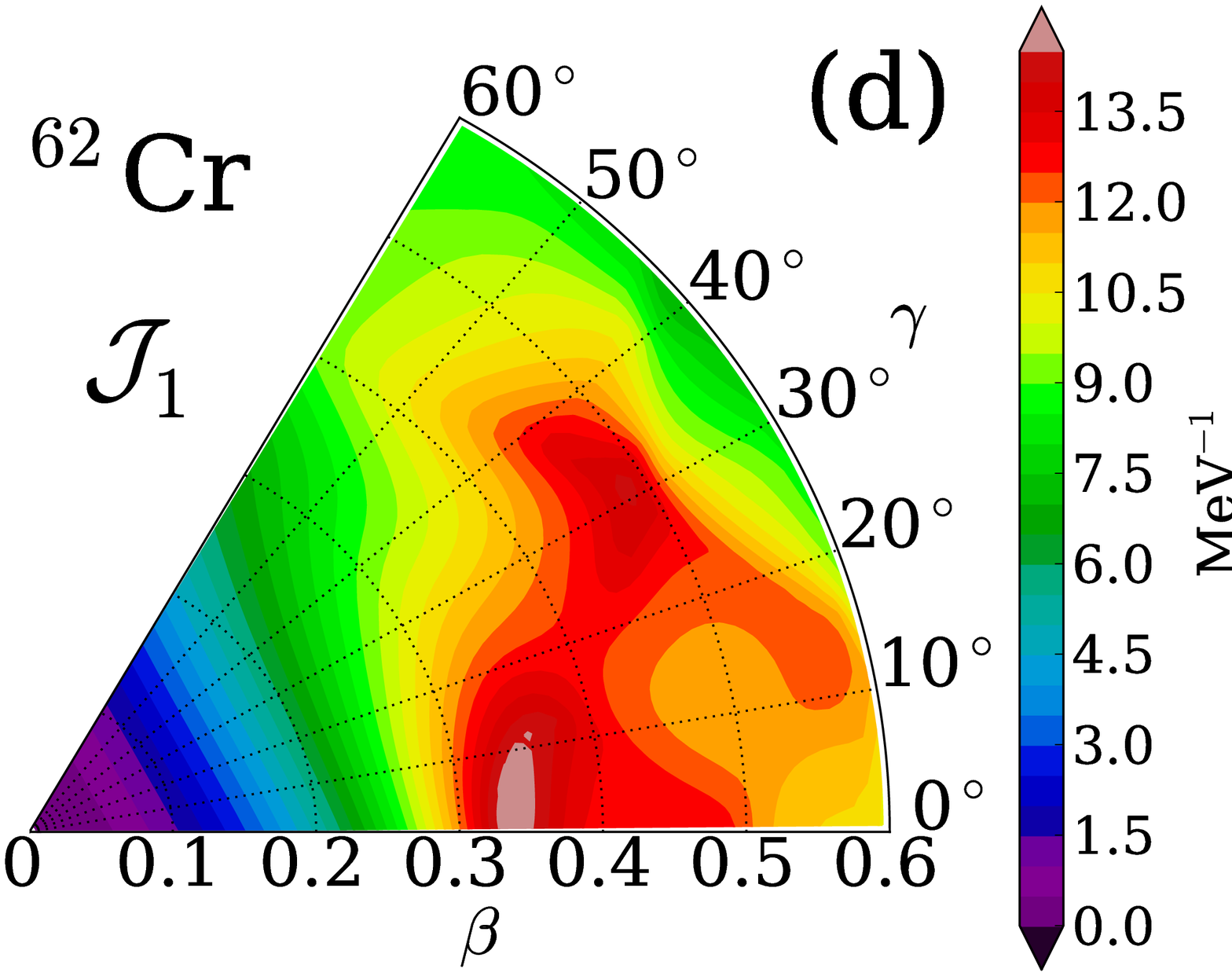}  
\end{center}
\caption{(Color online)
Neutron and proton monopole pairing gaps, 
$\Delta_{0}^{(n)}(\beta.\gamma)$, $\Delta_{0}^{(p)}(\beta.\gamma)$, 
vibrational inertial function $D_{\beta\beta}(\beta,\gamma)$, 
and rotational moment of inertia $\mathcal{J}_1(\beta,\gamma)$,  
calculated for $^{62}$Cr.
}
\label{fig:masses and gaps}
\end{figure}

In Fig. \ref{fig:masses and gaps}, we plot  
the neutron and proton monopole pairing gaps 
$\Delta_{0}^{(n)}(\beta,\gamma)$ and $\Delta_{0}^{(p)}(\beta,\gamma)$, 
the vibrational inertial function $D_{\beta\beta}(\beta,\gamma)$, 
and the rotational moment of inertia $\mathcal{J}_1(\beta,\gamma)$,  
calculated for $^{62}$Cr.
Figures~\ref{fig:masses and gaps} (c) clearly shows  
that the vibrational inertial function is well correlated 
with the magnitudes of the paring gaps:
$D_{\beta\beta}(\beta,\gamma)$ becomes small in the near spherical region 
where $\Delta_0^{(n)}$ and $\Delta_0^{(p)}$ take large values.
One might be concerned for complicated behaviors of $D_{\beta\beta}(\beta,\gamma)$ 
in the strongly deformed region. However, they hardly
affect low-lying states,
because the collective potential energy is very high there and 
contributions from this region to the vibrational wave 
functions are negligibly small. 
Figures~\ref{fig:masses and gaps} (d) clearly indicates   
that the rotational moment of inertia also has a strong correlation 
with the pairing gaps. It takes the maximum value  
in the prolate region around $\beta \simeq 0.35$.
Both the neutron and proton pairing gaps become small there 
due to the deformed shell gaps for $N=38$ and $Z=24$,
see Fig.~\ref{fig:single-particle energies}.  
In particular, the proton pairing gap almost vanishes.
It results in the increase of the moment of inertia.
As we shall see later, this enhancement promotes  
the localization of the vibrational wave functions 
in the $(\beta,\gamma)$ plane 
for excited states with non-zero angular momenta. 
The rotational and vibrational inertial functions for the other isotopes are 
qualitatively the same as those for $^{62}$Cr. 
The enhancement of the moments of inertia mentioned above grows gradually 
with increasing neutron number up to $N=40$.

\subsection{Yrast states in $^{58-66}$Cr}


\begin{figure}[h]
\begin{center}
\includegraphics[width=0.3\textwidth,keepaspectratio,clip,trim=0 0 0 0]{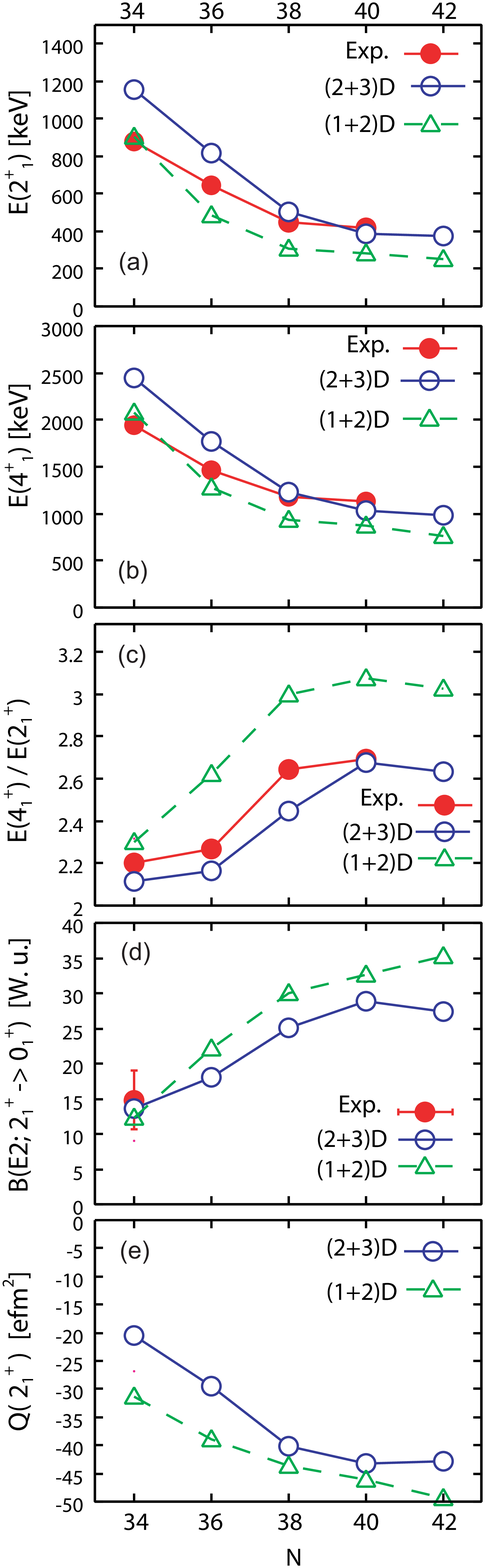} 
\end{center}
\caption{(Color online) 
(a) Excitation energies of the $2_1^+$ states for $^{58-66}$Cr.
(b) Excitation energies of the $4_1^+$ states.
(c) Ratios of $E(4_1^+)$ to $E(2_1^+)$.
(d) Reduced $E2$ transition probabilities $B(E2;2_1^+ \rightarrow 0_1^+)$ in Weisskopf units.
(e) Spectroscopic quadrupole moments of the $2_1^+$ states.
Experimental data are taken from Refs. \cite{Burger2005, Zhu2006, Aoi2009, Gade2010}.
}
\label{fig:EnergyAndE2}
\end{figure}

We show in Fig. \ref{fig:EnergyAndE2} the excitation energies of 
the $2_1^+$ and  $4_1^+$ states, their ratios $R_{4/2}$,
the $E2$ transition strengths $B(E2; 2_1^+ \rightarrow 0_1^+)$, and the spectroscopic 
quadrupole moments of the $2_1^+$ states, 
together with the available experimental data.
The decrease in the excitation energies of the $2_1^+$  and $4_1^+$ states toward $N=40$  
and the increase in their ratio from $N=36$ to $N=40$ are well described and 
indicate that the nature of the quadrupole collectivity gradually changes 
from vibrational to rotational as the neutron number increases.
However, the ratio $R_{4/2}$ at $N=40$ is still 2.68, 
which is considerably smaller than the rigid-rotor value 3.33.
The $B(E2)$ values and spectroscopic quadrupole moments $Q(2_1^+)$ also 
suggest the onset of deformation:
$B(E2)$ increases and the magnitude of the spectroscopic quadrupole 
moments, which has a negative sign indicating a prolate shape,
increase with increasing neutron number 
and both of them reach a maximum at $N=40$.
In Fig. \ref{fig:EnergyAndE2},
we also plot the results of the (1+2)D calculations, 
in which only the axially symmetric deformation is taken into account. 
Physical meaning of the differences between the (2+3)D and (1+2)D calculations
will be discussed in a subsequent subsection. 

\begin{figure*}[h]
\begin{center}
\includegraphics[height=0.15\textwidth,keepaspectratio,clip,trim=60 50 160 50]{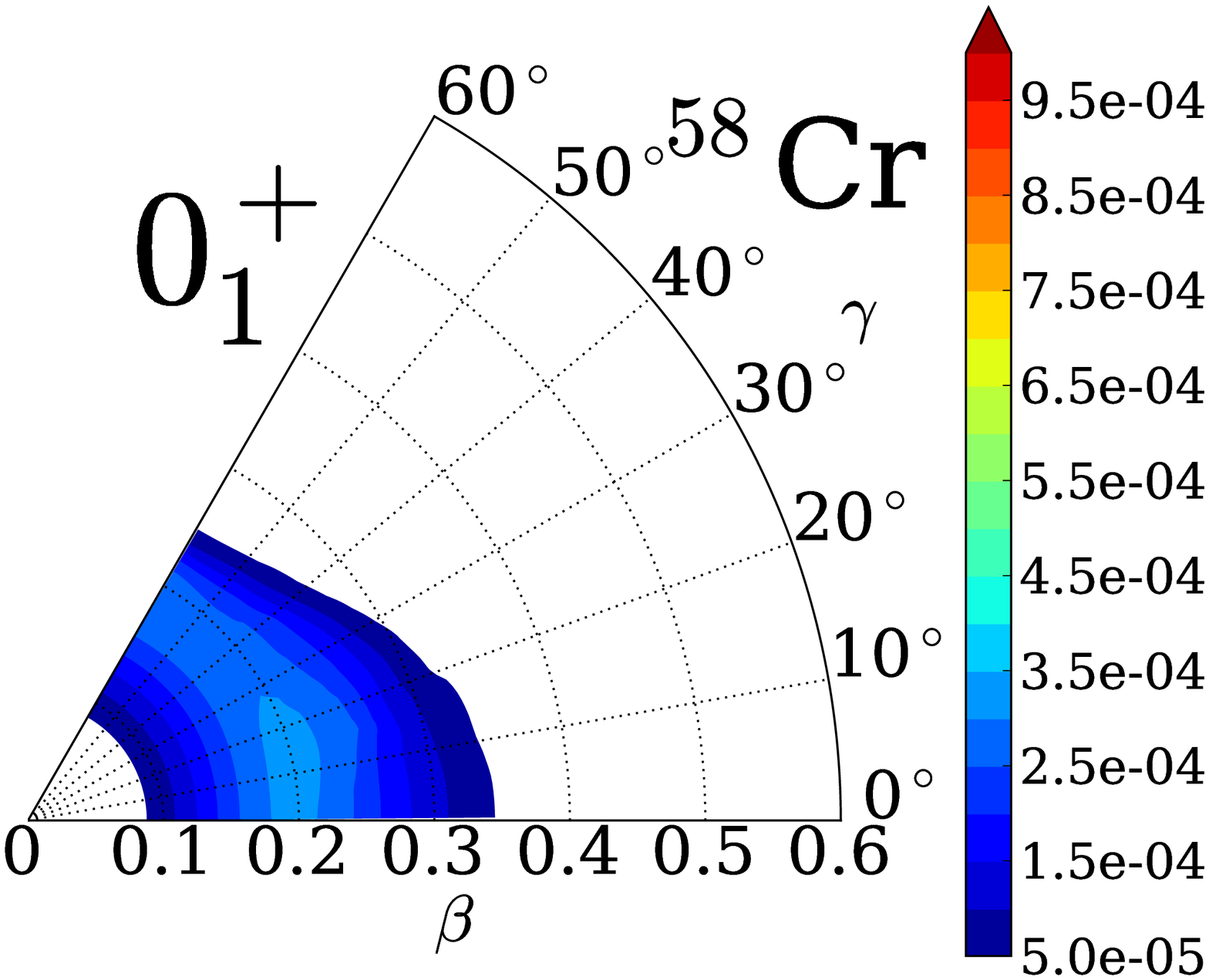}
\includegraphics[height=0.15\textwidth,keepaspectratio,clip,trim=60 50 160 50]{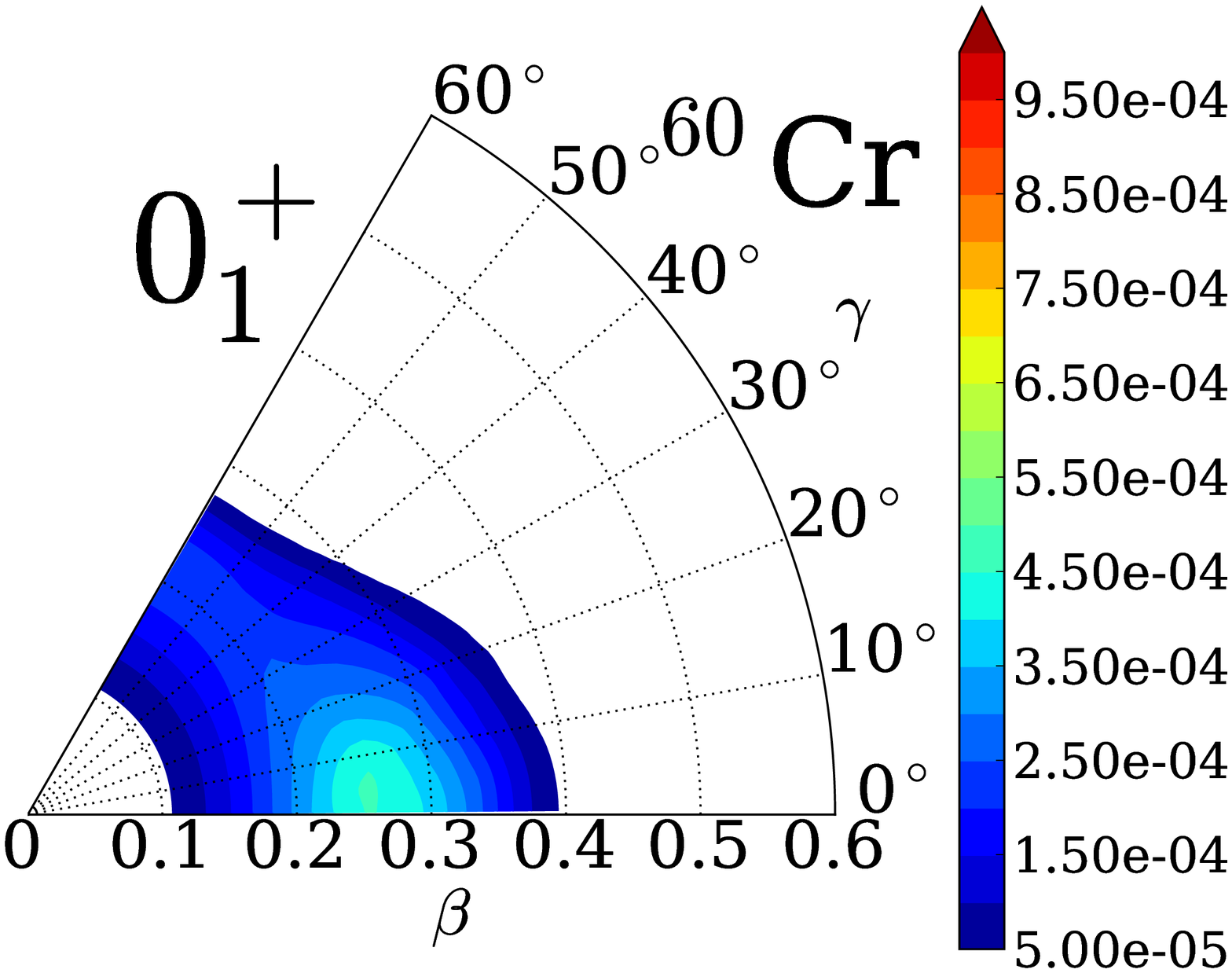} 
\includegraphics[height=0.15\textwidth,keepaspectratio,clip,trim=60 50 160 50]{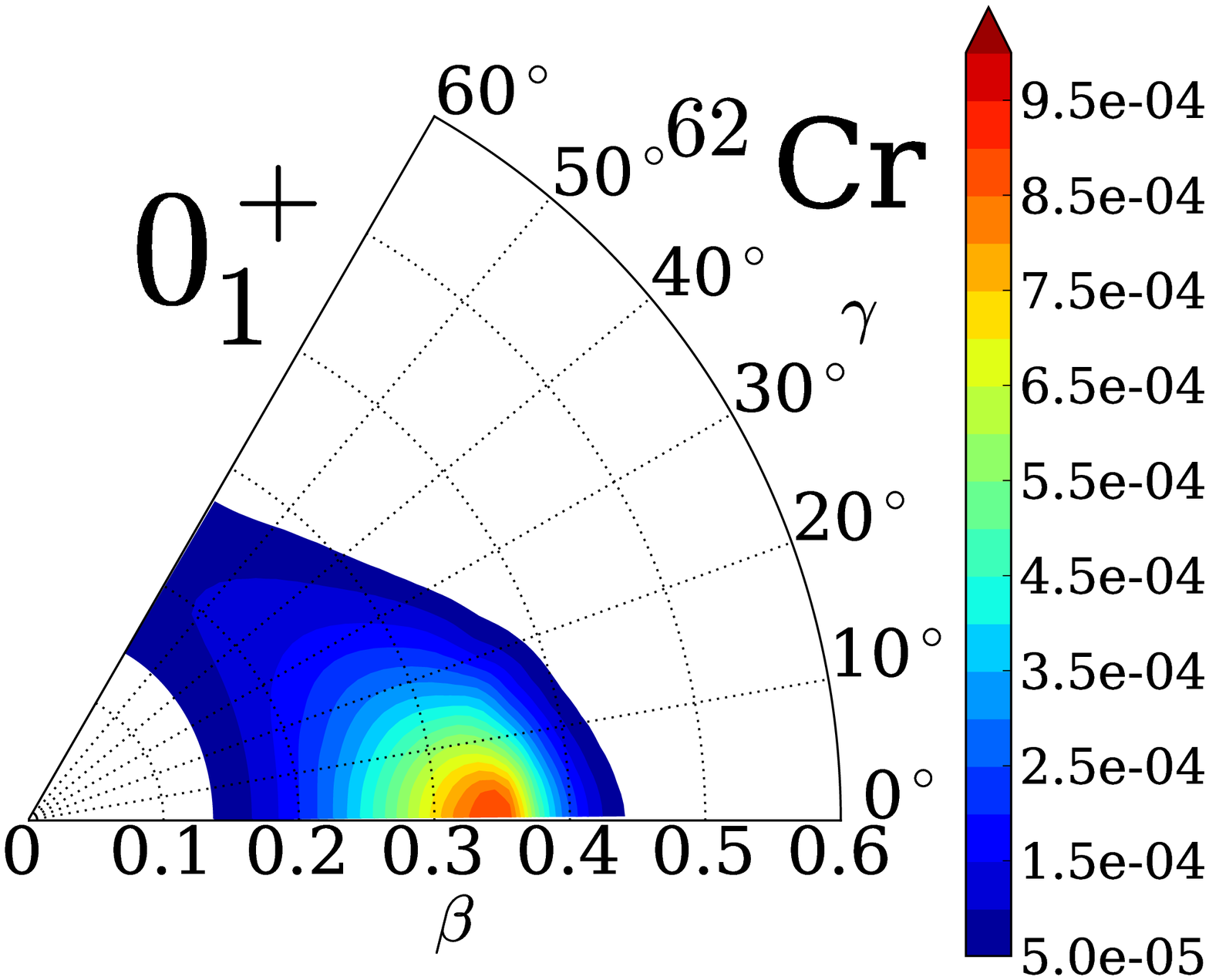} 
\includegraphics[height=0.15\textwidth,keepaspectratio,clip,trim=60 50 160 50]{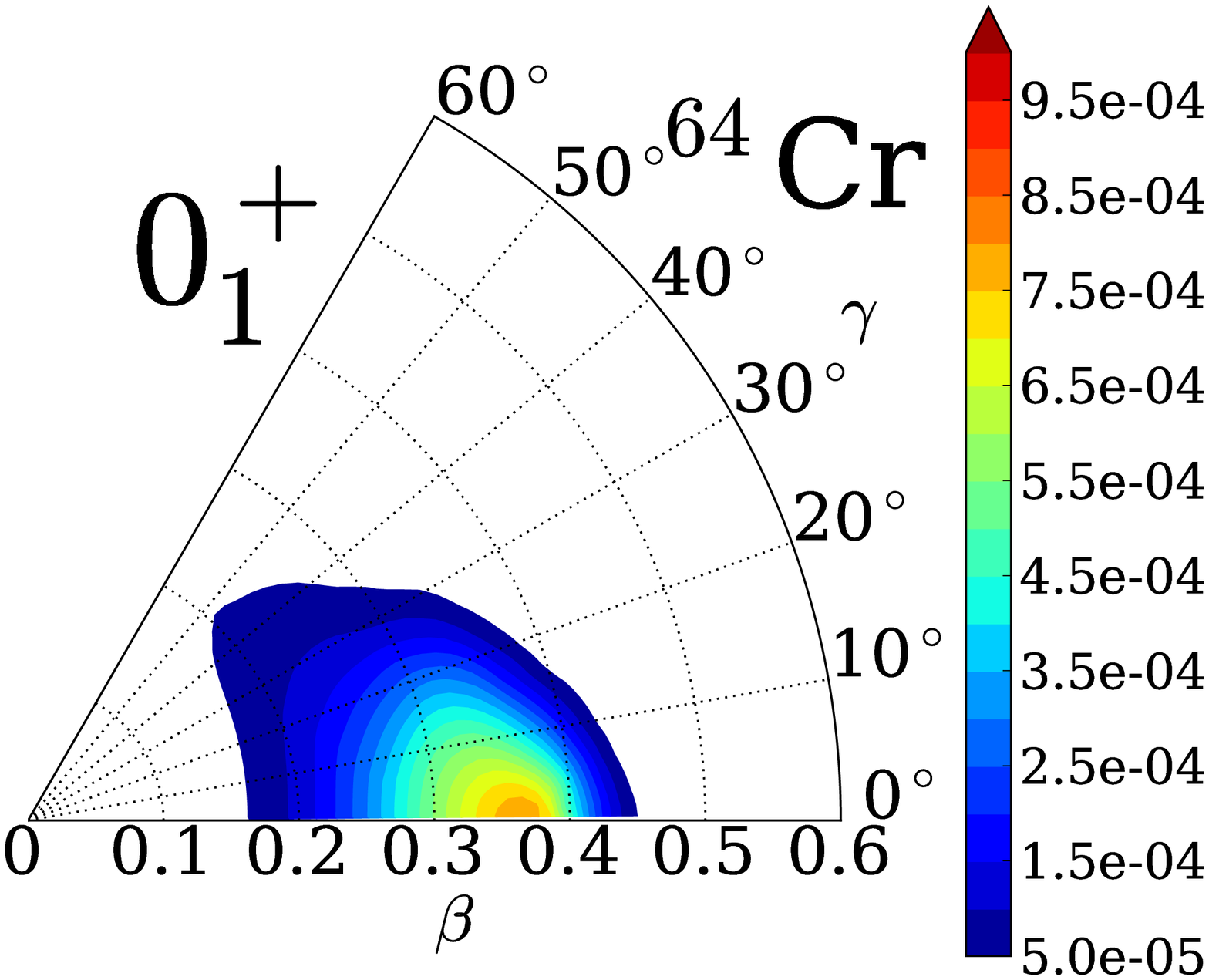} 
\includegraphics[height=0.15\textwidth,keepaspectratio,clip,trim=60 50   0 50]{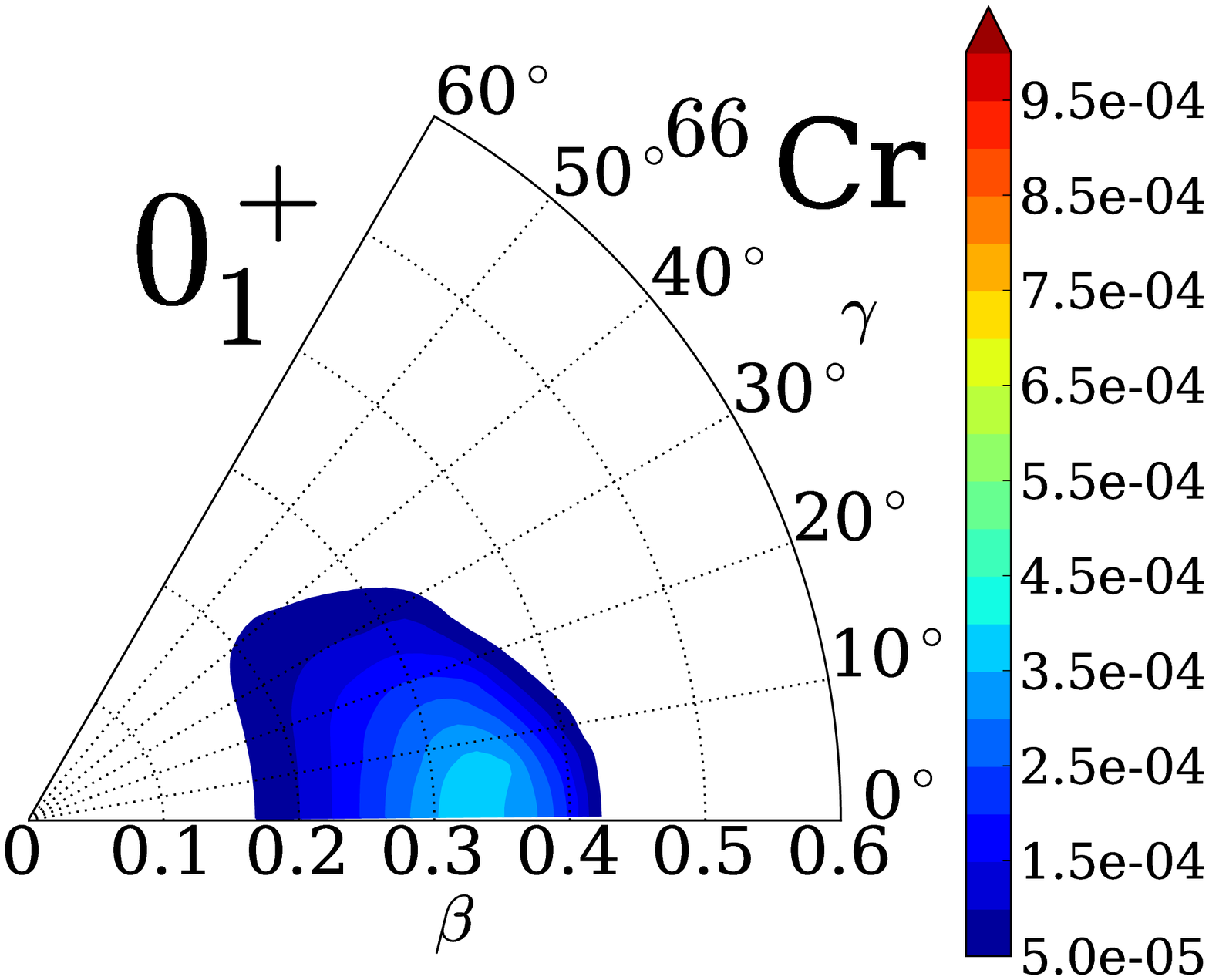} \\
\includegraphics[height=0.15\textwidth,keepaspectratio,clip,trim=60 50 160 50]{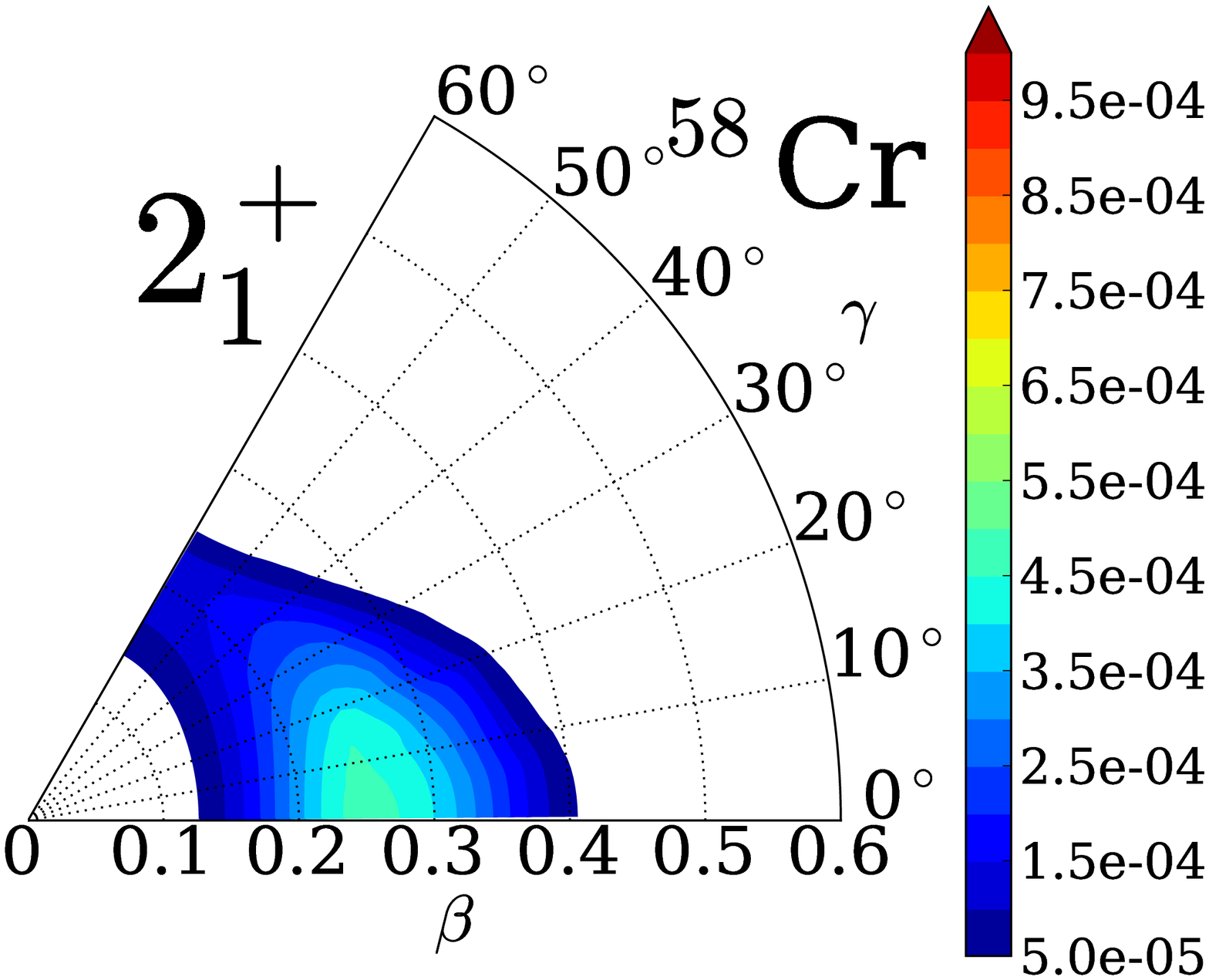}
\includegraphics[height=0.15\textwidth,keepaspectratio,clip,trim=60 50 160 50]{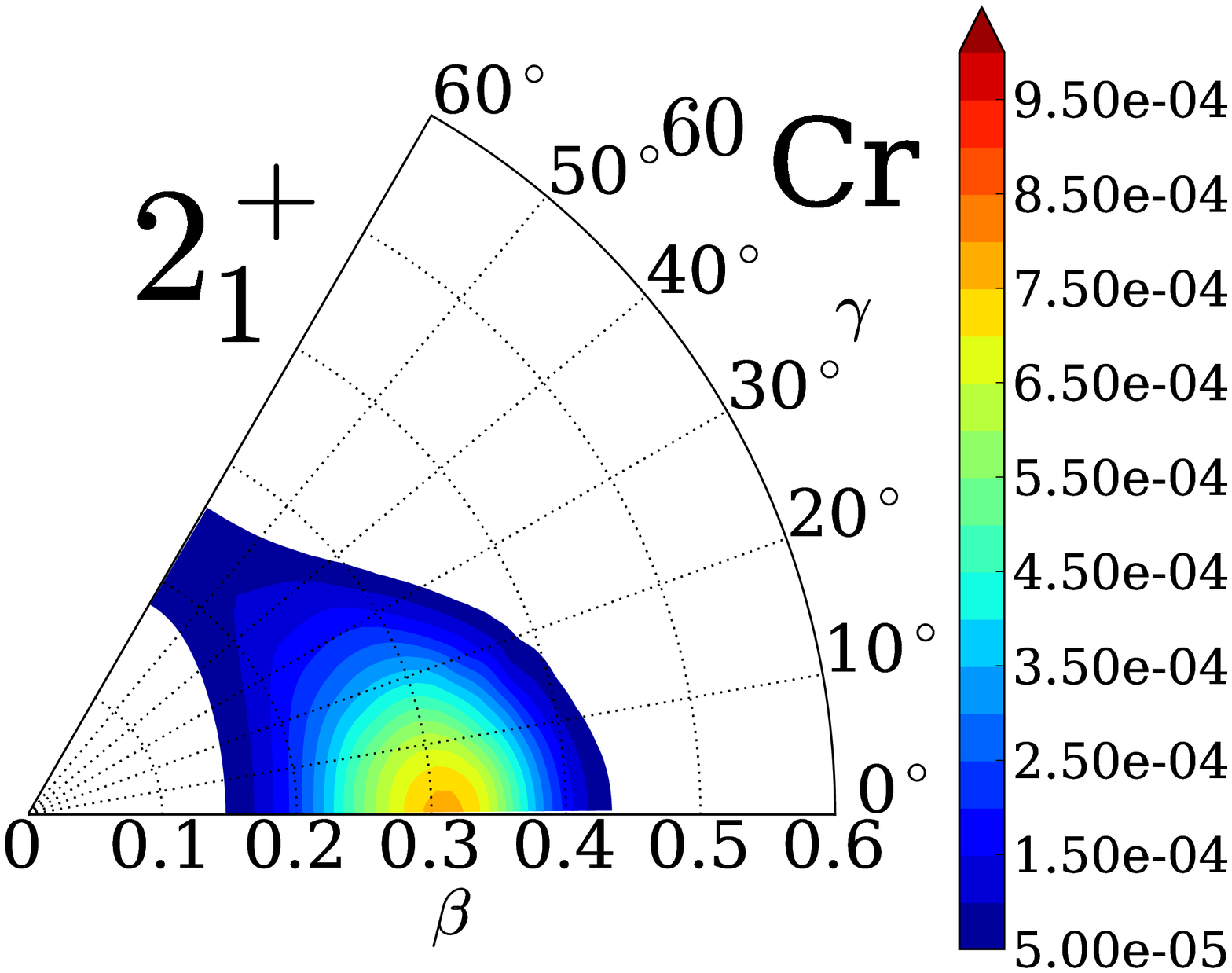}
\includegraphics[height=0.15\textwidth,keepaspectratio,clip,trim=60 50 160 50]{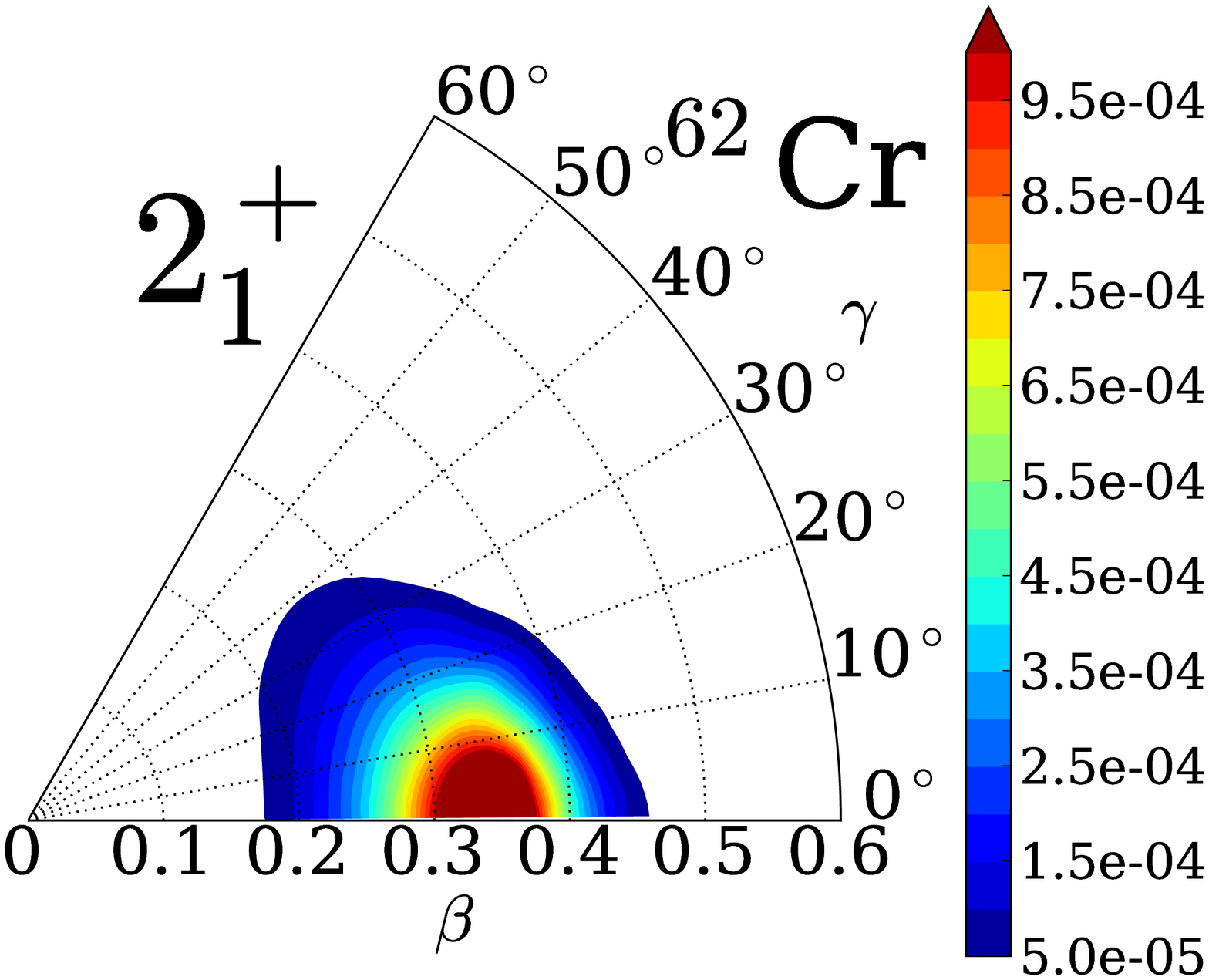} 
\includegraphics[height=0.15\textwidth,keepaspectratio,clip,trim=60 50 160 50]{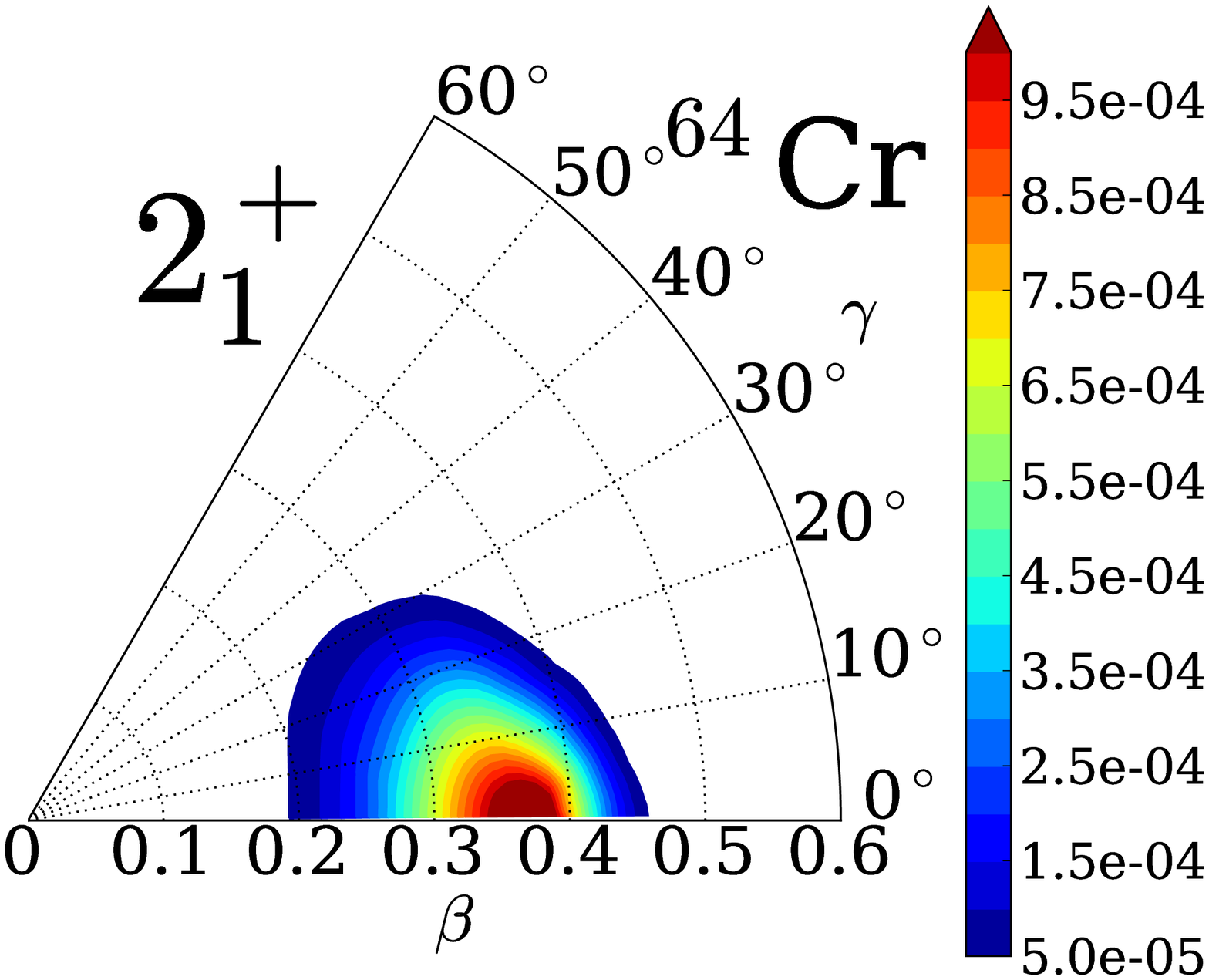}
\includegraphics[height=0.15\textwidth,keepaspectratio,clip,trim=60 50   0 50]{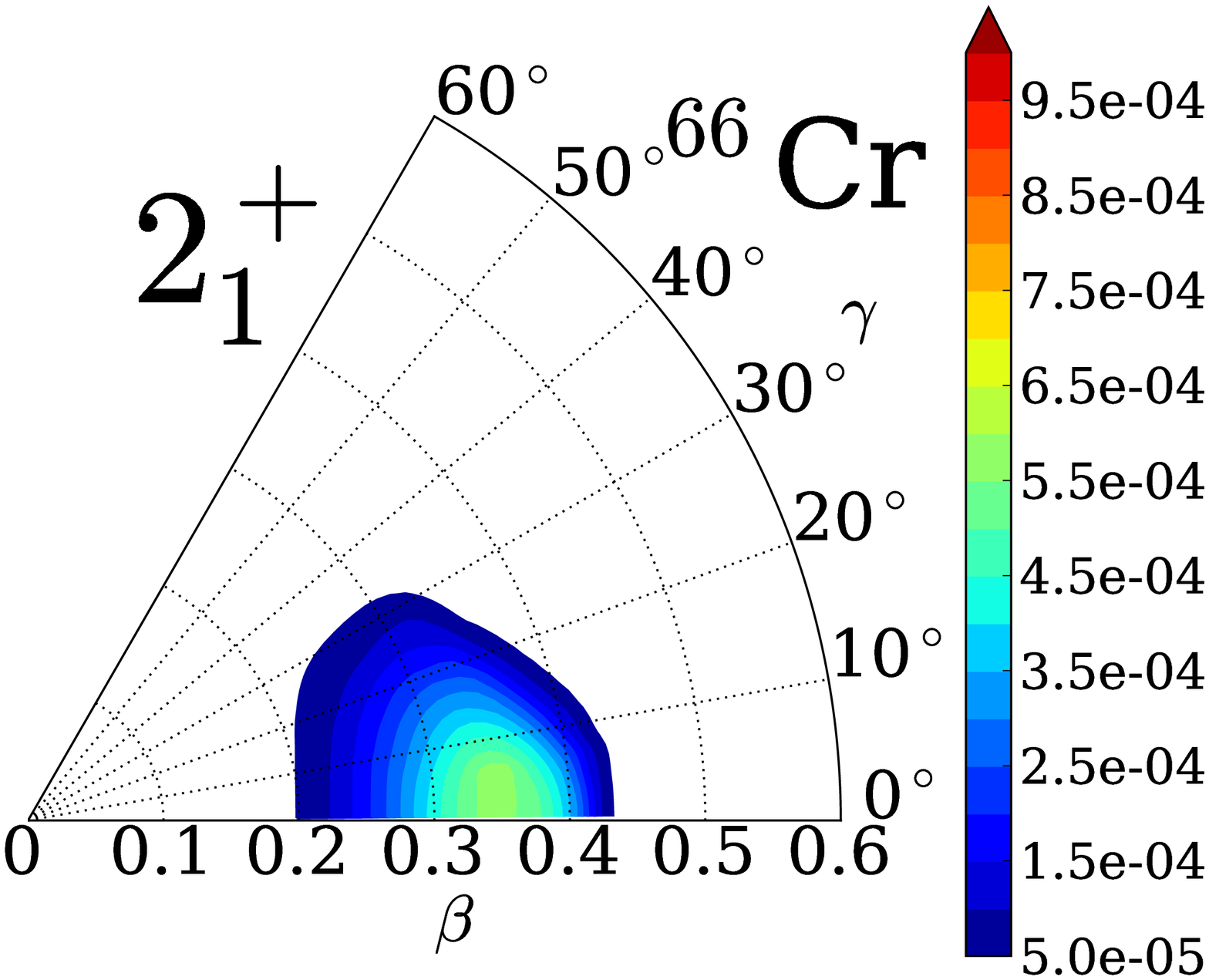} \\
\includegraphics[height=0.15\textwidth,keepaspectratio,clip,trim=60 50 160 50]{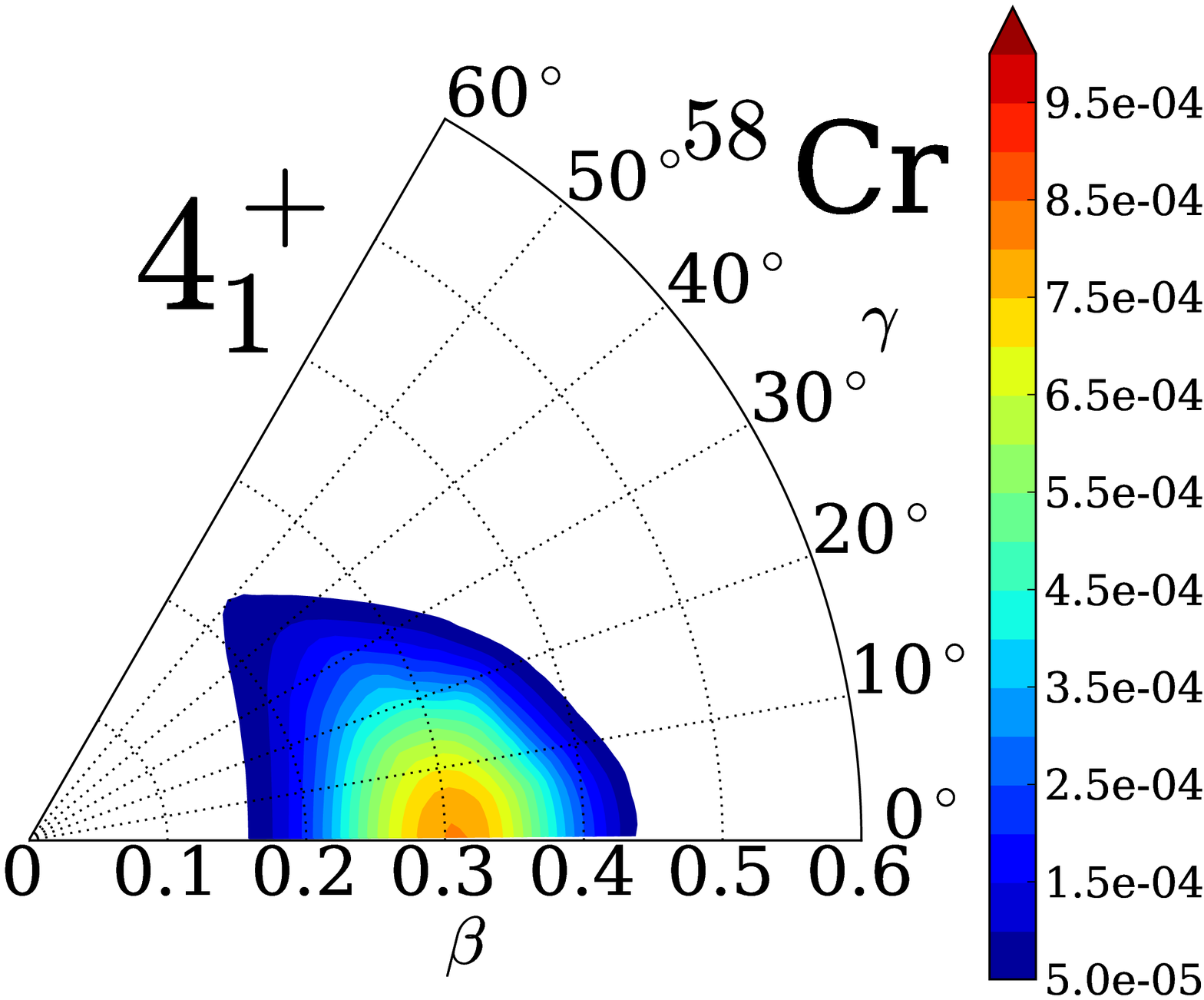}
\includegraphics[height=0.15\textwidth,keepaspectratio,clip,trim=60 50 160 50]{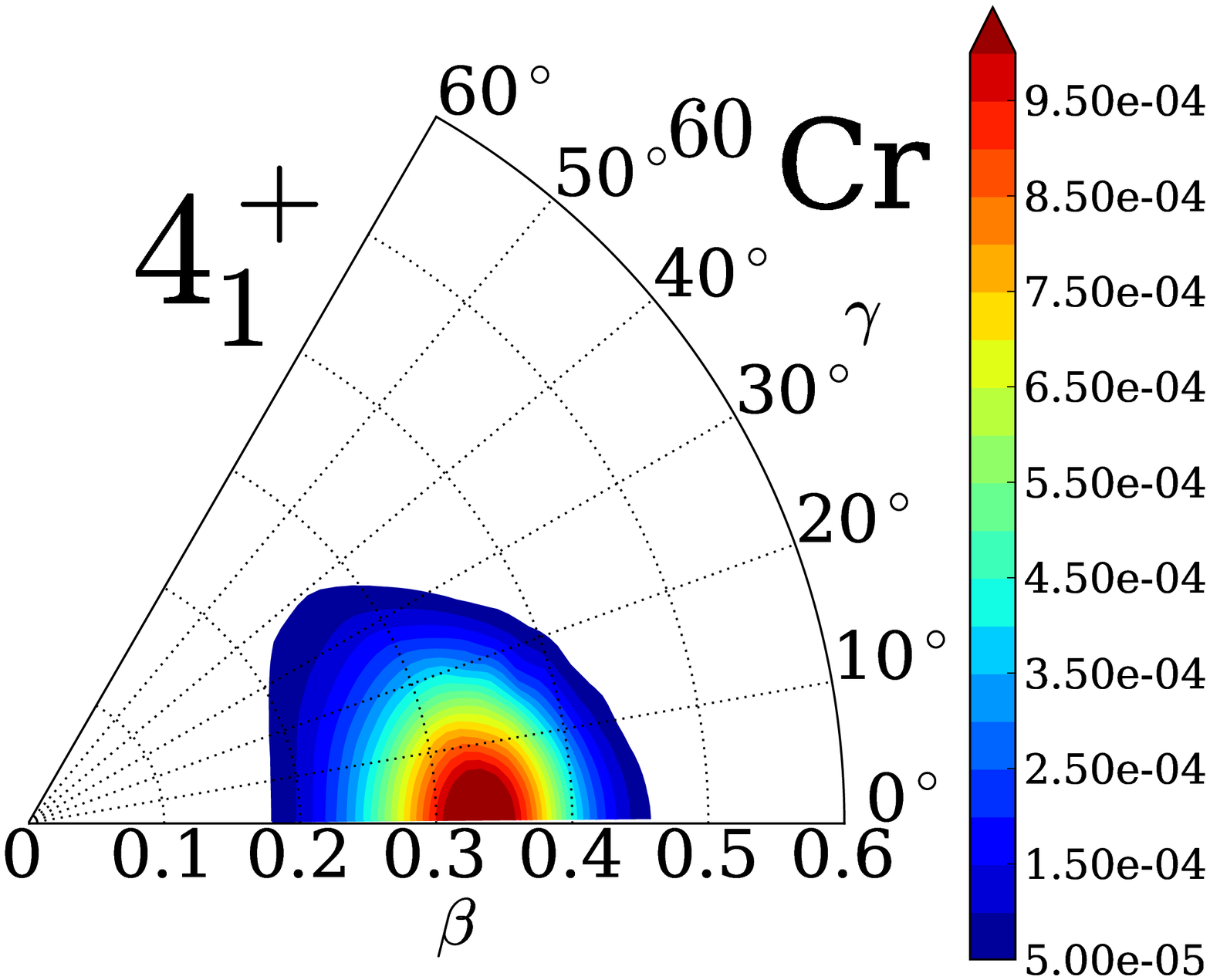}
\includegraphics[height=0.15\textwidth,keepaspectratio,clip,trim=60 50 160 50]{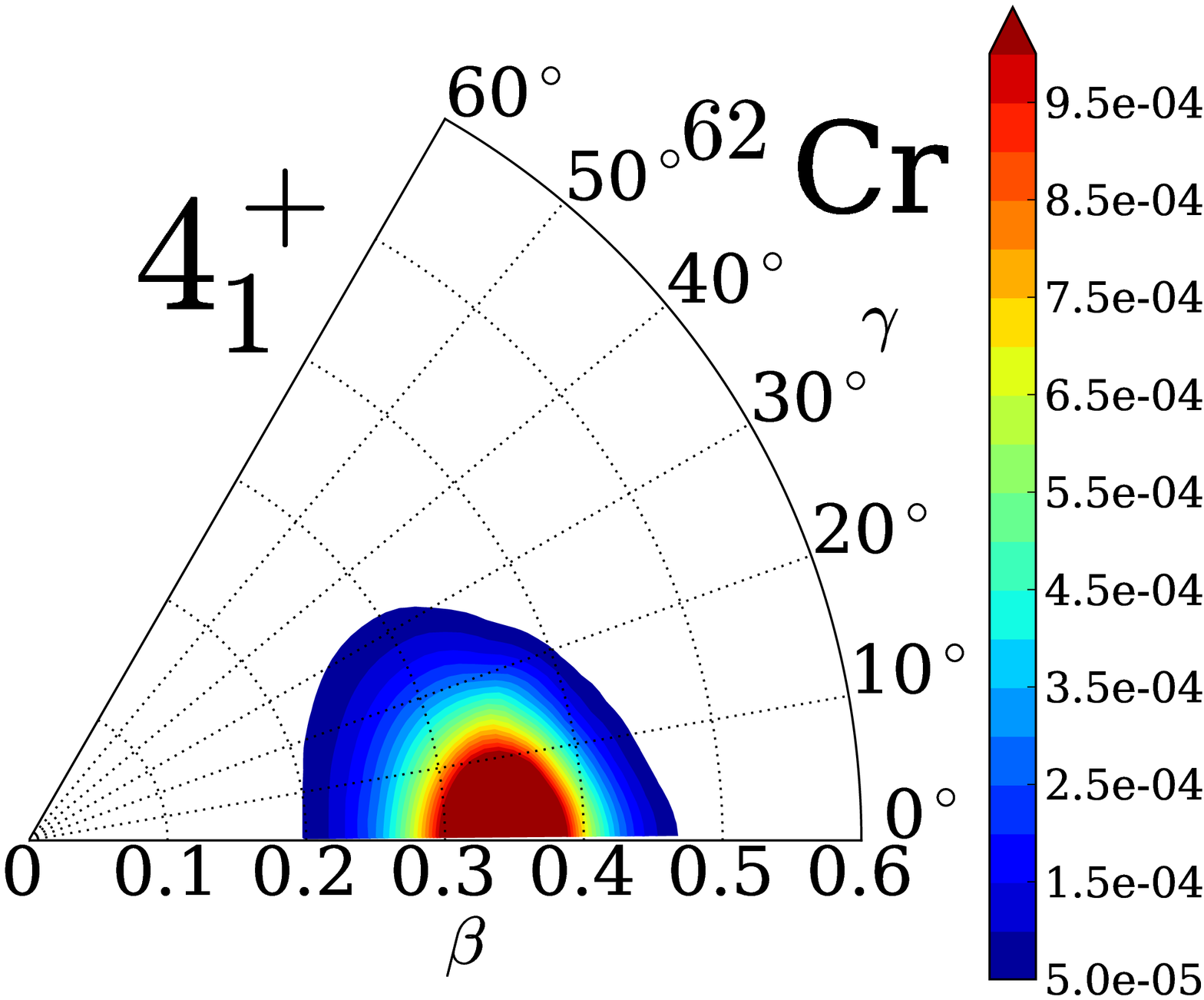} 
\includegraphics[height=0.15\textwidth,keepaspectratio,clip,trim=60 50 160 50]{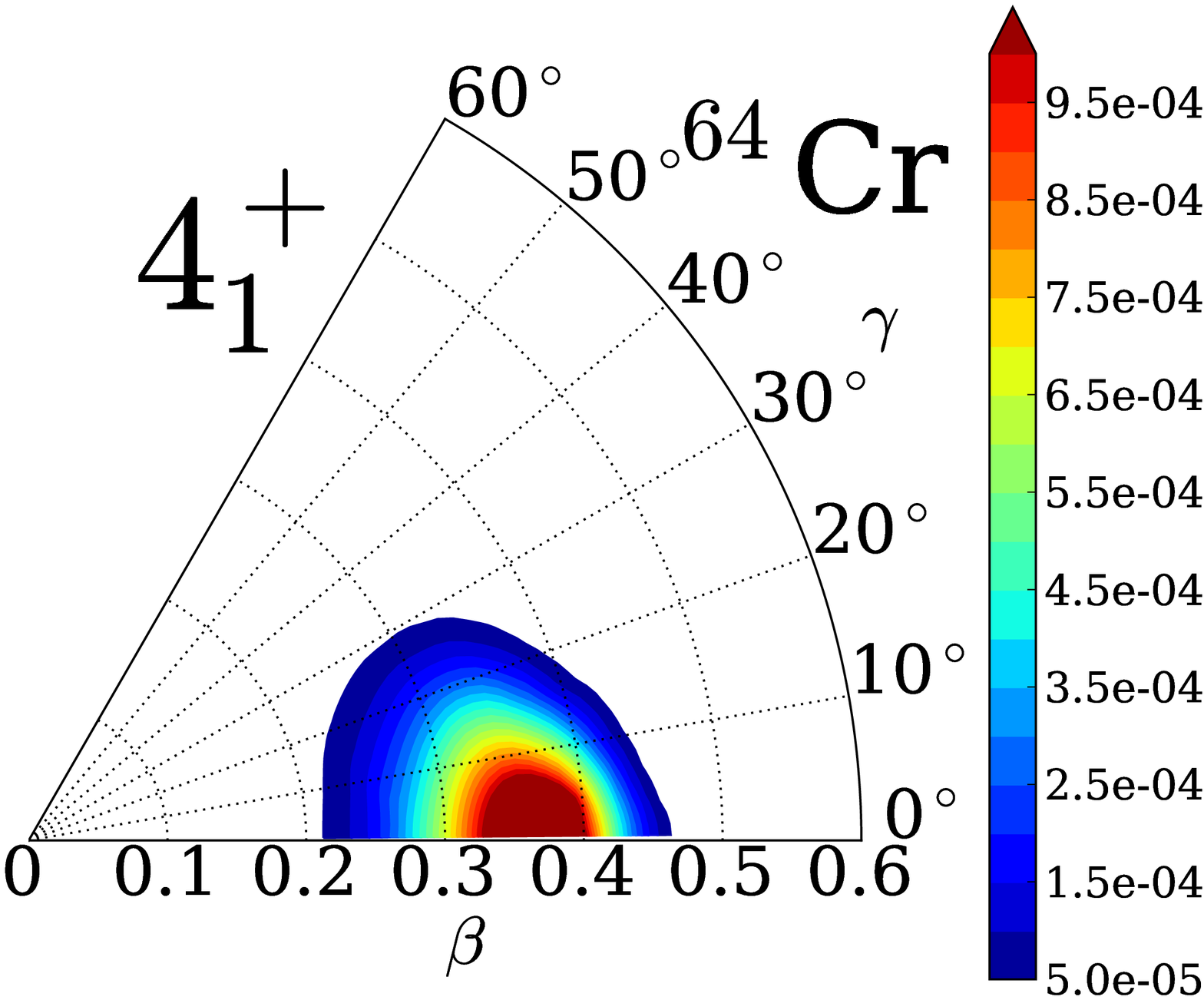}
\includegraphics[height=0.15\textwidth,keepaspectratio,clip,trim=60 50   0 50]{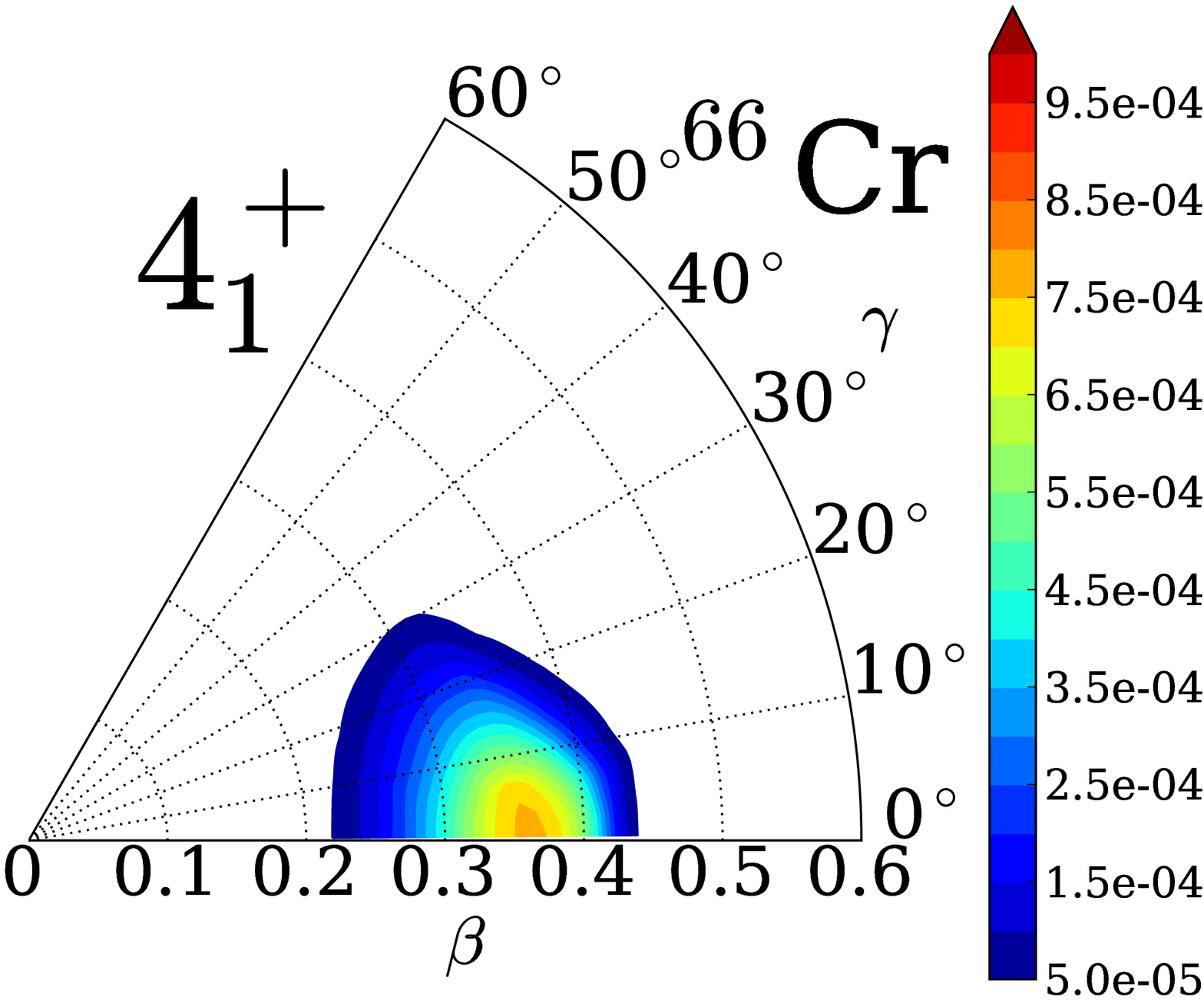}
\end{center}
\caption{(Color online) Squared vibrational wave functions multiplied by $\beta^4$, 
$\beta^4 \sum_K |\Psi_{\alpha IK}(\beta,\gamma)|^2$,  
for the $0_1^+, 2_1^+$ and $4_1^+$ states in $^{58-68}$Cr. }
\label{fig:2Dwf}
\end{figure*}

\begin{figure*}[h]
\begin{center}
\includegraphics[height=0.15\textwidth,keepaspectratio,clip,trim=60 50 160 50]{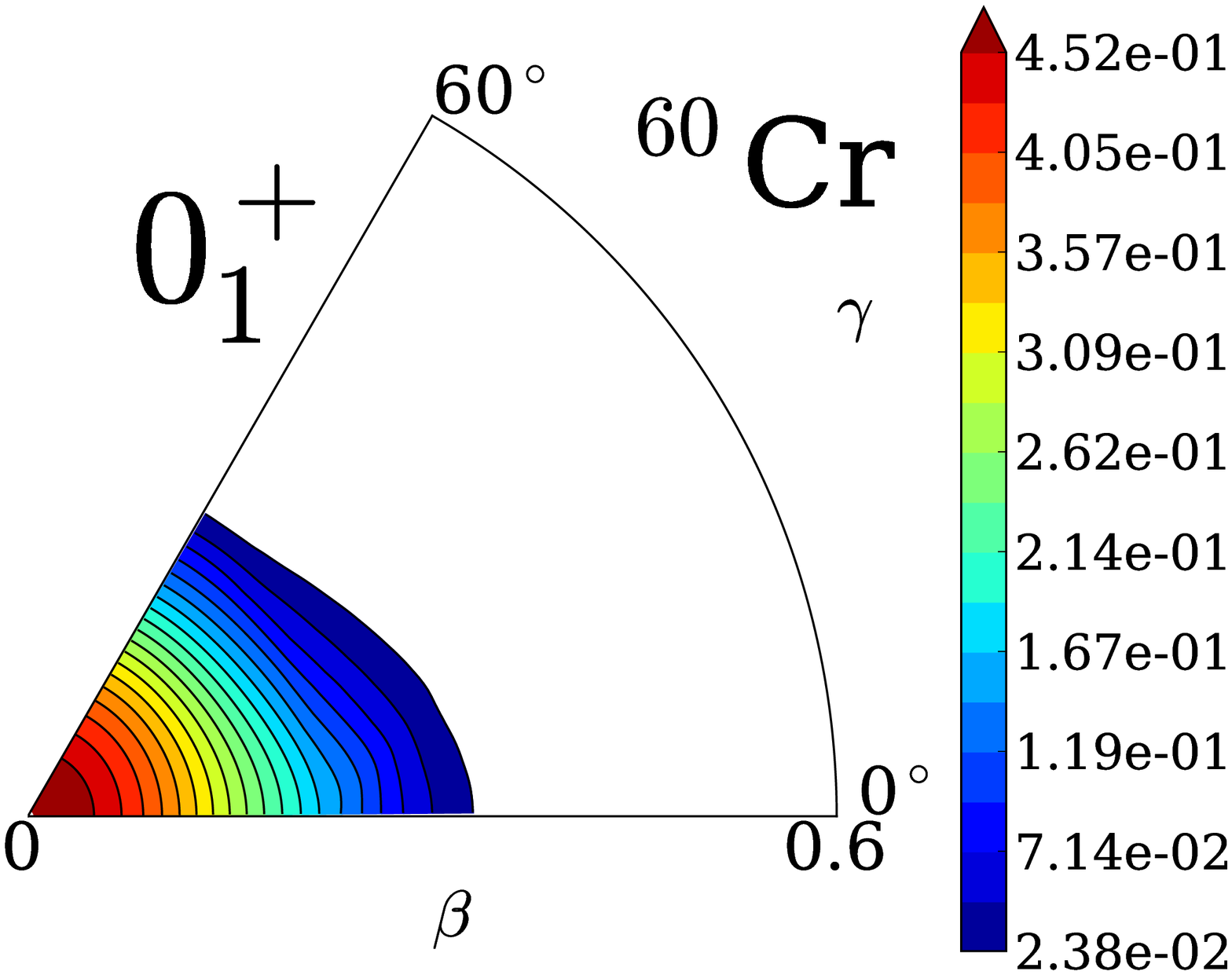}
\includegraphics[height=0.15\textwidth,keepaspectratio,clip,trim=60 50 160 50]{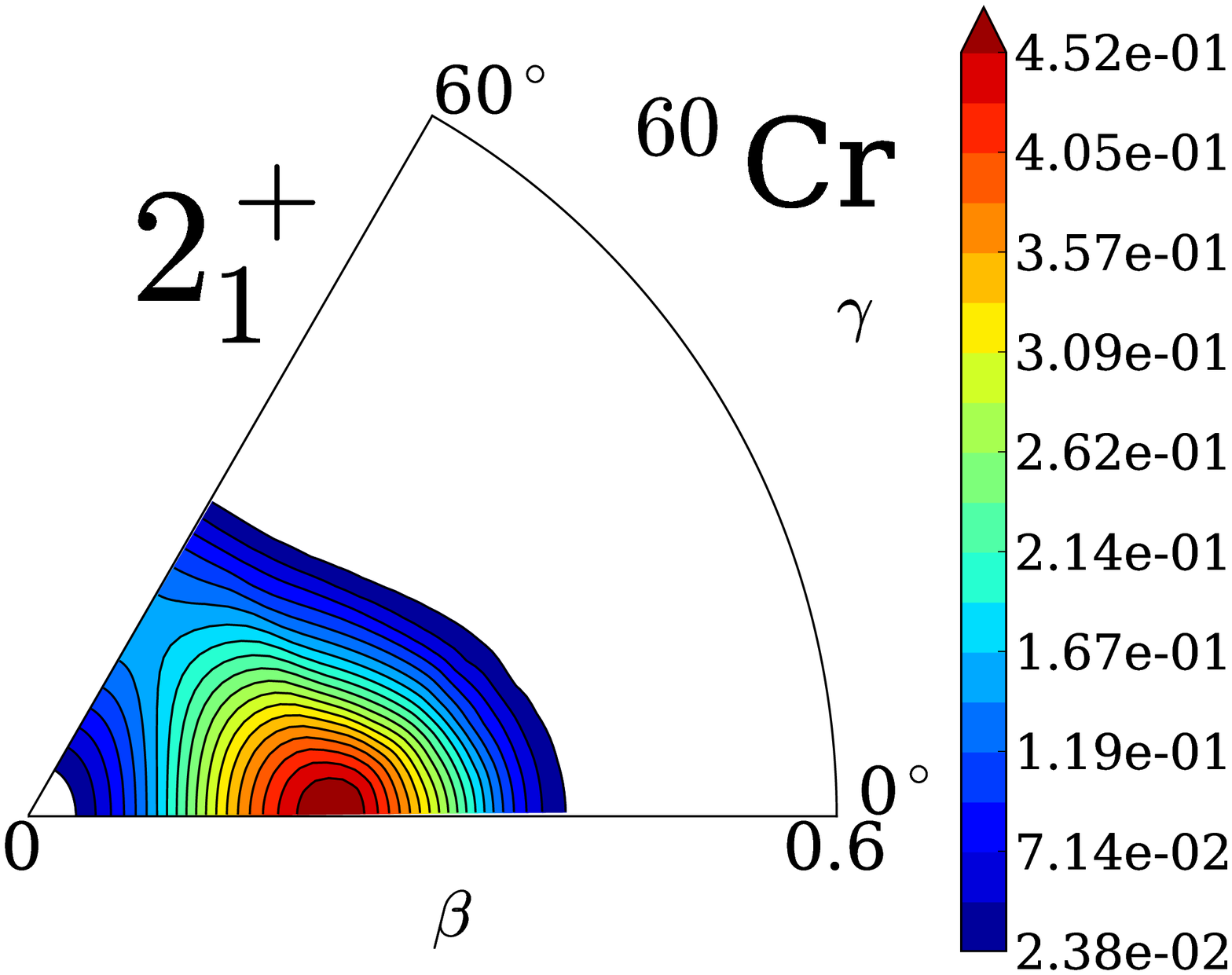}
\includegraphics[height=0.15\textwidth,keepaspectratio,clip,trim=60 50 160 50]{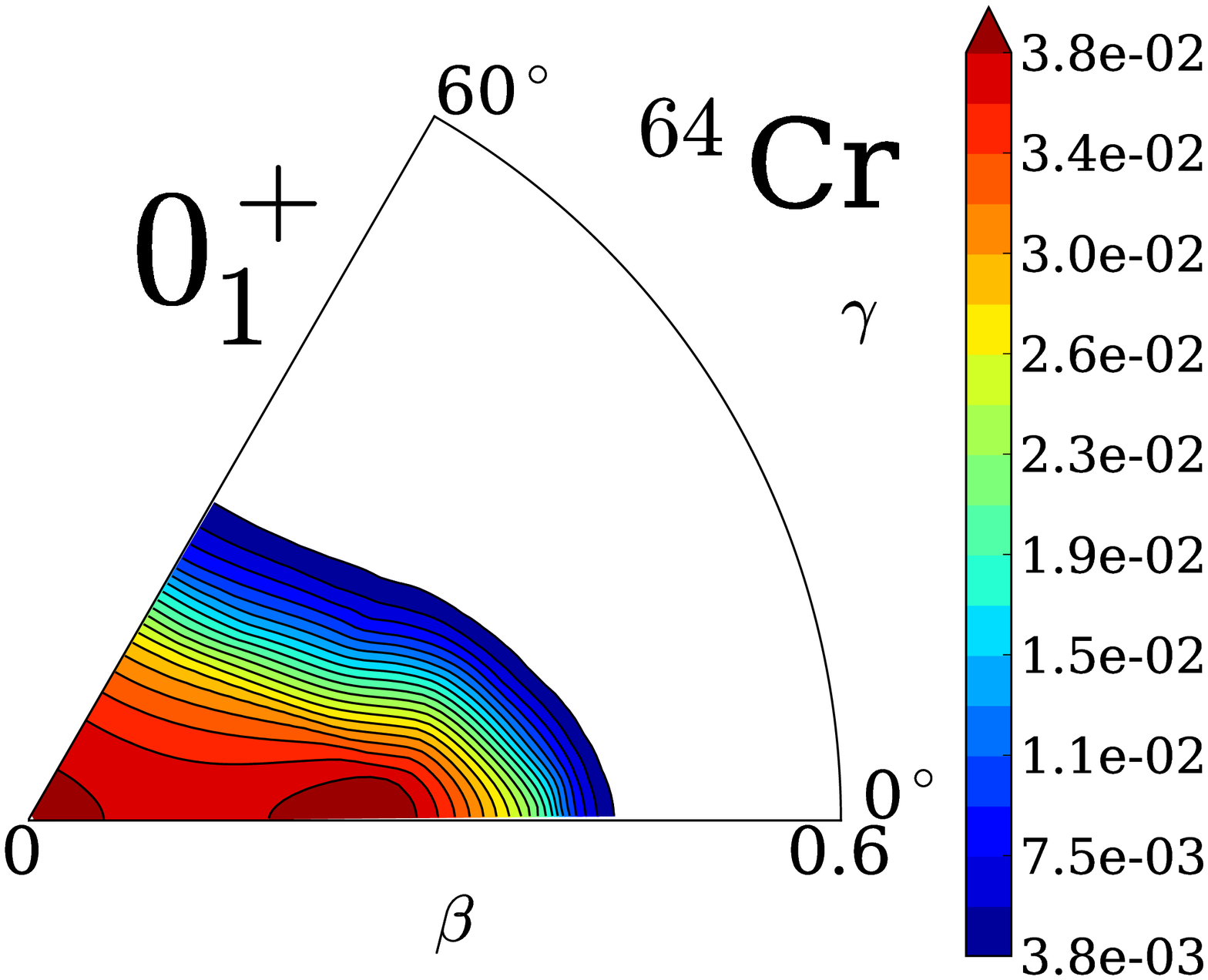}
\includegraphics[height=0.15\textwidth,keepaspectratio,clip,trim=60 50 160 50]{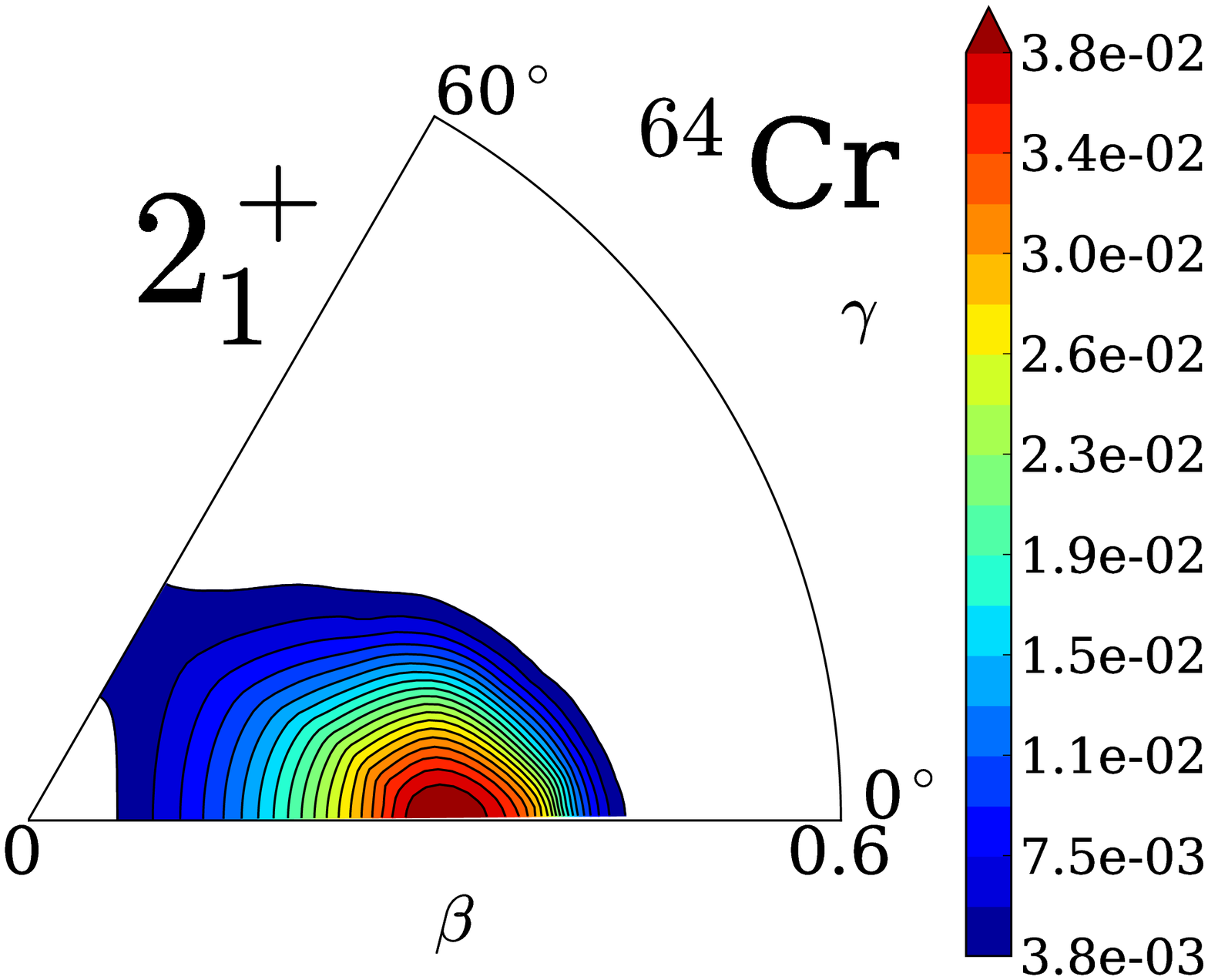} \\
\end{center}
\caption{(Color online) Vibrational wave function squared, 
$\sum_K |\Psi_{\alpha IK}(\beta,\gamma)|^2$,  
for the $0^+_1$ and  $2^+_1$states in $^{60}$Cr and $^{64}$Cr. 
The contour lines are drawn at every twentieth part of the maximum value.}
\label{non-weighted yrast wf}
\end{figure*}

\begin{figure*}[h]
\begin{center}
\includegraphics[width=0.7\textwidth,keepaspectratio,clip,trim=0 0 0 0]{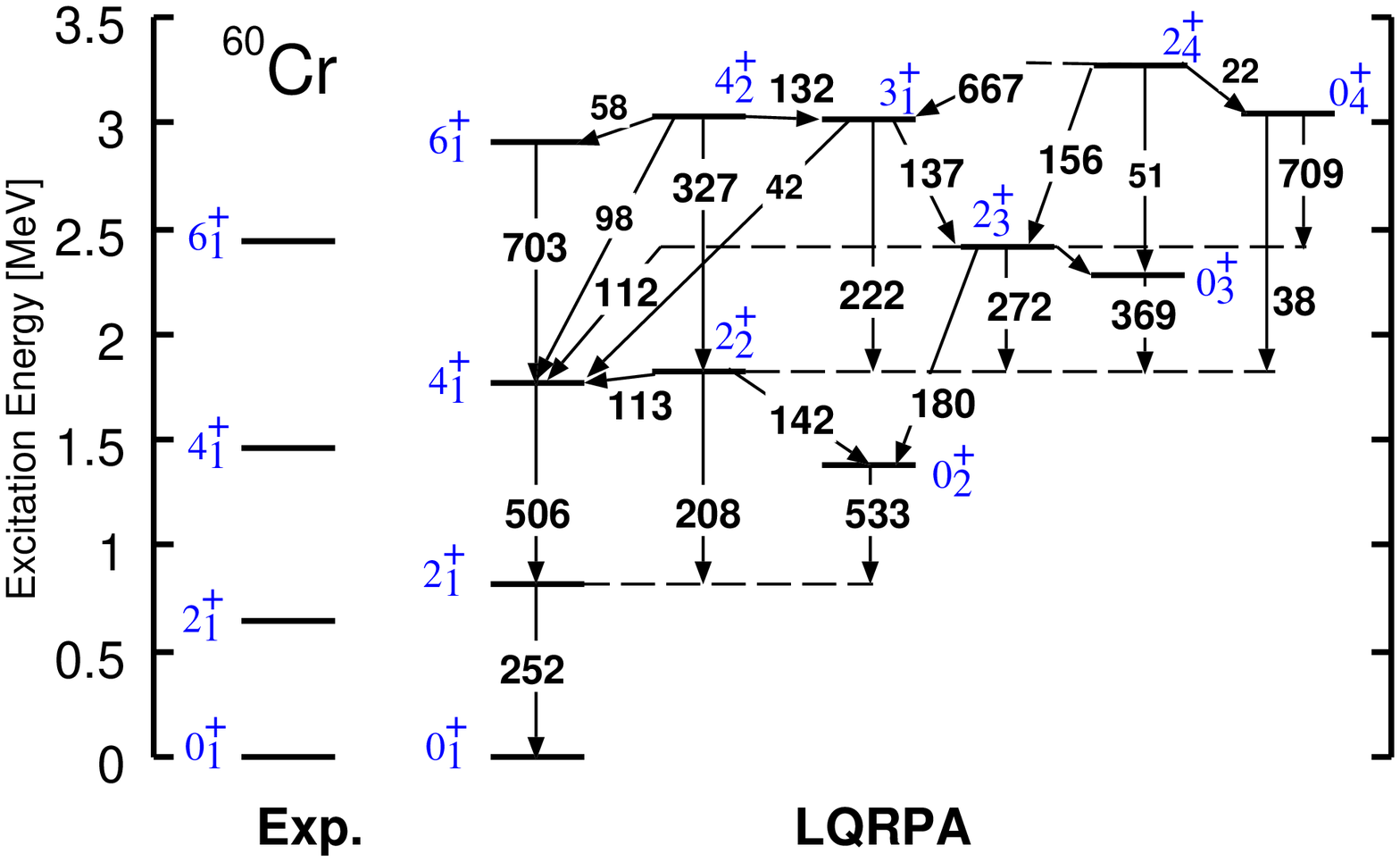} 
\end{center}
\caption{(Color online) Excitation energies and $B(E2)$ values for $^{60}$Cr 
in comparison with experimental data.
Values on arrows indicate $B(E2)$ in units of $e^2$fm$^4$. 
Only $B(E2)$ values larger than 1 Weisskopf unit are shown. 
Experimental data are taken from Ref. \cite{Zhu2006}.}
\label{fig:60CrSpectra}
\end{figure*}

\begin{figure*}[h]
\begin{center}
\includegraphics[width=0.7\textwidth,keepaspectratio,clip,trim=0 0 0 0]{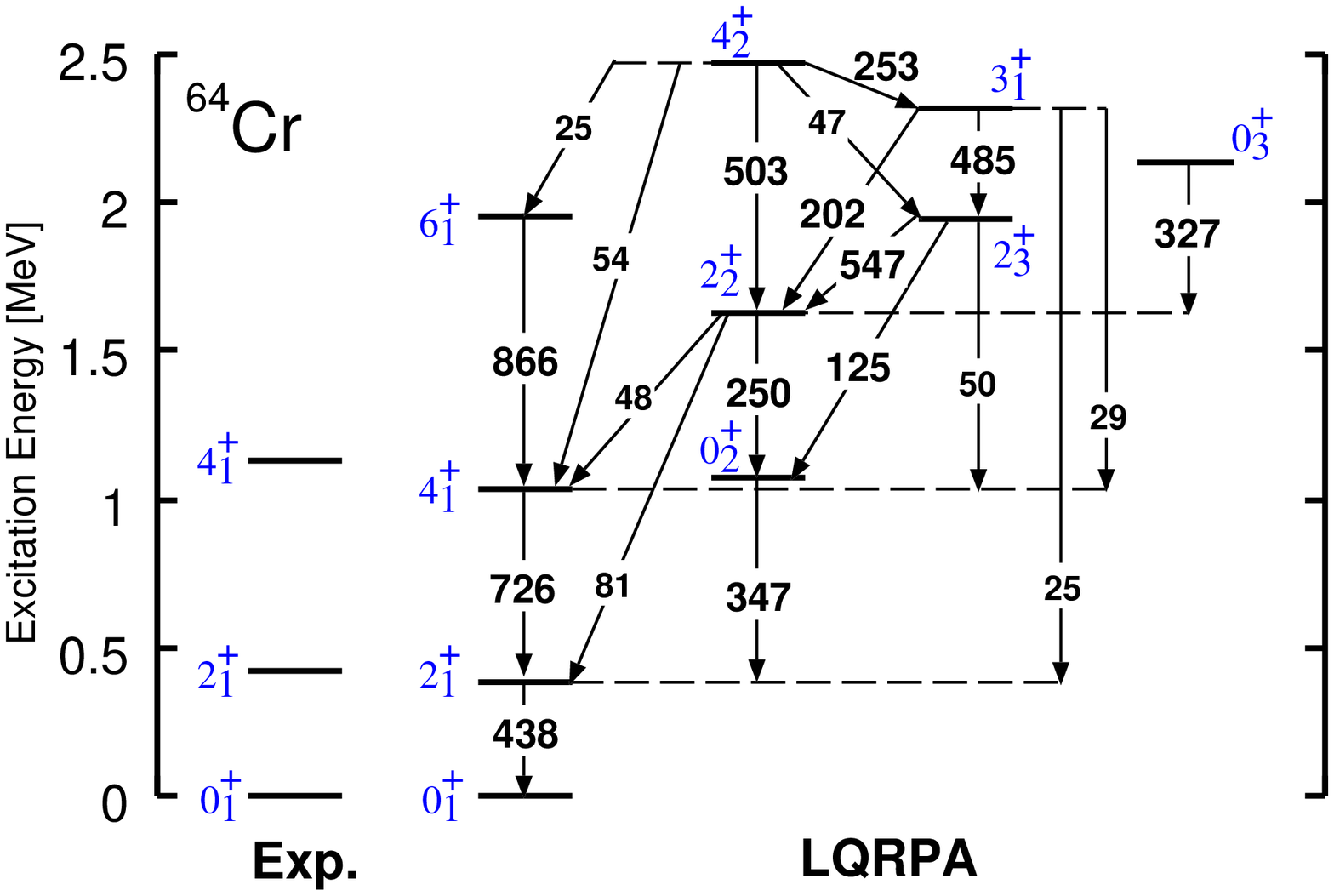} 
\end{center}
\caption{(Color online) Same as Fig.~\ref{fig:60CrSpectra} but for $^{64}$Cr. 
The experimental data are taken from Ref.~\cite{Gade2010}.
}
\label{fig:64CrSpectra}
\end{figure*}

\begin{figure*}[h]
\begin{center}
\includegraphics[height=0.15\textwidth,keepaspectratio,clip,trim=60 50 160 50]{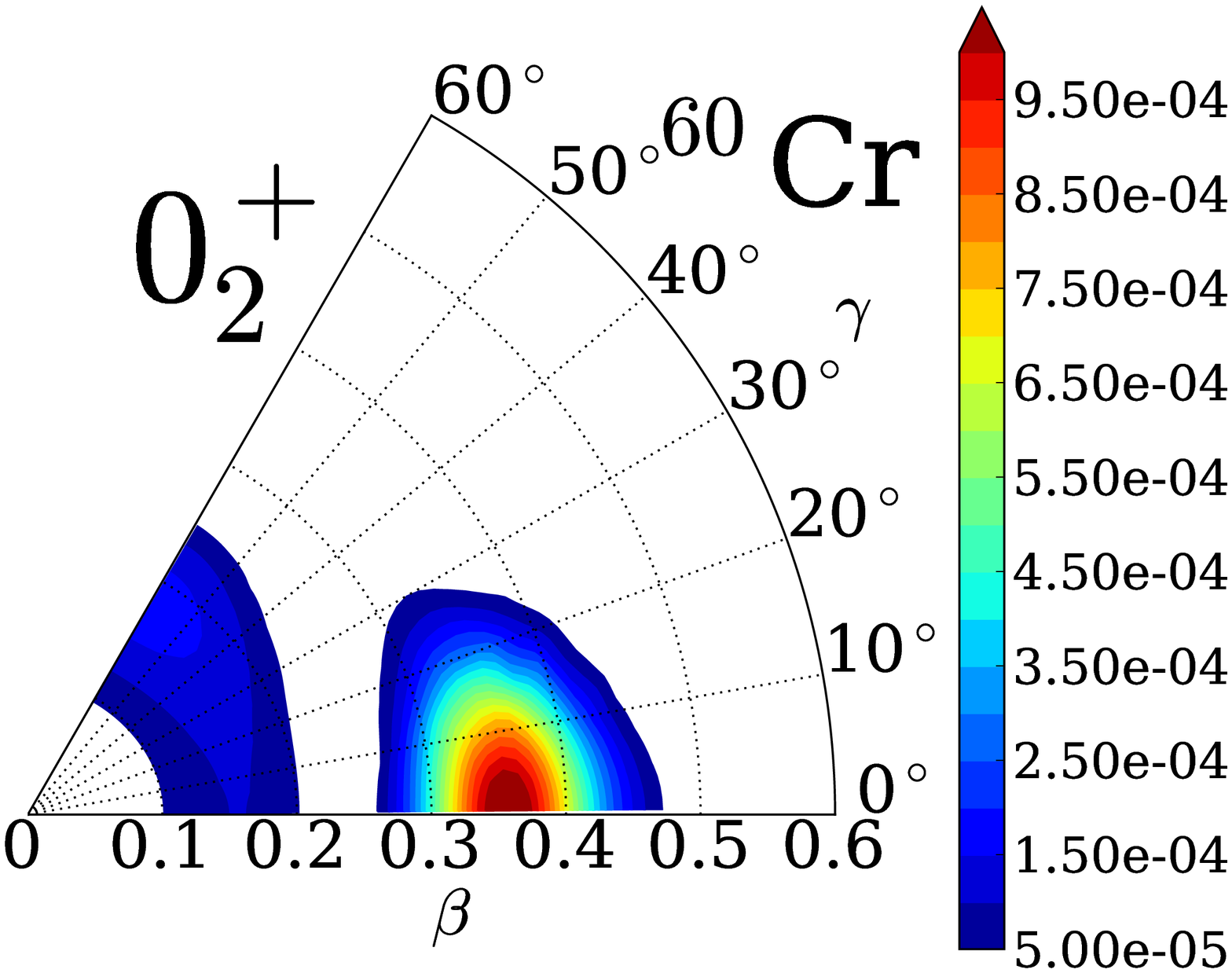}
\includegraphics[height=0.15\textwidth,keepaspectratio,clip,trim=60 50 160 50]{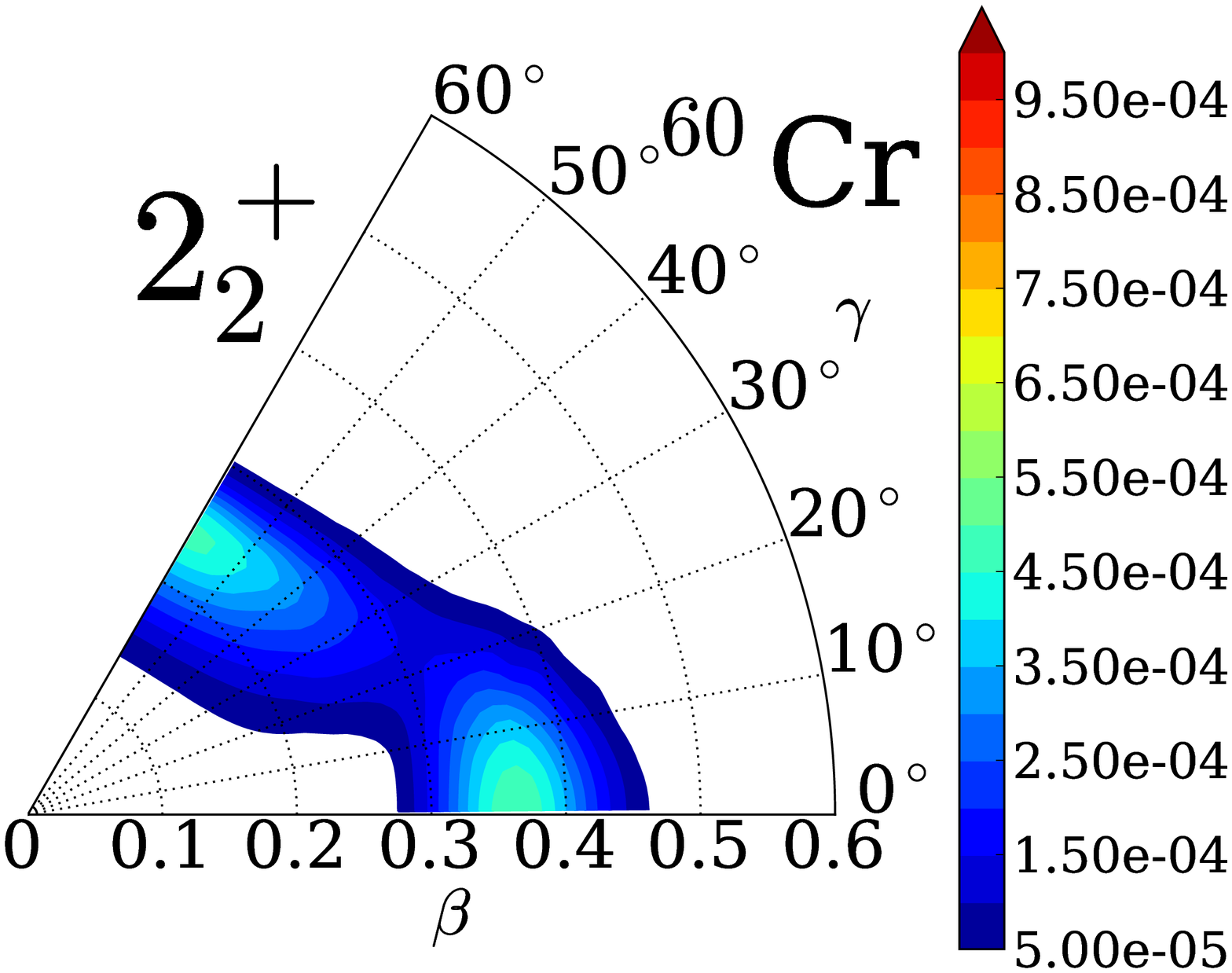}
\includegraphics[height=0.15\textwidth,keepaspectratio,clip,trim=60 50 160 50]{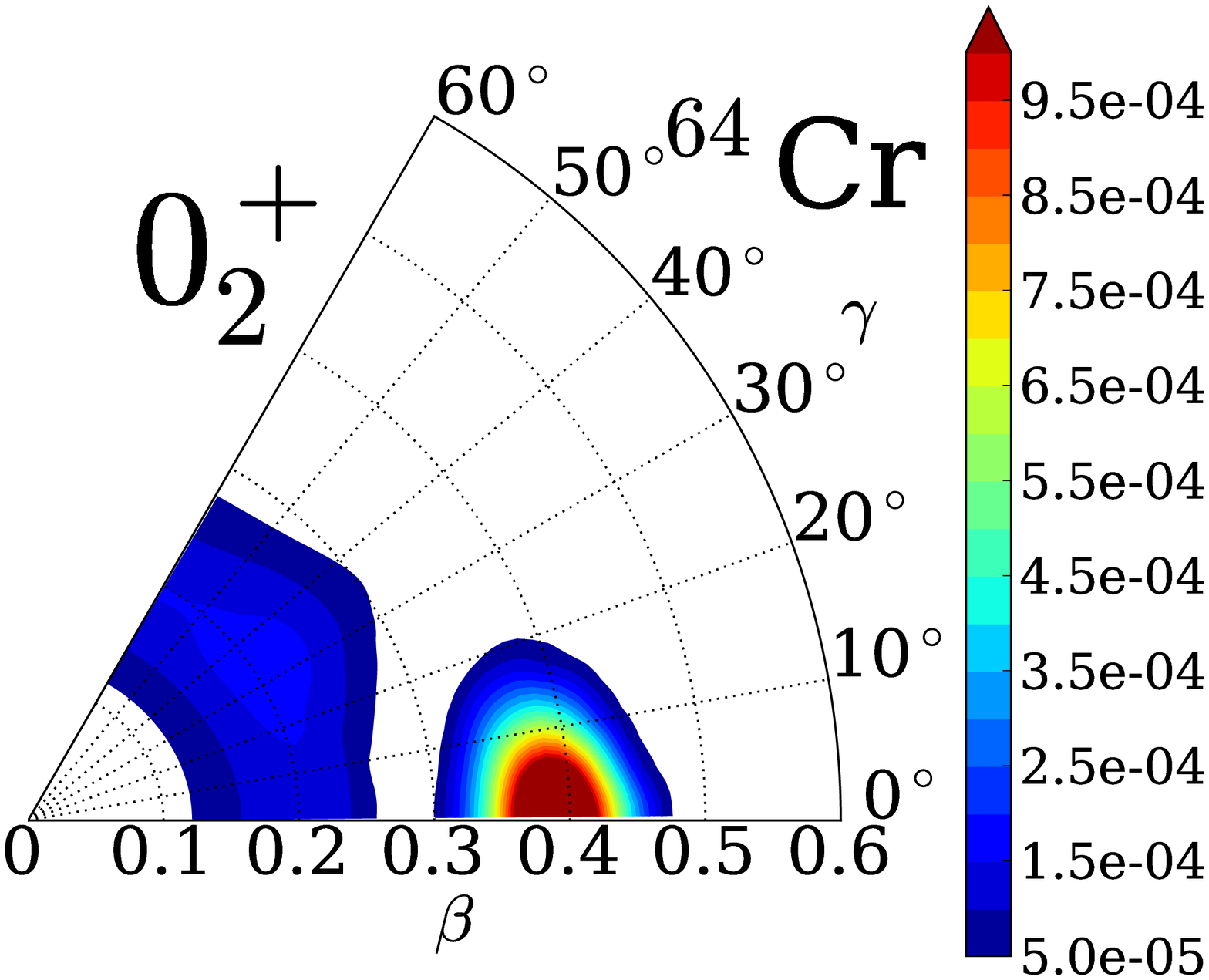}
\includegraphics[height=0.15\textwidth,keepaspectratio,clip,trim=60 50   0 50]{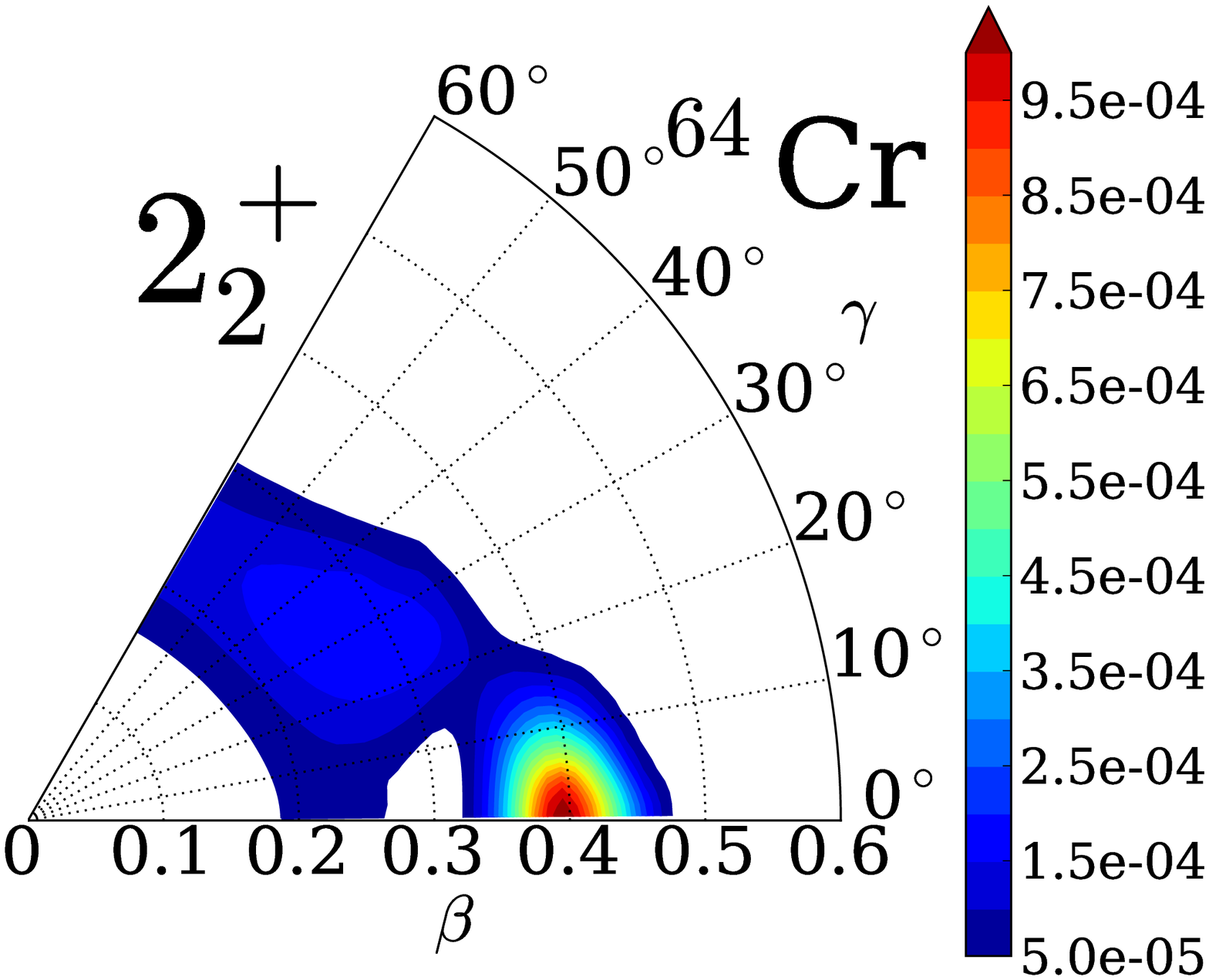} 
\end{center}
\caption{(Color online) Squared vibrational wave functions multiplied by $\beta^4$, 
$\beta^4 \sum_K |\Psi_{\alpha IK}(\beta,\gamma)|^2$,  
for the $0_2^+$ and $2_2^+$ states in $^{60}$Cr and $^{64}$Cr. }
\label{fig:yrare_b4wf}
\end{figure*}

\begin{figure*}[h]
\begin{center}
\includegraphics[height=0.15\textwidth,keepaspectratio,clip,trim=60 50 160 50]{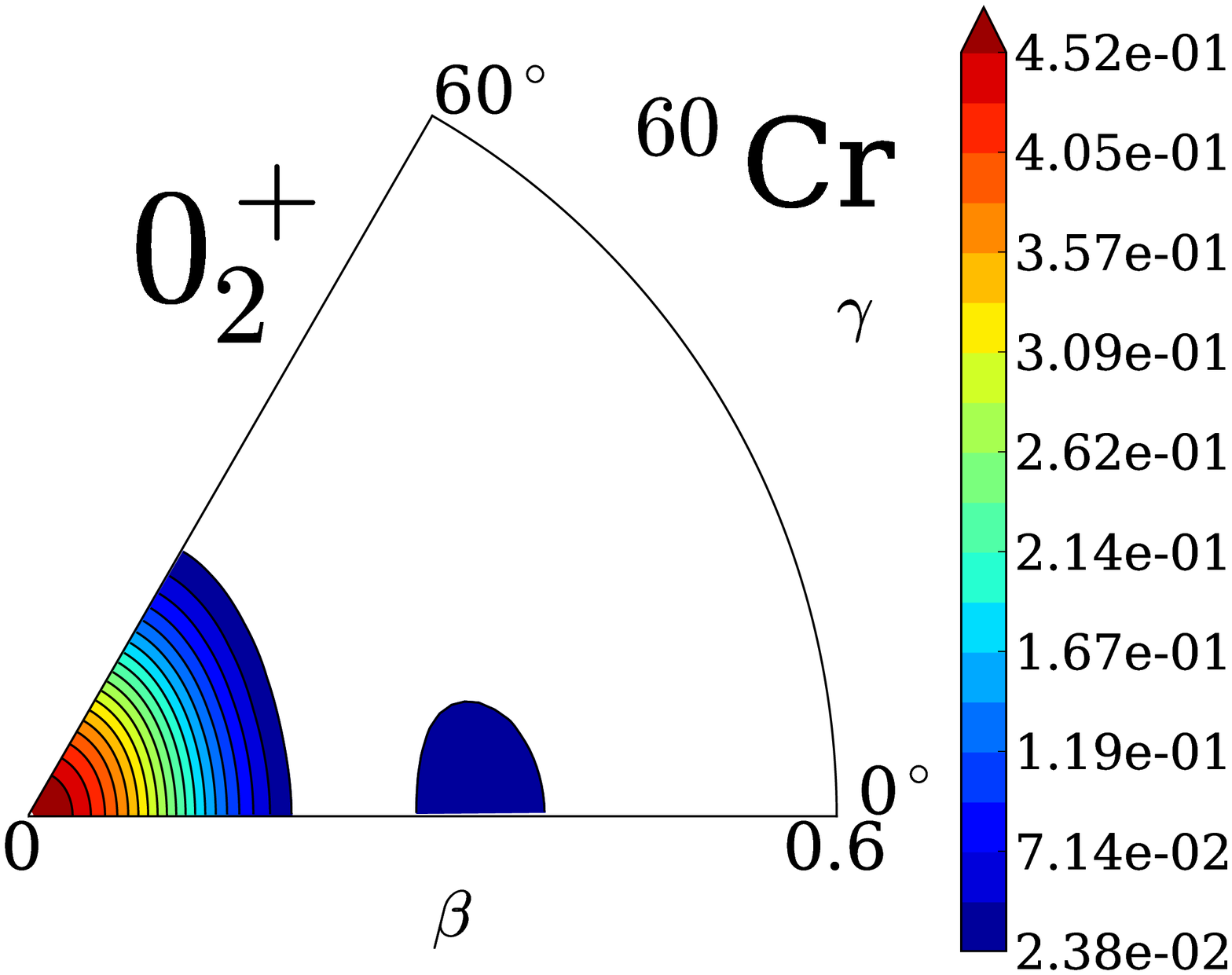}
\includegraphics[height=0.15\textwidth,keepaspectratio,clip,trim=60 50 160 50]{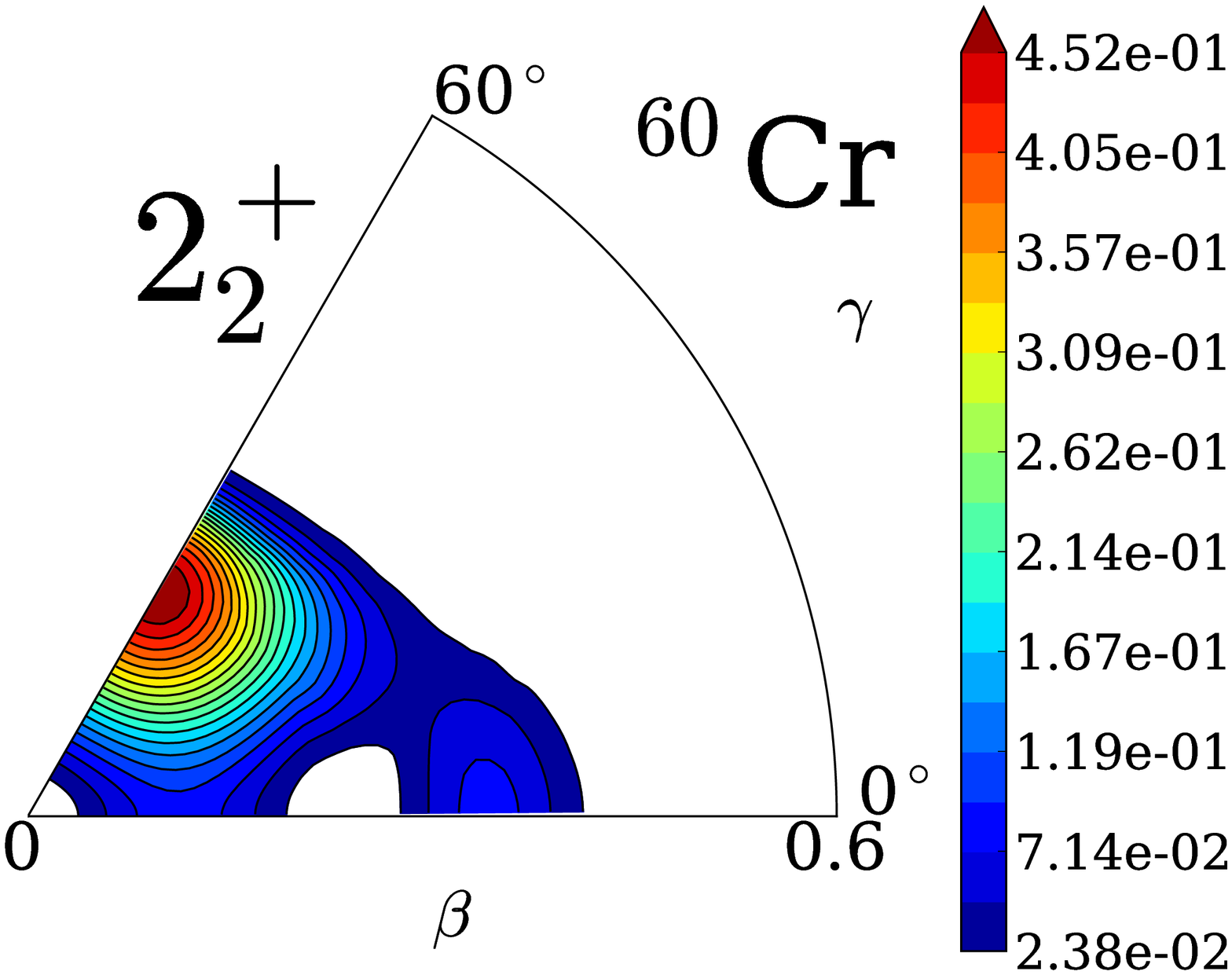}
\includegraphics[height=0.15\textwidth,keepaspectratio,clip,trim=60 50 160 50]{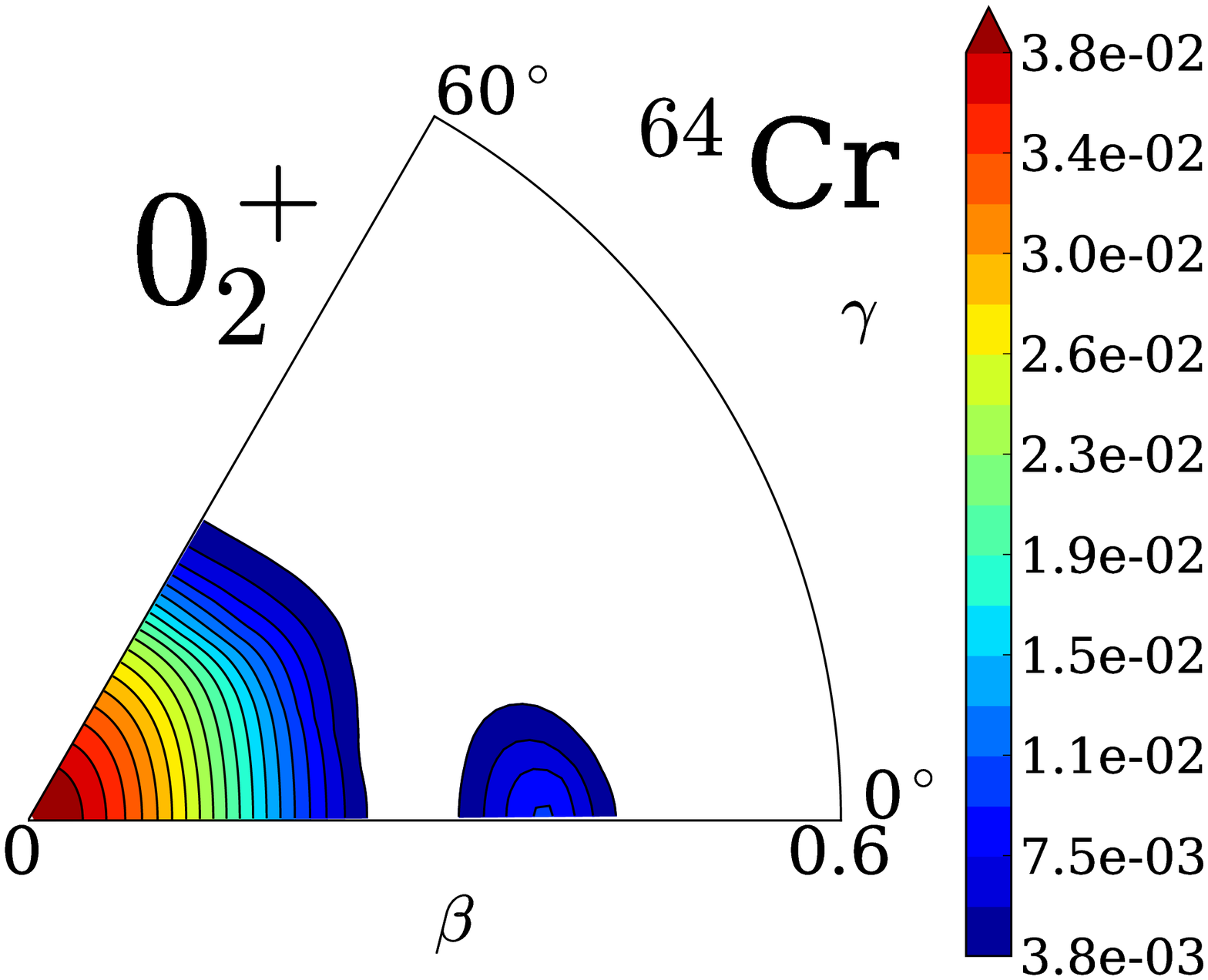}
\includegraphics[height=0.15\textwidth,keepaspectratio,clip,trim=60 50 160 50]{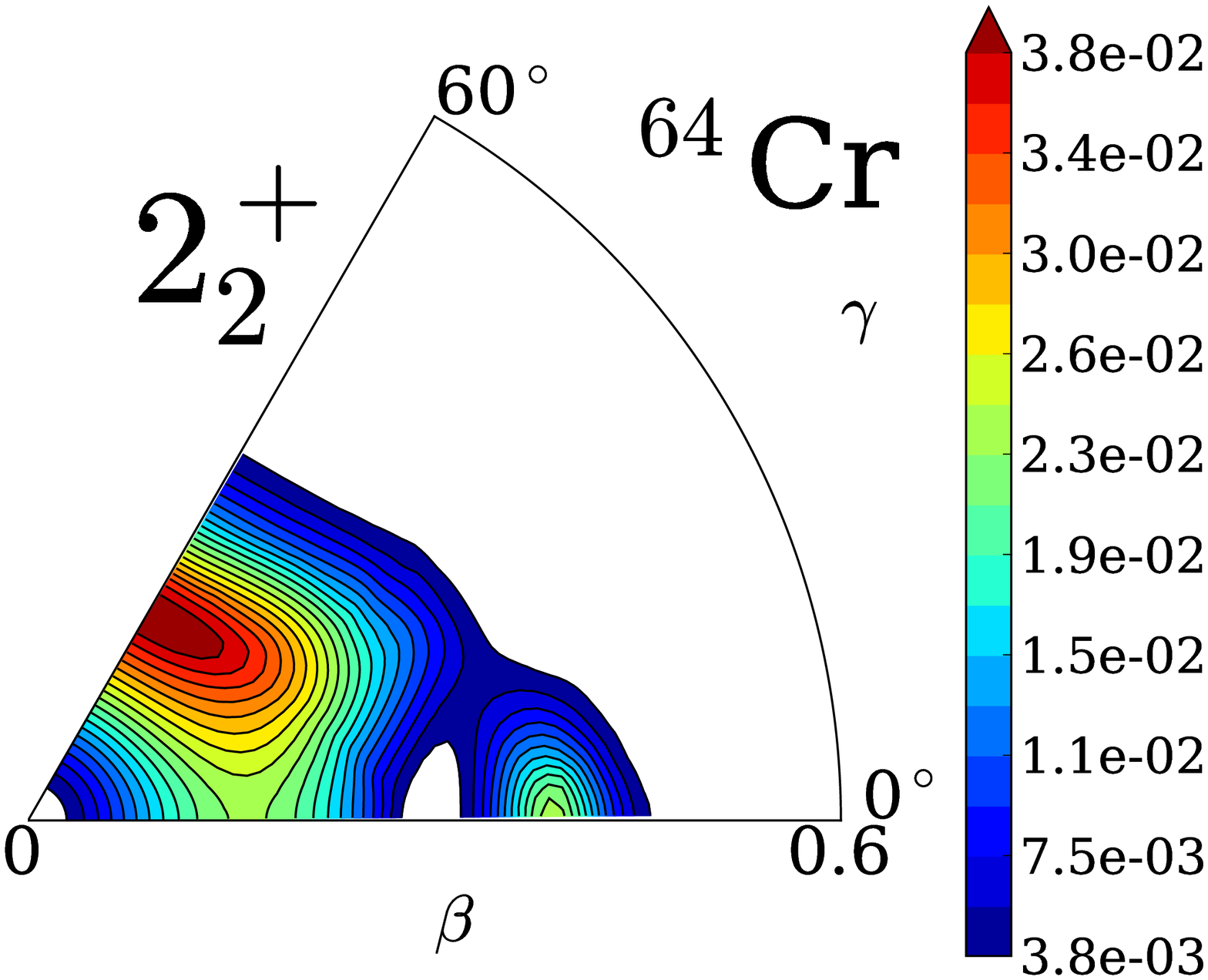}
\end{center}
\caption{(Color online) Same as Fig. \ref{non-weighted yrast wf} but for $0^+_2$ and  $2^+_2$ states. 
The contour lines are drawn at every twentieth part of the maximum value.}
\label{non-weighted yrare wf}
\end{figure*}



In Table I, we compare
the results for $^{64}$Cr with those for $^{66}$Fe.
Experimental data indicate that the quadrupole collectivity is stronger in $^{64}$Cr than in $^{66}$Fe:
the smaller $E(2_1^+)$ and the larger $R_{4/2}$ and $B(E2)$ values for $^{64}$Cr than those for $^{66}$Fe.
Our calculation reproduces these features quite well.


\begin{table}[b]
\caption{\label{tab:table1}%
Excitation energies of the $2_{1}^+$ state $E(2_1^+)$ in keV, 
the ratios $R_{4/2}$ of $E(4_1^+)$ to $E(2_1^+)$, 
and $B(E2;2_1^+ \rightarrow 0_1^+)$ in Weisskopf units for $^{64}$Cr
and  $^{66}$Fe.
Experimental data are taken from Ref.~\cite{Gade2010, Rother2011}.
}
\begin{ruledtabular}
\begin{tabular}{ccccccccccc}
           &             & Calc.       &        
           &             & Exp.       &         \\
           & $E(2_1^+)$  & $R_{4/2} $  & $B(E2)$   
           & $E(2_1^+)$  & $R_{4/2} $  & $B(E2)$ \\
\colrule
$^{64}$Cr  & 386 &  2.68   &  28.8    & 420 &  2.69  &    \\
$^{66}$Fe  & 685 &  2.29  &   15.5    & 573 &  2.47  & 21.0  \\
\end{tabular}
\end{ruledtabular}
\end{table}

We depict in Figs.~\ref{fig:2Dwf} and \ref{non-weighted yrast wf} 
the squared vibrational wave functions multiplied by $\beta^4$ for 
the $0_1^+,2_1^+$ and $4_1^+$ states in $^{58-66}$Cr and 
those without the $\beta^4$ factor for the $0_1^+$ and $2_1^+$ states
in $^{60}$Cr and $^{64}$Cr, respectively. 
The $\beta^4$ factor comes from the volume element 
and carries its dominant $\beta$ dependence 
(see Eqs.~(\ref{eq:normalization}) and (\ref{eq:metric})).  
The wave functions look quite different between the two cases. 
For instance, while the non-weighted $0_1^+$ wave function for $^{60}$Cr  
shown in Fig. \ref{non-weighted yrast wf} distributes around the spherical shape, 
the ${\beta^4}$ factor changes it to the arcuate pattern seen   
in Fig. \ref{fig:2Dwf}. 
In $^{58}$Cr and $^{60}$Cr, the $\beta^4$-weighted $0_1^+$ wave functions  
exhibit arcuate distributions around $\beta=0.2$ covering the entire $\gamma$ region.
Closely looking, one finds that,  
while the distribution for $^{58}$Cr is almost uniform in the $\gamma$ direction, 
it is slightly leaning to the prolate side for $^{60}$Cr. 

With increasing neutron number, 
the $0_1^+$ wave function localizes more and more on the prolate side, 
reflecting the deepening of the prolate minima 
(see the collective potential in Fig. \ref{fig:potential}). 
In $^{62}$Cr, the $0_1^+$ wave function still spreads over the entire $\gamma$ region, 
although it has a clear concentration on the prolate side.  
In $^{64}$Cr, one can see a distinct peak around the prolate potential minimum,
and the $0_1^+$ wave function is most localized at $^{64}$Cr.
The vibrational wave functions clearly indicate the shape transition 
from spherical to prolate along the isotopic chain.

For all these isotopes, one can see that the prolate peak grows with increasing 
angular momentum.
This is due to the enhancement of the moments of inertia on the prolate side 
we have already seen in Fig. \ref{fig:masses and gaps}.
Even in $^{58}$Cr whose ground state is rather spherical,
the  $2_1^+$ and $4_1^+$ states are weakly localized on the prolate side, 
which results in the finite spectroscopic quadrupole moment 
shown in Fig. \ref{fig:EnergyAndE2}. 
In $^{64}$Cr, 
the $2^+_1$ and $4^+_1$ wave functions are well localized on the prolate side, 
although the ground state wave function still exhibit 
non-negligible shape fluctuation in the $\gamma$ direction.
Due to the growth of localization of the wave functions, 
higher angular momentum states acquire more rotor-like character than the ground state. 
This fact can be quantified by calculating the ratio 
\begin{equation}
R_{6/4/2} \equiv (E(6_1^+)-E(2_1^+))/(E(4_1^+)-E(2_1^+)). 
\end{equation}
For instance, $R_{6/4/2}=2.42$ for $^{64}$Cr, 
which is fairly close to the rigid-rotor value 2.57, 
although the calculated $R_{4/2}$ is 2.67 
which is far from the rigid-rotor value 3.33. 
These results clearly indicate the importance of dynamical effects 
of rotation on the nuclear shape.

Lenzi et al.~\cite{Lenzi2010} evaluated the intrinsic quadrupole moments  
$Q_{\rm{int}}(I)$ for the yrast states of $^{62-66}$Cr using 
the spectroscopic quadrupole moments $Q(I)$ obtained in their shell-model calculation    
and the well-known relation between them for the axially symmetric deformation with $K$=0. 
The resulting $Q_{\rm{int}}(I)$ stay approximately constant 
along the yrast sequences in $^{62,64,66}$Cr, 
and they interpreted this as a fingerprint of a rigid rotor behavior.
We have evaluated $Q_{\rm{int}}(I)$  
in the same way as Lenzi et al. but using our calculated $Q(I)$.
The resulting $Q_{\rm{int}}(I)$ values are similar to those of Lenzi et al.
We feel, however, that this fact is insufficient to conclude that 
$^{62-66}$Cr are good rotors
because the $Q(I)$ are average values that are insensitive to the shape fluctuations. 
We need to examine the properties of non-yrast states which are sensitive to 
shape fluctuation effects. 
We also note that $Q(I)$ does not carry direct information about
the ground state, and that, according to our calculation, the ground-state vibrational wave function is
significantly different from those of the other yrast states with $I \neq 0$.

\subsection{Non-yrast states in $^{58-66}$Cr}

To understand the nature of quadrupole collectivity, 
it is important to examine the properties of the excited bands 
including their interband transitions to the ground band,      
although they have not been observed experimentally yet. 
As typical examples of the calculated results, we display 
in Figs. \ref{fig:60CrSpectra} and \ref{fig:64CrSpectra} 
the excitation spectra and the $B(E2)$ values of 
the low-lying states in $^{60}$Cr and $^{64}$Cr. 

Let us first discuss the $^{60}$Cr case. 
We notice that the calculated excitation spectrum exhibits 
some features characteristic of the 5D harmonic oscillator (HO) limit;
approximately equal level spacing in the ground band,  
approximate degeneracy of the $4_1^+$ and $2_2^+$ states,
nearly equal values of  $B(E2; 0_2^+ \rightarrow 2_1^+)$ and 
$B(E2; 4_1^+ \rightarrow 2_1^+)$, which are about twice of 
$B(E2; 2_1^+ \rightarrow 0_1^+)$, etc.
On the other hand, we also notice significant deviations from that limit. 
First, the $0_2^+$ state is considerably lower than the $4_1^+$ and $2_2^+$ states. 
Second, the $E2$ transitions forbidden in the HO limit are sizable; 
e.g., those from the $2_2^+$ state to the $4_1^+$ and $0_2^+$ states 
are fairly large. Third, the $B(E2)$ value from the $2_2^+$ state to the $2_1^+$ state 
is less than half of those from the $4_1^+$ and $0_2^+$ states.

To examine the origin of these anharmonicities, 
let us look into the vibrational wave functions of the excited states.
The $\beta^4$-weighted and non-weighted vibrational wave functions  
of the $0^+_2$ and $2^+_2$ states are displayed 
in Figs. \ref{fig:yrare_b4wf} and \ref{non-weighted yrare wf}, respectively.  
The $0_2^+$ wave function exhibits two components: 
one around the spherical shape and the other around $\beta=0.35$. 
Although it has a $\beta$-vibrational feature, i.e., a node in the $\beta$ direction,
it also exhibits a considerable deviation from the 5D HO limit, in which 
the deformed component concentrating on the prolate side would spread uniformly over the $\gamma$ direction.
We can see a deviation from the 5D HO limit also in the $2_2^+$ state.
The $\beta^4$-weighted $2_2^+$ wave function 
spreads from the prolate to the oblate sides. 
However, the non-weighted wave function reveals that 
it also has the $\beta$-vibrational component.  
In fact, this state is a superposition of 
the large-amplitude $\gamma$-vibrational component 
spreading over the entire $\gamma$ region   
and the $\beta$-vibrational component. 
In the 5D HO limit, the $2_3^+$ wave function has a node in the $\beta$ direction, 
while the $2_2^+$ wave function has no node.  
The calculated $2_2^+$ wave functions indicate significant mixing of these components. 
Thus, the low-lying states in $^{60}$Cr may be characterized as  
a quadrupole vibrational spectrum with strong anharmonicity.   

Let us proceed to the $^{64}$Cr case. 
We immediately notice some features different from $^{60}$Cr. 
First, the approximate degeneracy of the $4_1^+$ and $2_2^+$ states 
seen in $^{60}$Cr is completely lifted here.
Second, the $E2$ transitions within the ground band are 
much stronger than those in $^{60}$Cr.
Third, two low-lying excited bands appear: 
one consisting of the $0_2^+, 2_2^+$ and $4_2^+$ states (excited band I), 
and the other consisting of the $2_3^+, 3_1^+$, $4_3^+$ states 
(excited band II, the $4_3^+$ state not shown here is at 2.84 MeV).
One might be tempted to interpret these excited bands in terms of 
the conventional concept of the $\beta$ and $\gamma$ bands 
built on a well-deformed prolate ground state, 
but, in fact, they are markedly different from them. 
First, there is a strong mixing of the $\beta$- and $\gamma$-vibrational components,  
as seen from strong interband $E2$ transitions between the two excited bands.
Second, the calculated ratio of the excitation energies relative to $E(0_2^+)$, 
$(E(4_2^+)-E(0_2^+))/(E(2_2^+)-E(0_2^+))$, is 2.51,  
which is far from the rigid-body value.
Third, the $K$-mixing effects are strong, e.g., 
the $K=0$ $(K=2)$ components of the $2_2^+$ ($2_3^+$) 
and $4_2^+$ ($4_3^+$) wave functions are at most 60\%.  
To sum up, although the prolate deformation is appreciably developed 
in the low-lying states of $^{64}$Cr,  
the large-amplitude shape fluctuations play a dominant role and 
lead to the strong $\beta-\gamma$ coupling and 
significant interband $E2$ transitions.

\begin{figure}[h]
\begin{center}
\includegraphics[width=0.3\textwidth,keepaspectratio,clip,trim=0 0 0 0]{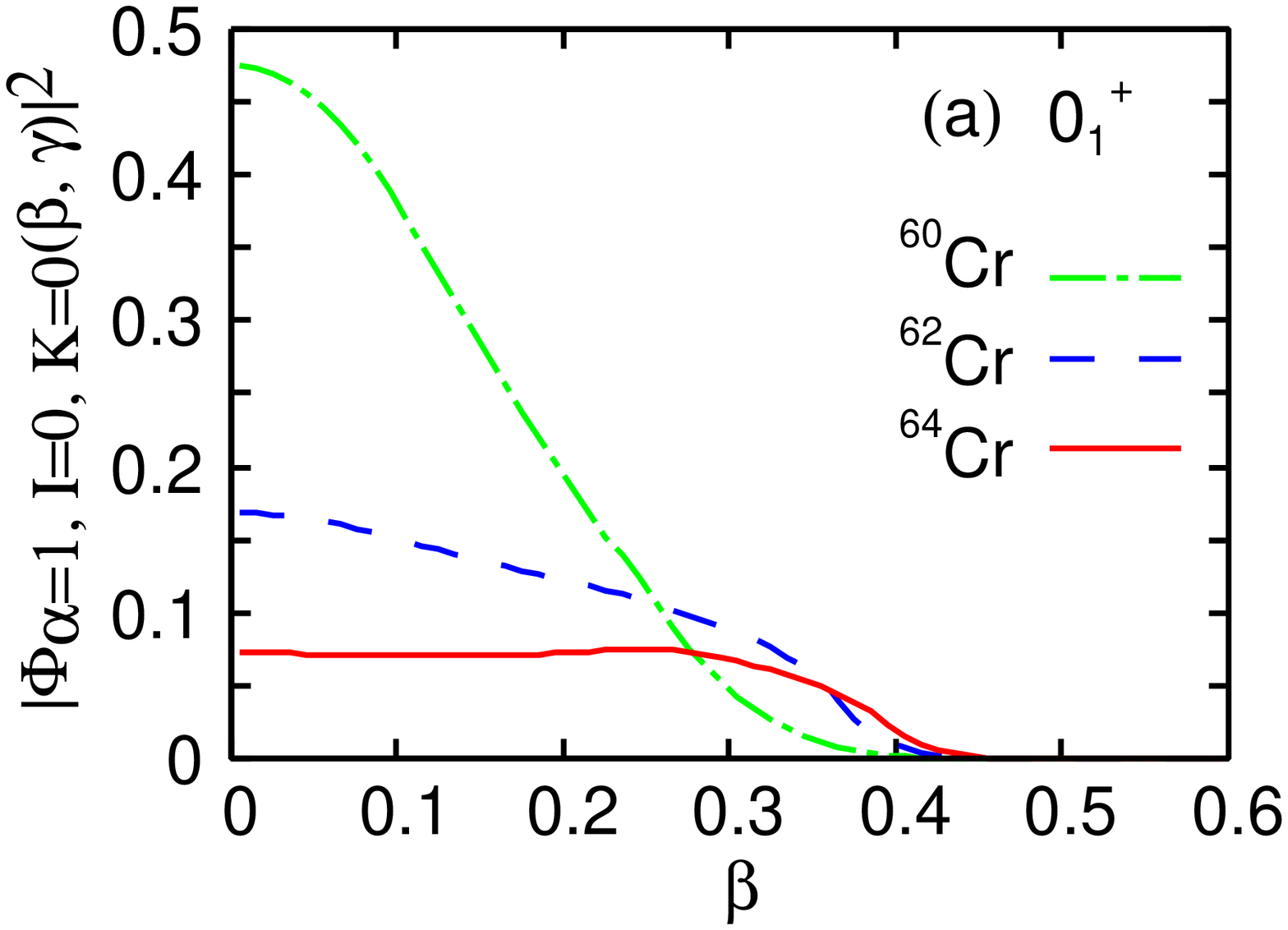} 
\includegraphics[width=0.3\textwidth,keepaspectratio,clip,trim=0 0 0 0]{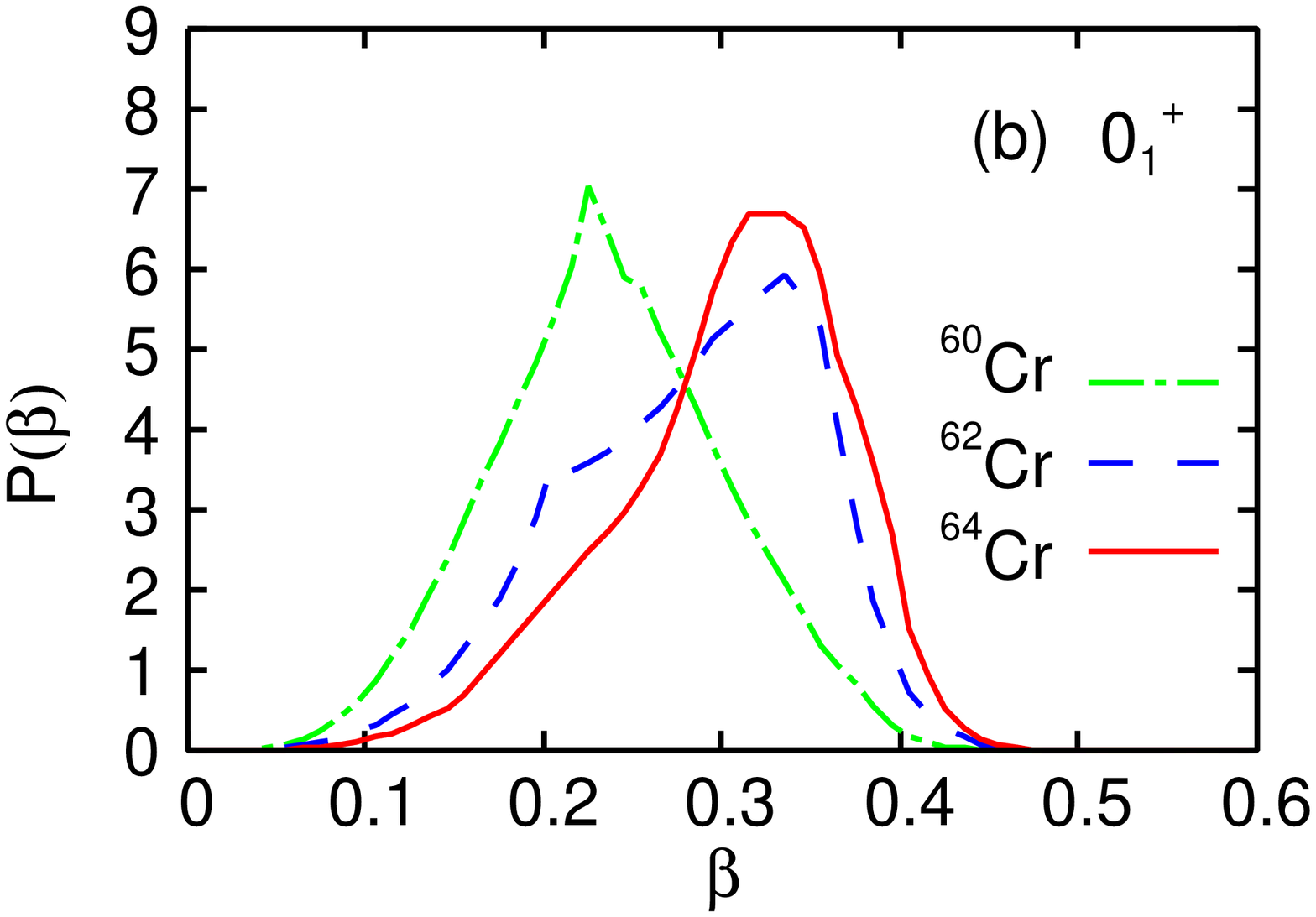} 
\includegraphics[width=0.3\textwidth,keepaspectratio,clip,trim=0 0 0 0]{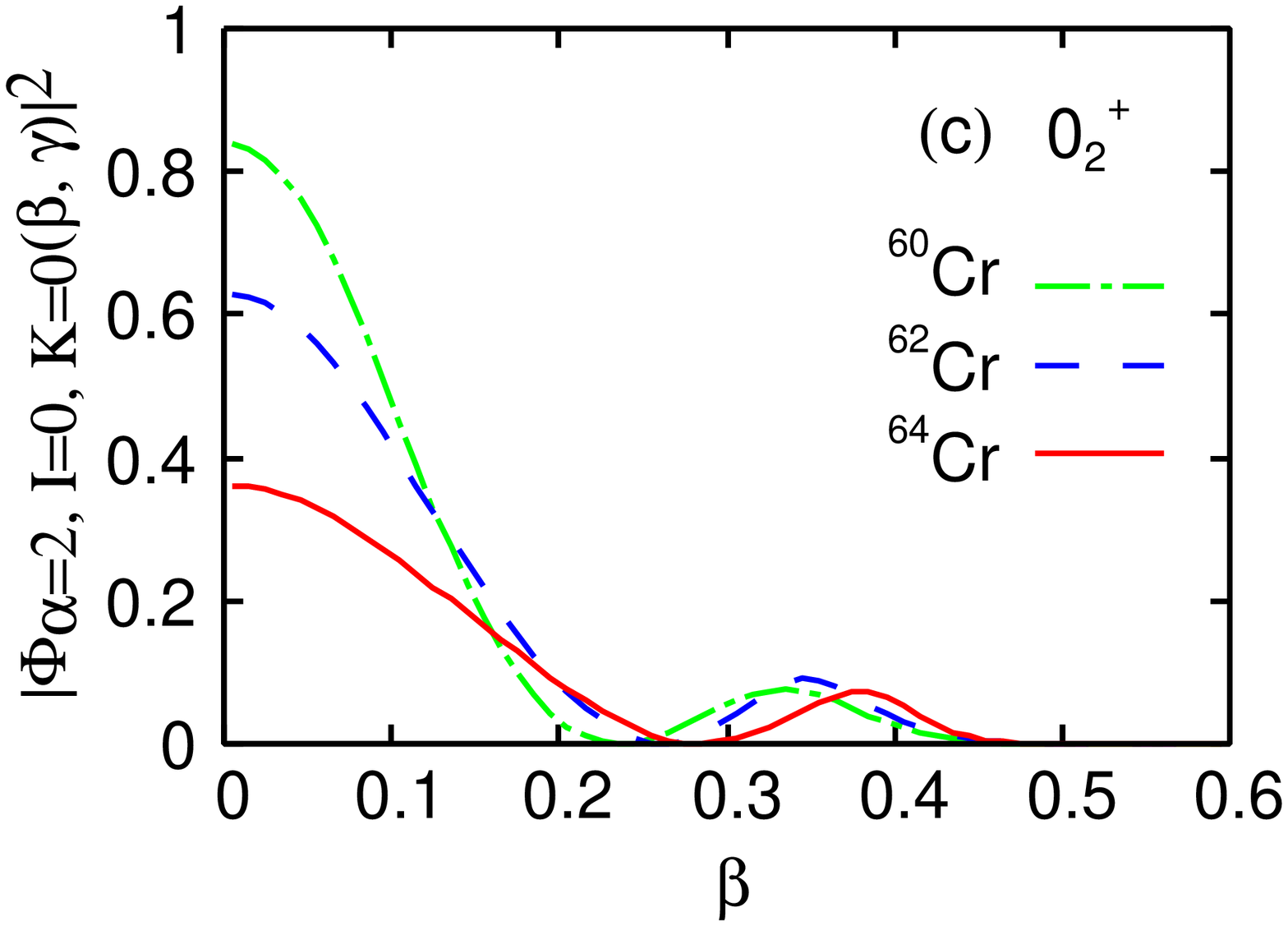} 
\includegraphics[width=0.3\textwidth,keepaspectratio,clip,trim=0 0 0 0]{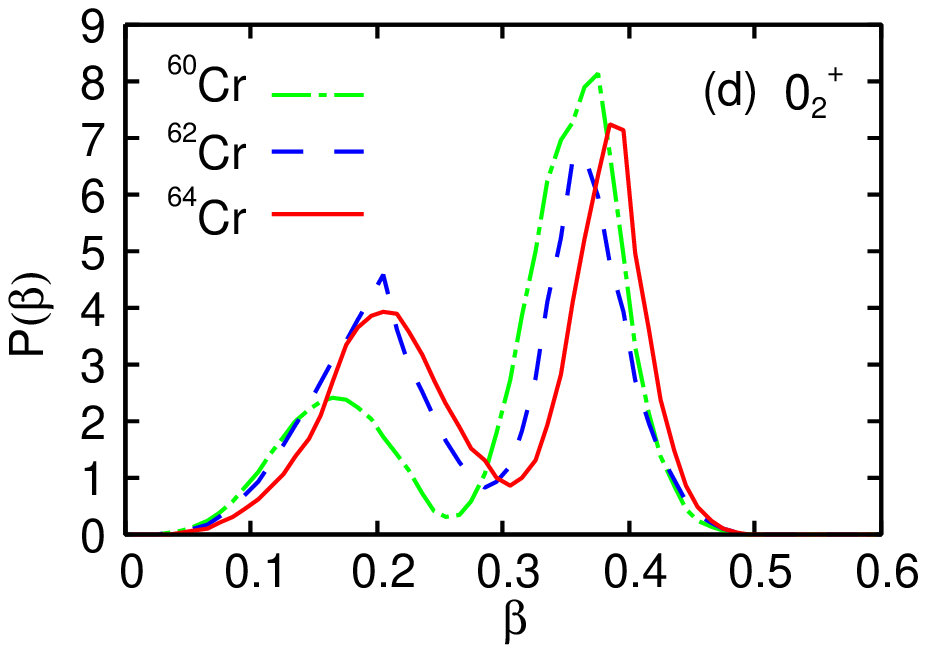} 
\end{center}
\caption{(Color online)
(a) Vibrational wave function squared 
$|\Phi_{\alpha=1, I=0,K=0}(\beta,\gamma=0.5^{\circ})|^2$ 
of the ground states in $^{60-64}$Cr.
(b) Probability densities integrated over $\gamma$, 
$P(\beta)=\int d\gamma |\Phi_{\alpha=1,
I=0,K=0}(\beta,\gamma)|^2|G(\beta,\gamma)|^{1/2}$. 
(c) Same as (a) but for the $0_2^+$ states.
(d) Same as (b) but for the $0_2^+$ states.
}
\label{fig:wfandPrb}
\end{figure}

In Fig.~\ref{fig:wfandPrb}, 
we plot the vibrational wave functions at $\gamma=0.5^{\circ}$
and the probability density $P(\beta)$ of finding a shape 
with a specific value of $\beta$ 
for the ground and excited $0^+$ states in $^{60-64}$Cr.
Note that the probability density vanishes at the spherical shape
because of the $\beta^4$ factor in the volume element.
It is seen that, 
while the ground-state wave function for $^{60}$Cr distributes 
around the spherical shape,  
those for $^{62}$Cr and $^{64}$Cr extend from the spherical to deformed regions 
with $\beta \simeq 0.4$ (see Fig.~\ref{fig:wfandPrb}(a)). 
Accordingly, the peak of the probability distribution 
moves toward larger $\beta$ in going from $^{60}$Cr to $^{64}$Cr 
(see Fig.~\ref{fig:wfandPrb}(b)).  
Concerning the excited $0^+$ states, 
their vibrational wave functions exhibit two peaks; 
a large peak at the spherical shape and a small peak at a prolate shape 
(see Fig.~\ref{fig:wfandPrb}(c)). 
In the probability distribution displayed in Fig.~\ref{fig:wfandPrb}(d), 
the spherical peaks moves to the $\beta \approx 0.2$ region   
and the peaks at $\beta=0.35-0.4$ in turn become prominent.

\begin{figure}[h]
\includegraphics[width=0.3\textwidth,keepaspectratio,clip]{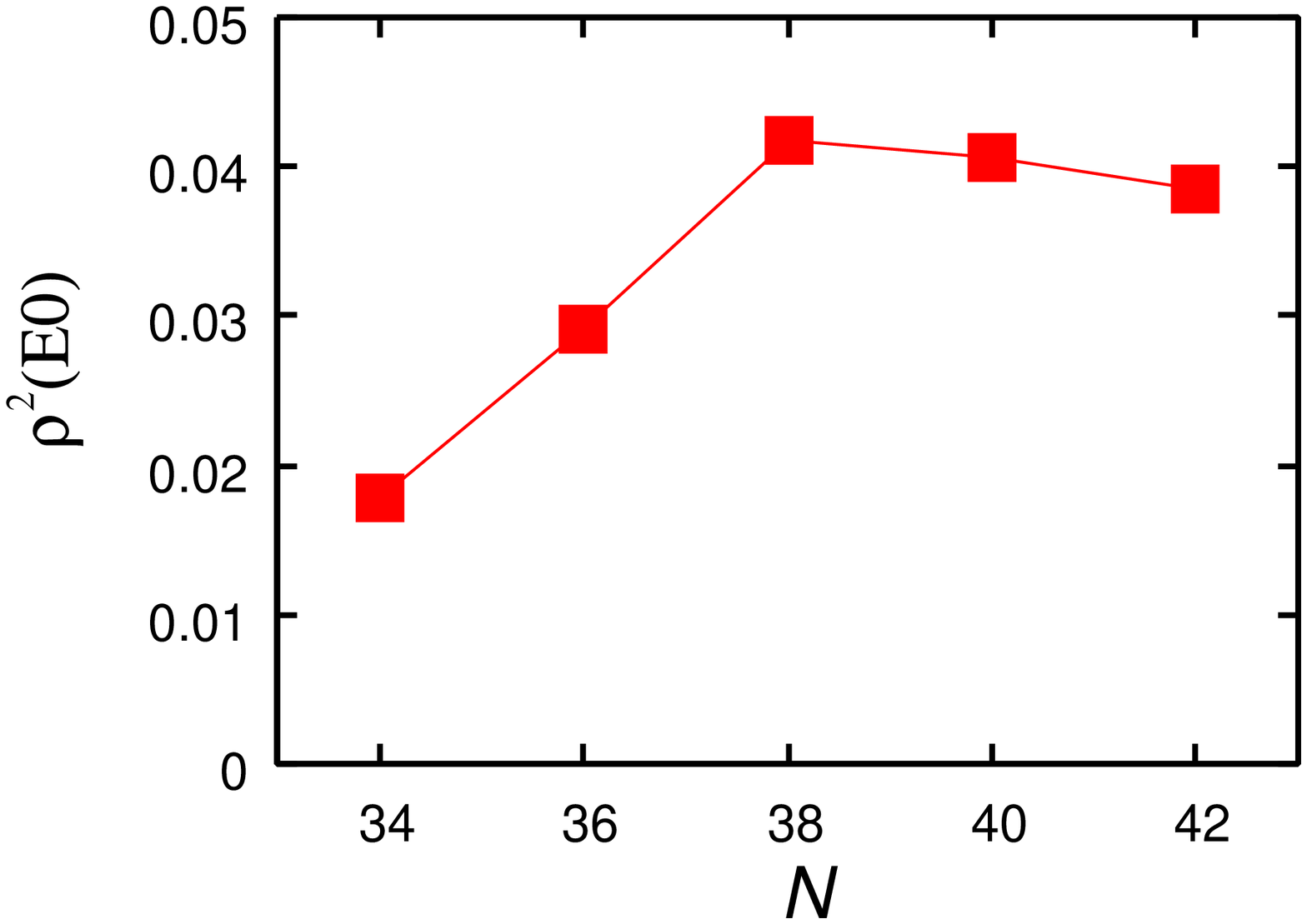} 
\caption{(Color online)
$E0$ transition strengths $\rho^2(E0; 0^+_2 \to 0^+_1)$ calculated for $^{58-66}$Cr.
}
\label{E0 transitions}
\end{figure}

The above results indicate that 
large-amplitude shape fluctuations play an important role 
both in the ground and excited $0^+$ states. 
The growth of the shape fluctuations leads to an enhancement of the 
calculated $E0$ transition strengths $\rho^2(E0; 0^+_2 \to 0^+_1)$ 
in going from $^{58}$Cr to $^{62-66}$Cr,  
as displayed in Fig.~\ref{E0 transitions}.

\subsection{Comparison with the (1+2)D calculation}

To examine the role of dynamical shape fluctuations 
breaking the axial symmetry,
we compare the above result of the (2+3)D calculation 
with the (1+2)D calculation where the axial symmetry is imposed.  
The procedure of the latter calculation is summarized in Appendix.  

The obtained excitation energies, $E2$ transitions 
and spectroscopic quadrupole moments are shown in Fig.~\ref{fig:EnergyAndE2}. 
Obviously, the results of the (1+2)D calculation are more rotational 
than the (2+3)D results for all the isotopes; 
the $R_{4/2}$ ratios, the $B(E2)$ values and the magnitude of $Q(2_1^+)$ 
are larger while the excitation energies are smaller.  
The difference is most remarkable in the $R_{4/2}$ ratios.
This implies that the neglect of the triaxial degree o    f freedom
leads to spectra much closer to the rotational limit.
In fact, the (1+2)D calculation considerably overestimates 
the experimental $R_{4/2}$ values for $^{60-64}$Cr. 
The reason of this overestimation is easily understood in the following way. 
Let us first consider the $\beta$-rigid limit  
where $\beta$ is fixed at a certain  non-zero value, $\beta=\beta_0$. 
In this limit, the (1+2)D calculation obviously gives 
the axial rigid-rotor value 3.33 for $R_{4/2}$. 
On the other hand, in the (2+3)D calculation,  
the shape fluctuation in the $\gamma$ degree of freedom is more or less always present.
If the collective potential is flat in the $\gamma$ direction, 
as in the Wilets-Jean model \cite{Wilets1956},   
$R_{4/2}$ will be 2.50. 
In general, $R_{4/2}$ will take a value between 2.50 and 3.33 in the $\beta$-rigid limit,  
depending on the softness of the collective potential 
toward the $\gamma$ direction (see Figs. 3 and 4 in Ref. \cite{Sato2010}). 
In reality, as a quantum system, the nuclear shape always 
fluctuates in both the $\beta$ and $\gamma$ directions. 
The experimental data that $R_{4/2} \approx 2.68$  
even in $^{64}$Cr clearly indicate that such shape fluctuations cannot be ignored. 
In the $^{60}$Cr case, 
the collective potential is quite soft in the $\beta$ direction 
and the (2+3)D calculation yields  $R_{4/2} \approx 2.17$.  
It is noteworthy that, even in such a situation, 
the $2_1^+$ and $4^+_1$ vibrational wave functions 
exhibit a localization about a prolate shape  (see Fig.~\ref{fig:2Dwf}).  
This is due to the dynamical effect minimizing  
the rotational energy for a given angular momentum.

\subsection{Similarities and differences with the Mg isotopes around $N=20$}

In Ref.~\cite{Lenzi2010}, Lenzi et al. discussed similarities of the 
neutron-rich Cr isotopes near $N=40$ and 
the neutron-rich Mg isotopes around $N=20$.  
Indeed, we have also found such similarities in our calculation. 
First of all, the growth of quadrupole collectivity in going from 
$^{60}$Cr to $^{64}$Cr is similar to that from $^{30}$Mg to $^{32}$Mg. 
In Fig. \ref{non-weighted yrast wf}, while the ground state wave function in $^{60}$Cr distributes 
around the spherical shape, they are considerably extended to the prolately 
deformed region in $^{64}$Cr. 
The $2_1^+$ wave function has a peak on the prolate side in $^{60}$Cr 
and it shifts to larger $\beta$ in $^{64}$Cr.
These features are similar to those seen in going from $^{30}$Mg to $^{32}$Mg 
in our calculation \cite{Hinohara2011}.  
Concerning the excited $0_2^+$ states in $^{60}$Cr and $^{64}$Cr, 
as shown in Figs.~\ref{fig:wfandPrb}(c) and \ref{fig:wfandPrb}(d),
both vibrational wave functions exhibit a two-hump structure. 
Similar two-hump structures of the excited $0^+$ states have been 
obtained also in our calculation for $^{30}$Mg and $^{32}$Mg.

On the other hand, we have also found significant differences 
between the $^{64}$Cr region and the $^{32}$Mg region. 
First of all, the $K$ mixing is strong in the excited bands  
in the Cr isotopes, whereas it is weak in the Mg region.
The shape fluctuations toward the $\gamma$ direction
and the effect of the $\beta-\gamma$ coupling
are larger in the Cr isotopes than in Mg.
This can be clearly seen,
for instance, in the $2_2^+$ wave functions 
displayed in Figs.~\ref{fig:yrare_b4wf} and \ref{non-weighted yrare wf}.

\section{Conclusions}
\label{Conclusions}

In this paper, we have investigated the nature of the quadrupole collectivity 
in the low-lying states of neutron-rich Cr isotopes $^{58-66}$Cr 
by solving the 5D collective Schr\"odinger equation.
The vibrational and rotational inertial functions and the collective potential 
in the 5D quadrupole collective Hamiltonian are microscopically derived 
with use of the CHFB+LQRPA method.
The calculated inertial functions include the contributions from 
the time-odd components of the moving mean field.
The results of calculation are in good agreement with the available experimental data.
The prolate deformation remarkably develops along the isotopic chain from $N=36$ to 40. 
It is not appropriate, however, to characterize the low-lying state of 
Cr isotopes around $^{64}$Cr in terms of the prolate rigid-rotor model: 
the excitation spectra are still transitional  
and the large-amplitude shape fluctuations dominate in their low-lying states.
The calculated excited bands exhibit strong couplings between 
the $\beta$ and $\gamma$ vibrational degrees of freedom.  
For close examination of the nature of quadrupole collectivity 
in these nuclei, experimental exploration of their excited bands is strongly desired. 

\section*{Acknowledgments}
The numerical calculations were carried out
on SR16000 at Yukawa Institute for Theoretical Physics in Kyoto University
and RIKEN Integrated Cluster of Clusters (RICC) facility.
This work is supported by
KAKENHI (Nos. 21340073, 23540294, and 23740223).

\appendix
\section*{Appendix}
\label{Appendix}

Here we summarize the procedure of the (1+2)D calculation 
to which the (2+3)D calculation reduces when the axial symmetry 
is imposed on the intrinsic shape. 
The dynamical variables in this reduced model are $\beta$ and 
two Euler angles $(\theta_1, \theta_2)$, 
which describe the magnitude of the quadrupole deformation and  
the rotational motion perpendicular to the symmetry axis, respectively. 
The classical collective Hamiltonian of the (1+2)D model is given by    
\begin{equation}
\mathcal{H}_{\mathrm{coll}}=\dfrac{1}{2}\mathcal{M}_{\beta}(\beta)\dot{\beta}^{2}
+\dfrac{1}{2}\sum_{k=1,2}\mathcal{J}(\beta)\omega_{k}^2+V(\beta), 
\label{cla_hami}
\end{equation}
where $\omega_{i}$ are rotational frequencies related to the time derivative 
of the Euler angles.
We quantize it according to the Pauli prescription, and then
the collective Schr\"odinger equation reads
\begin{align}
\left[\hat T_{\rm vib } + \sum_{k=1,2}\frac{\hat I_k^2}{2\mathcal{J}(\beta)} +V \right]
\Psi_{\alpha IM}(\beta,\theta_{1},\theta_{2})= \notag \\
E_{\alpha I}\Psi_{\alpha IM}(\beta,\theta_{1},\theta_{2}),
\label{hami}
\end{align}
with
\begin{align}
\hat{T}_{\mathrm{vib}}=&
-\dfrac{1}{2M_{\beta}(\beta)}\dfrac{\partial^{2}}{\partial \beta^{2}} \notag \\
& +\dfrac{1}{2M_{\beta}(\beta)}
\left[
\dfrac{1}{2M_{\beta}(\beta)}\dfrac{\partial M_{\beta}(\beta)}{\partial \beta}
-\dfrac{1}{\mathcal{J}(\beta)}\dfrac{\partial \mathcal{J}(\beta)}{\partial \beta}
\right]
\dfrac{\partial}{\partial \beta}.
\end{align}
The collective wave function can be written as a product of    
the vibrational and rotational wave functions:
\begin{equation}
\Psi_{\alpha IM}(\beta,\Omega)=\Phi_{\alpha I}(\beta)\langle \Omega|IM0\rangle.
\end{equation}
The collective Schr\"odinger equation in the intrinsic frame is 
\begin{equation}
\left \{ 
\hat{T}_{\mathrm{vib}} + \dfrac{I(I+1)}{2\mathcal{J}(\beta)}+V(\beta)
\right \} 
\Phi_{\alpha I}(\beta)=E_{\alpha I}\Phi_{\alpha I}(\beta), 
\label{coll_Sch}
\end{equation}
and the ortho-normalization condition is given by 
\begin{equation}
\int d\beta \Phi^{*}_{\alpha I}(\beta)\Phi_{\alpha^{\prime} I}(\beta)|G(\beta)|^{1/2}
=\delta_{\alpha \alpha^{\prime}}
\end{equation}
with $|G(\beta)|=M_{\beta}(\beta)\mathcal{J}^{2}(\beta)$. 

For the (1+2)D calculation with the P+Q model presented in subsection 3.3, 
we have used the same parameters as used in the (2+3)D calculation.
In Ref.~\cite{Yoshida2011}, 
the (1+2)D collective Schr\"odinger equation (\ref{coll_Sch}) was solved 
in the range $-0.4 \leq \beta \leq 0.6$ without respecting 
the boundary condition that the vibrational wave functions 
should satisfy at the spherical point (see Ref.~\cite{Kumar1967}). 
In this paper, we have solved the collective Schr\"odinger equation 
in the range $0 \leq \beta  \leq 0.6$
respecting the boundary condition at $\beta=0$. 
A detailed account on this point will be given in a future publication. 


%

\end{document}